\documentclass[12pt]{article}

\ifx\pdfoutput\undefined
\usepackage[dvips,bookmarks]{hyperref}
\else
\usepackage{hyperref}
\fi
\hypersetup{colorlinks=false,bookmarksopen,bookmarksnumbered,citecolor=blue,
   pdfstartview=FitH}

\usepackage[pdftex]{graphicx}	
\usepackage{latexsym}

\usepackage{amssymb,amsfonts,amsmath}
\usepackage{graphicx} 
\usepackage{indentfirst}

 \usepackage{bbm}

\parskip=\medskipamount

\newcommand {\cC}{{\cal C}}
\newcommand {\cD}{{\cal D}}
\newcommand {\cE}{{\cal E}}

\newcommand {\cH}{{\cal H}}

\newcommand {\cJ}{{\cal J}}
\newcommand {\cK}{{\cal K}}
\newcommand {\cL}{{\cal L}}
\newcommand {\cM}{{\cal M}}
\newcommand {\cN}{{\cal N}}
\newcommand {\cO}{{\cal O}}

\newcommand {\cR}{{\cal R}}
\newcommand {\cS}{{\cal S}}
\newcommand {\cT}{{\cal T}}
\newcommand {\cU}{{\cal U}}
\newcommand {\cV}{{\cal V}}

\newcommand {\cX}{{\cal X}}

\newcommand {\cZ}{{\cal Z}}

\newcommand{\bL}{{\bf L}}

\newcommand{\bR}{{\bf R}}

\def\a{\alpha}

\def\b{\beta}
\def\c{\chi}
\def\d{\delta}
\def\e{\epsilon}
\def\f{\phi}
\def\g{\gamma}

\def\j{\psi}

\def\l{\lambda}
\def\m{\mu}
\def\n{\nu}
\def\o{\omega}
\def\p{\pi}
\def\q{\theta}
\def\r{\rho}
\def\s{\sigma}
\def\t{\tau}

\def\x{\xi}
\def\z{\zeta}
\def\D{\Delta}
\def\F{\Phi}
\def\J{\Psi}
\def\L{\Lambda}
\def\O{\Omega}

\def\S{\Sigma}
\def\U{\Upsilon}
\def\X{\Xi}

\def\rd{{\rm d}}
\def\ri{{\rm i}}
\def\re{{\rm e}}

\def\rb{{\rm b}}
\def\ra{{\rm a}}
\def\rc{{\rm c}}

\newcommand{\ve}{\varepsilon}                            
\newcommand{\cDB}{{\bar\cD}}                            

\newcommand{\ab}{{\a\b}}

\newcommand{\pa}{\partial}                           
\newcommand{\hf}{\frac12}

\newcommand{\vf}{\varphi}

\newcommand{\be}{\begin{equation}}
\newcommand{\ee}{\end{equation}}
\newcommand{\bea}{\begin{eqnarray}}
\newcommand{\eea}{\end{eqnarray}}
\newcommand{\non}{\nonumber}
\newcommand{\1}{{\underline{1}}}
\newcommand{\2}{{\underline{2}}}

\def\dt#1{{\buildrel {\hbox{\LARGE .}} \over {#1}}}    

\newcommand{\bm}[1]{\mbox{\boldmath$#1$}}

\def\double #1{#1{\hbox{\kern-2pt $#1$}}}

\newcommand{\qb}{{\bar{\theta}}}

\newif\ifdtup

\newcommand{\bsubeq}{\begin{subequations}}
\newcommand{\esubeq}{\end{subequations}}

\newcommand{\bai}{{\bar i}}
\newcommand{\baj}{{\bar j}}
\newcommand{\bak}{{\bar k}}
\newcommand{\bal}{{\bar l}}

\newcommand{\bau}{{\bar 1}}
\newcommand{\bad}{{\bar 2}}

\newcommand{\rL}{{\rm L}}
\newcommand{\rR}{{\rm R}}

\newcommand{\mub}{{{\bar{\mu}}}}

\newcommand{\eol}{\notag \\}
\newcommand{\veps}{\varepsilon}

\numberwithin{equation}{section}
\newcommand{\tsD}{{\nabla}}
\newcommand{\HC}{{\mathrm{c.c.}}}

\newcommand{\nrho}{{\veps}}

\begin{document}
\begin{titlepage}
\begin{flushright}
Nikhef-2012-020\\
October, 2012\\
\end{flushright}
\vspace{5mm}

\begin{center}
{\Large \bf 
Nonlinear sigma models with  AdS supersymmetry\\
 in three dimensions
}\\ 
\end{center}

\begin{center}

{\bf
Daniel Butter${}^{ab}$, Sergei M. Kuzenko${}^{a}$
and
Gabriele Tartaglino-Mazzucchelli${}^{a}$
} \\
\vspace{5mm}

\footnotesize{
${}^{a}${\it School of Physics M013, The University of Western Australia\\
35 Stirling Highway, Crawley W.A. 6009, Australia}}  
~\\
\texttt{gabriele.tartaglino-mazzucchelli@uwa.edu.au}
\vspace{2mm}

\footnotesize{
${}^{b}${\it Nikhef Theory Group, Science Park 105, 1098 XG Amsterdam, The Netherlands}
}\\
\texttt{dbutter@nikhef.nl}
\vspace{2mm}

\end{center}

\begin{abstract}
\baselineskip=14pt
In three-dimensional anti-de Sitter (AdS) space, there exist several realizations 
of $\cN$-extended supersymmetry, 
which are traditionally labelled by two non-negative integers $ p\geq q$ such that $p+q=\cN$. 
Different choices of $p$ and $q$, with $\cN$ fixed, prove to lead to different restrictions on 
the target space geometry of supersymmetric nonlinear $\s$-models. We classify  all possible types
of hyperk\"ahler target spaces 
for the cases $\cN=3$  and $\cN = 4 $ by making use of two different realizations for the most general  
$(p,q)$ supersymmetric $\s$-models: 
(i)  off-shell formulations in terms  of  $\cN=3$ and $\cN=4$ projective supermultiplets; and
(ii) on-shell formulations in terms of covariantly chiral scalar superfields in (2,0) AdS superspace.
Depending on the type of $\cN=3,\, 4$ AdS supersymmetry, nonlinear $\s$-models can support 
one of the following target space geometries: 
(i) hyperk\"ahler cones; (ii) non-compact hyperk\"ahler manifolds with a U(1) isometry group 
which acts non-trivially  on the two-sphere of complex structures;  
(iii) arbitrary hyperk\"ahler manifolds including compact ones.
The option (iii) is realized only in the case of 
critical (4,0) AdS supersymmetry.

As an application of the (4,0) AdS techniques developed,
we also construct the most general nonlinear $\s$-model in Minkowski space with  
a non-centrally extended $\cN=4$ Poincar\'e supersymmetry.
Its target space is a hyperk\"ahler cone (which is characteristic of $\cN=4$ superconformal 
$\s$-models), but the $\s$-model is massive. The Lagrangian includes a positive potential 
constructed in terms of the homothetic conformal Killing vector the target space is endowed with. 
This mechanism of mass generation differs from the standard one which corresponds to 
a $\s$-model with the ordinary  $\cN=4$ Poincar\'e supersymmetry 
and which makes use of a tri-holomorphic Killing vector. 
\end{abstract}

\vfill

\vfill
\end{titlepage}

\newpage
\renewcommand{\thefootnote}{\arabic{footnote}}
\setcounter{footnote}{0}

\tableofcontents{}
\vspace{1cm}
\bigskip\hrule


\section{Introduction}
\setcounter{equation}{0}

In maximally symmetric spacetimes of dimension $3\leq d\leq 5$, 
rigid supersymmetry with six ($d=3$)
or eight  ($d=3, 4, 5$) supercharges requires hyperk\"ahler geometry for
the target spaces of nonlinear $\s$-models (more exotic 
geometries can originate in two spacetime dimensions \cite{GHR}).
Arbitrary hyperk\"ahler target spaces are allowed by ordinary Poincar\'e supersymmetry
corresponding to Minkowski spacetime \cite{A-GF}.\footnote{In three dimensions, 
there exist non-central extensions  of the $\cN \geq 4$ Poincar\'e 
superalgebras \cite{Nahm} which have no higher-dimensional analogs. These superalgebras 
have appeared in various string- and field-theoretic applications, 
see \cite{Blau:2001ne,Itzhaki:2005tu,LM,Gomis:2008cv,Hosomichi:2008qk,Bergshoeff:2008ta} 
and references therein. 
The non-central extension of the $\cN=4$ Poincar\'e supergroup
originates geometrically as the isometry group of the  deformed $\cN=4$ Minkowski superspace
introduced in  \cite{KLT-M12}. 
As will be shown below,  this supersymmetry type requires the target space of any $\s$-model  
to be a hyperk\"ahler cone.} 
In particular, there exists a one-to-one correspondence between massless
$\cN=2$ supersymmetric $\s$-models in four dimensions (4D) 
and hyperk\"ahler manifolds, see \cite{HKLR} for a nice derivation of this result.
The situation is completely different in the cases of 4D $\cN=2$ and 5D $\cN=1$ 
anti-de Sitter (AdS) supersymmetries, which enforce nontrivial restrictions 
on the hyperk\"ahler target spaces of supersymmetric $\s$-models   
\cite{BKsigma1,BKsigma2,BaggerXiong,BaggerLi}.
Since these restrictions are identical in  $\rm AdS_4$ \cite{BKsigma1,BKsigma2} 
and  $\rm AdS_5$ \cite{BaggerXiong,BaggerLi}, 
it suffices to discuss the former case only. 
The most general $\cN=2$ supersymmetric 
 $\s$-model in $\rm AdS_4$ was constructed in \cite{BKsigma1,BKsigma2}
 using a formulation in terms of $\cN=1$ covariantly chiral superfields.
As demonstrated in \cite{BKsigma1,BKsigma2}, 
the target space of such a $\s$-model is  a non-compact hyperk\"ahler manifold 
possessing a special Killing vector field which generates an
SO(2) group of rotations on the two-sphere of complex structures and necessarily
leaves one of them, $\mathbb J$,  invariant. 
This implies that each of the complex structures that are orthogonal 
to $\mathbb J$ is characterized by  an exact K\"ahler two-form, and therefore the target space 
is non-compact. The existence of such hyperk\"ahler spaces was pointed out
twenty-five years ago by Hitchin {\it et al.} \cite{HitchinKLR} without 
addressing their physical significance for supersymmetric
 $\s$-models in AdS.

${}$From the point of view of supersymmetry, the space $\rm AdS_3$ 
is rather special as compared with $\rm AdS_4$ and $\rm AdS_5$. 
Here $\cN$-extended AdS supersymmetry exists 
in several incarnations which are labelled by two non-negative integers $ p\geq q$ such that 
$p+q=\cN$. 
The reason is that the 3D anti-de Sitter  group is reducible, 
$$\rm SO_0(2,2) \cong \Big( SL(2, {\mathbb R}) \times SL( 2, {\mathbb R}) \Big)/{\mathbb Z}_2~,$$ 
and so are its supersymmetric extensions,  
${\rm OSp} (p|2; {\mathbb R} ) \times  {\rm OSp} (q|2; {\mathbb R} )$, 
which are known as $(p,q)$ AdS supergroups. 
This implies that  there are several versions of $\cN$-extended AdS supergravity 
\cite{AT},  known as the  $(p,q)$ AdS supergravity theories. These theories can naturally be 
described in superspace using the off-shell formulation for $\cN$-extended conformal supergravity 
\cite{HIPT,KLT-M11}. The supergeometry of $\cN$-extended conformal supergravity 
developed in \cite{KLT-M11}
allows maximally symmetric backgrounds with non-zero covariantly 
constant curvature, which were classified in  \cite{KLT-M12} and called the $(p,q)$ AdS superspaces.
These superspaces have  $\rm AdS_3$ as the  bosonic body, and their  isometry groups 
are generated by the $(p,q) $ AdS superalgebras \cite{KLT-M12}.\footnote{A general
setting to determine the (conformal) isometries of a given curved superspace background 
in  off-shell supergravity was developed long ago in \cite{BK}, with the goal of constructing 
rigid supersymmetric theories in curved superspace. Later on,
it was applied to various supersymmetric backgrounds in five, four and three dimensions
\cite{KT-M,KT-M-4D-2008,KT-M11,KLT-M12}. 
More recently, an equivalent construction in the component  approach, 
albeit  specifically restricted  to supersymmetry transformations, 
has gained considerable prominence 
\cite{FS,Jia:2011hw,Samtleben:2012gy,Klare:2012gn,DFS,KMTZ} .}
It turns out that different choices of $p$ and $q$, for fixed $\cN =p+q$, 
lead to supersymmetric field theories with drastically different properties.
This has been demonstrated for the cases $\cN=2$ and $\cN=3$ 
by studying the nonlinear $\s$-models with (2,0) and (1,1) 
AdS supersymmetry  \cite{KT-M11} and with (3,0) and (2,1) AdS
supersymmetry  \cite{KLT-M12} respectively. 
The main goal of the present paper is to provide a thorough 
study of the last nontrivial case $\cN=4$ allowing nonlinear $\s$-models of sufficiently general 
functional form.\footnote{For extended supersymmetry with $\cN>4$ the target space geometries
are highly restricted. In particular, the target spaces of superconformal $\s$-models with $\cN>4$ 
are flat \cite{Cecotti:2010dg}. }  
Specifically, we construct the most general 
nonlinear $\s$-models with (4,0), (3,1) and (2,2) AdS supersymmetries. 
We also provide a formulation in terms of $\cN=2$ chiral superfields for the most general 
nonlinear $\s$-models with (3,0) and (2,1) AdS supersymmetry. 
Our analysis is based on the use of two different realizations for 
$\cN=3$ and $\cN=4$ supersymmetric nonlinear $\s$-models in $\rm AdS_3$:
(i) off-shell formulations in terms  of  $\cN=3$ and $\cN=4$ projective 
supermultiplets\footnote{The projective-superspace techniques are ideal for 
$\s$-model applications in various dimensions. Originally, the projective superspace approach 
\cite{KLR,LR1,LR2}  (see also \cite{KLT}) was introduced as a method to construct 
off-shell 4D $\cN=2$ super-Poincar\'e invariant theories in the superspace 
${\mathbb R}^{4|8} \times {\mathbb C}P^1$ pioneered by Rosly \cite{Rosly}
(the same superspace is used within the harmonic superspace approach \cite{GIKOS,GIOS}). 
The projective superspace approach  was extended to conformal supersymmetry
\cite{K-comp,K-conf,KPT-MvU} and supergravity in various dimensions
\cite{KT-M_5D,KT-M_5Dconf,KLRT-M_4D-1,KLRT-M_4D-2,GTM_2D_SUGRA,KLT-M11,LT-M-2012},  
more than  twenty years after 
the original publication on self-interacting $\cN=2$ tensor multiplets \cite{KLR}. }
\cite{KLT-M11,KLT-M12}; and 
(ii) on-shell formulations in terms of covariantly chiral scalar superfields in (2,0) AdS superspace
and (1,1) AdS superspace (the latter formulation exists in special cases only).
We will also heavily build on the results of \cite{K-duality,KPT-MvU,BKLT-M}.
The realization (ii) proves to be a very convenient tool to study the target space geometry
(this is similar to the $\cN=1$ superfield formulation for $\cN=2$ supersymmetric $\s$-models 
in four dimensions \cite{HKLR}).  
Therefore, we would like to briefly discuss the reduction from $\cN>2$ to (2,0) AdS superspace.

Consider a supersymmetric field theory formulated in a given $(p,q)$ AdS superspace 
with $\cN = p+q\geq 3$ and $p\geq q$. Such a  dynamical system can always be reformulated 
as a supersymmetric theory realized in the $(2,0)$ AdS superspace, with $(p+q -2)$ 
supersymmetries hidden. This observation will be important for the subsequent analysis, 
and therefore we would like to elaborate on it.
The conceptual possibility for a $(p,q) \to (2,0)$ AdS reformulation 
follows from the explicit structure of 
the algebra of $(p,q)$ AdS covariant derivatives 
\bea 
\cD_A =(\cD_a , \cD^I_\a) = E_A + \hf\O_{A}{}^{\b\g}\cM_{\b\g} + \hf\F_A{}^{KL}\cN_{KL}
~, \qquad I=1,\dots, \cN~,
\eea
given in \cite{KLT-M12}:
\bsubeq\label{alg-AdS}
\bea
\{\cD_\a^I,\cD_\b^J\}&=&
2\ri\d^{IJ}\cD_{\a\b}
-4\ri S^{IJ}\cM_{\a\b}
+\ri\ve_{\a\b}\Big(
X^{IJKL}
-4S^{K}{}^{[I}\d^{J]L}
\Big)
\cN_{KL}
~,
\label{alg-AdS-1}
\\
{[}\cD_{a},\cD_\b^J{]}
&=&
S^{J}{}_{K}(\g_a)_\b{}^\g\cD_{\g}^K
~,
\label{alg-AdS-3/2}
\\
{[}\cD_{a},\cD_b{]}
&=&
-4\,S^2\cM_{ab}
~,
\label{alg-AdS-2}
\eea
\esubeq
with $\cM_{\a\b}= \cM_{\b\a}$ (or, using the three-vector notation,  $\cM_{ab}=-\cM_{ba}$)
the Lorentz generators and $\cN_{KL}= - \cN_{LK}$ 
the ${\rm SO}(\cN) $ generators.
In general,  $S^{IJ} = S^{JI}$
is a non-vanishing  covariantly constant tensor such that
$S=\sqrt{(S^{IJ}S_{IJ})/\cN} >0$ is a positive constant parameter of unit mass dimension. 
It can be chosen, by applying a local 
${\rm SO}(\cN) $ transformation,  in the diagonal form 
\bea
S^{IJ}=S\,{\rm{diag}}(\,  \overbrace{+1,\cdots,+1}^{p} \, , \overbrace{-1,\cdots,-1}^{q=\cN-p} \,)~.
\label{diag-S}
\eea
The other component of the SO$(\cN)$ curvature, $X^{IJKL} = X^{[IJKL]}$, is a 
completely antisymmetric and covariantly constant tensor which  
may exist only in the cases $q=0$ and $p\geq4$. If present, this tensor
in the gauge \eqref{diag-S}
  obeys  the quadratic constraint 
\bea
X_N{}^{IJ[K}X^{LPQ]N} =0 ~.
\eea
Splitting each ${\rm SO}(\cN)$ index as $I = ({\hat i}, {\hat I} )$, where ${\hat i}=1,2$ and 
${\hat I} = 3, \dots, \cN$, 
one may see from  \eqref{alg-AdS} that the operators $(\cD_a , \cD^{\hat i}_\a)$ 
form a closed subalgebra,
\bsubeq \label{111.4}
\bea
\{\cD_\a^{\hat i} ,\cD_\b^{\hat j}\}&=&
2\ri\d^{{\hat i}{\hat j}}\cD_{\a\b}
-4\ri S \d^{{\hat i}{\hat j}}\cM_{\a\b}
-4\ri   S\ve_{\a\b} \tilde{\cN}^{{\hat i}{\hat j}} ~,
\\
{[}\cD_{a},\cD_\b^{\hat j}{]}
&=&
S (\g_a)_\b{}^\g\cD_{\g}^{\hat j}
~,
\\
{[}\cD_{a},\cD_b{]}
&=&
-4\,S^2\cM_{ab}~,
\eea
\esubeq
where we have defined a modified SO(2) generator
\bea
\tilde{\cN}^{{\hat i}{\hat j}} := \cN^{{\hat i}{\hat j} } 
-\frac{1}{4S} X^{{\hat i}{\hat j}\hat K \hat L} \cN_{\hat K \hat L}~, \qquad
{\big [} \tilde{\cal N}{}^{{\hat k}{\hat l}},\cD_{\a}^{{\hat i}}{\big]} =2\d^{{\hat i}[{\hat k}}\cD_\a^{{\hat l}]}
   ~,\qquad
{\big [} \tilde{\cal N}{}^{{\hat k}{\hat l}},\cD_{a}{\big]} =0 ~.
\eea
The anti-commutation relations \eqref{111.4} correspond to the (2,0) AdS superspace  
\cite{KT-M11,KLT-M12} (in a real basis for the spinor covariant derivatives).  
Due to \eqref{111.4}, 
the SO$(\cN)$ connection corresponding to the covariant derivatives
$(\cD_a , \cD^{\hat i}_\a)$ can be reduced to an SO(2) connection associated with 
$\tilde{\cN}^{{\hat i}{\hat j}}$ by applying an SO($\cN)$  gauge transformation.\footnote{In such a 
gauge, the covariantly constant tensor  $X^{{\hat i}{\hat j}\hat K \hat L}$ becomes constant with 
respect to $(\cD_a , \cD^{\hat i}_\a)$.}
The (2,0) AdS superspace can be embedded into our $(p,q)$ AdS superspace 
as a surface defined by 
\bea
\q^\m_{\hat I} =0~, \qquad \hat I = 3, \dots, \cN~,
\eea 
for a certain local parametrization  of the superspace Grassmann variables 
$\q^\m_I =(\q^\m_{\hat i}, \q^\m_{\hat I})$.  

Now consider the case $ p+q\geq 3$ and $p\geq q>0$, and therefore $X^{IJKL}=0$, 
and assume that $S^{IJ}$ is given in the form
 \bea
 S^{IJ}=S\,{\rm{diag}}(\, +1, -1, \overbrace{+1,\cdots,+1}^{p-1} \, , \overbrace{-1,\cdots,-1}^{q-1} \,)~,
 \eea
 compare with \eqref{diag-S}.
Then any supersymmetric field theory  in the $(p,q)$ AdS superspace can be reformulated as 
a theory in the (1,1) AdS superspace. Indeed, if we  split again each ${\rm SO}(\cN)$ index as 
$I = ({\hat i}, {\hat I} )$, where ${\hat i}=1,2$ and 
${\hat I} = 3, \dots, \cN$, 
the anti-commutation relations   \eqref{alg-AdS} imply that the operators $(\cD_a , \cD^{\hat i}_\a)$ 
obey a closed algebra of the form:
\bsubeq \label{1.8}
\bea
\{\cD_\a^{\hat i} ,\cD_\b^{\hat j}\}&=&
2\ri\d^{{\hat i}{\hat j}}\cD_{\a\b}
+4\ri S(-1)^{\hat i} \d^{{\hat i}{\hat j}}\cM_{\a\b}
~,
\\
{[}\cD_{a},\cD_\b^{\hat j}{]}
&=&
-S (-1)^{\hat j}(\g_a)_\b{}^\g\cD_{\g}^{\hat j}~,
\\
{[}\cD_{a},\cD_b{]}
&=&
-4\,S^2\cM_{ab}~.
\eea
\esubeq
This algebra corresponds to the (1,1) AdS superspace \cite{KT-M11,KLT-M12}
(in a real basis for the spinor covariant derivatives).  

Another interesting possibility occurs in the case $p+q\geq 4$ and $X^{IJKL}=0$. 
It may be seen that any supersymmetric field theory  in the $(p,q)$ AdS superspace 
can be reformulated as a theory in the (3,0) or (2,1)  AdS superspace.
This will be further explored in the main body of the present paper.

This paper is organized as follows. Section 2 provides review material on the 
three-dimensional nonlinear $\s$-models with (2,0) and (1,1) AdS supersymmetry. 
The $\s$-models with (3,0) AdS supersymmetry are studied in sections 3 and 4. 
In section 3, we start from the most general off-shell (3,0) supersymmetric $\s$-model
and then reformulate it in terms of chiral superfields in the (2,0) AdS superspace. 
In section 4  we start from a general nonlinear $\s$-model 
in the (2,0) AdS superspace and derive the conditions on the target space geometry 
for the $\s$-model to possess (3,0) AdS supersymmetry. 
Sections 5 and 6 extend the analysis of sections 3 and 4 to the case of 
(2,1) AdS supersymmetric $\s$-models. Various aspects of the $\cN=4$ AdS superspaces 
are discussed in section 7. Section 8 is devoted to the projective-superspace techniques
to formulate off-shell $\cN=4$ supersymmetric $\s$-models in AdS${}_3$ as well as 
to carry out their reduction  to $\cN=2$ and $\cN=3$ AdS superspaces. 
In section 9 we demonstrate that any nonlinear $\s$-model with ($p,q)$ AdS supersymmetry 
for $p+q=3$, in fact,  possesses a larger $(p',q')$ AdS supersymmetry with $p' + q' =4$ and 
$p' \geq  p $, $q' \geq q$. Sections 10 and 11 concern the formulation of the most general
$\s$-model with (4,0) AdS supersymmetry in terms of chiral superfields on the (2,0) AdS superspace.
In section 12, the results of the two previous sections are used to construct the most general 
$\s$-model with the non-centrally extended $\cN=4$ Poincar\'e supersymmetry. 
A brief discussion of the results obtained is given in section 13. 
The main body of the paper is accompanied by four technical appendices. 


\section{Nonlinear sigma models with four supercharges}
\setcounter{equation}{0}

As discussed in the introduction, for $\cN=p+q\geq 3$ and $p\geq q$, the most  general nonlinear 
$\s$-models with $(p,q)$ AdS supersymmetry can be realized in the (2,0) AdS superspace. 
If in addition $X^{IJKL} =0$ and $ q>0$, 
there also exists a realization in the (1,1) AdS superspace. It is therefore of special importance to 
understand the structure of supersymmetric field theories in the (2,0) and (1,1) AdS superspaces. 
The off-shell nonlinear $\s$-models in AdS$_3$ 
with (2,0) and (1,1) supersymmetry were thoroughly studied in \cite{KT-M11}.
In this section we review the $\s$-model results of \cite{KT-M11}.\footnote{The most general 
$\s$-model couplings
to (1,1) and (2,0) AdS supergravity theories were constructed  in  \cite{KT-M11} from first 
principles. These results generalize those obtained earlier \cite{IT,DKSS} 
within the Chern-Simons approach \cite{AT}.}

\subsection{Nonlinear sigma models in (2,0) AdS superspace}
\label{NLSM20}

The geometry of (2,0) AdS superspace is encoded in its  covariant derivatives
\bea
\cD_A = (\cD_a , \cD_\a, \bar \cD^\a)= {E}_A{}^M \pa_M +\hf {\O}_A{}^{cd}\cM_{cd}
+ \ri \,{ \F}_A \cJ
\eea 
obeying the following (anti) commutation relations: 
\bsubeq \label{20AdSsuperspace}
\bea
& \{\cD_\a,\cD_\b\} =  \{\bar \cD_\a,\bar \cD_\b\}
=0
~,\qquad
\{\cD_\a,\bar \cD_\b\}
=
-2\ri \cD_{\a\b}
-4\ri \,S\, \ve_{\a\b} \cJ
+4\ri \,S\, \cM_{\a\b} ~, ~~~
\label{AdS_(2,0)_algebra_1}
\\
& {[}\cD_{a},\cD_\b{]}
=
\,S\, (\g_a)_\b{}^\g\cD_{\g}~, 
\qquad
 {[}\cD_{a},\bar \cD_\b{]}
=
\,S\, (\g_a)_\b{}^\g\bar \cD_{\g}~,
\label{AdS_(2,0)_algebra_2}  \\
& {[}\cD_a,\cD_b{]} = -4\, S^2\, \cM_{ab}~.
\label{AdS_(2,0)_algebra_3} 
\eea
\esubeq
The generator $\cJ$ in \eqref{20AdSsuperspace}
corresponds to the gauged $R$-symmetry group, ${\rm U}(1)_R$, 
and acts on the covariant derivatives as
\bea
  {[}\cJ,\cD_\a{]}=\cD_\a~,\qquad
    {[}\cJ,\bar \cD _\a{]}=-\bar \cD_\a   ~.
 \eea
The constant parameter $S$ in \eqref{20AdSsuperspace} is a square root of  the curvature of 
AdS${}_3$. Unlike \eqref{111.4},  in this section we use the complex basis for the $\cN=2$ 
spinor covariant derivatives introduced in \cite{KLT-M11}.
 
The isometries of the (2,0) AdS superspace are described  
by Killing vector fields, 
$\t=\t^a\cD_a+\t^\a\cD_\a+\bar{\t}_\a\bar \cD^\a$, obeying the  equation
\bea
\Big{[}\t+\ri t\cJ+\hf t^{bc}\cM_{bc},\cD_A\Big{]}=0~,
\label{2_0-Killing_iso-def}
\eea
for some parameters $t$ and $t^{ab}$. Choosing  $\cD_A=\cD_a$ in
\eqref{2_0-Killing_iso-def} gives
\begin{subequations}  \label{Killing_2.6}
\bea
\cD_a t &=& 0~, \label{Killing_2.6a}\\
\cD_a \t_b&=&   t_{ab}  ~, \label{Killing_2.6b} \\
 \cD_a \t^\b &=& - S \t^\g (\g_a )_\g{}^\b~, \label{Killing_2.6c} \\
\cD_a t^{bc} &=& 4S^2 (\d_a{}^b \t^c - \d_a{}^c \t^b) ~. \label{Killing_2.6d}
\eea
\end{subequations}
Eq. \eqref{Killing_2.6b} implies the standard Killing equation 
\bea
\cD_{a} \t_{b } +\cD_b \t_a=0~,
\eea
while \eqref{Killing_2.6c} is a Killing spinor equation. From  \eqref{Killing_2.6b}
and \eqref{Killing_2.6d} it follows that 
\bea
\cD_a \cD_b \t_c = 4S^2 (\eta_{ab} \t_c - \eta_{ac}\t_b)~.
\eea
Next, choosing $\cD_A=\cD_\a$ in
\eqref{2_0-Killing_iso-def} gives
\begin{subequations}\label{Killing_2.8}
\bea
\cD_\a \bar \t_\b &=& 0~, \label{Killing_2.8a}\\
\cD_\a t &=&  4S\bar \t_\a  ~, \label{Killing_2.8b} \\
 \cD_\a t^{\b \g}&=& - 4\ri S (\d_\a{}^\b \bar \t^\g + \d_\a{}^\g \bar \t^\b  ) ~, \label{Killing_2.8c} \\
\cD_\a \t^{\b \g} &=& -2\ri  (\d_\a{}^\b \bar \t^\g + \d_\a{}^\g \bar \t^\b) ~, \label{Killing_2.8d}\\
\cD_\a  \t^\b &=& \hf t_\a{}^\b +S \t_\a{}^\b +\ri \d_\a{}^\b t ~. \label{Killing_2.8e}
\eea
\end{subequations}
These equations have a number of nontrivial implications including the following:
 \bsubeq \label{2_0-Killing_iso}
\bea
&
\cD_{(\a}\t_{\b\g)}=\cD_{(\a} t_{\b\g)}=0~, 
\label{2_0-Killing_iso_2} \\
&\cD_{(\a}\t_{\b)}=-\bar \cD_{(\a}\bar{\t}_{\b)}=\dfrac{1}{2} t_{\a\b}+ S \t_{\a\b} ~,  
\label{2_0-Killing_iso_4} \\
&  \t_\a=
\dfrac{\ri}{6} 
\bar \cD^\b\t_{\a\b}=\dfrac{\ri}{12S}\bar \cD^\b t_{\a\b}~,
\label{2_0-Killing_iso_1}
\\
&
\cD_\g\t^\g
=
-\bar \cD^\g\bar{\t}_\g
=2\ri t ~.
\label{2_0-Killing_iso_3}
\eea
\esubeq
It follows from the above equations that the Killing superfields $\t^\a$, $t$ and $t^{ab}$ 
are given in terms of the vector parameter $\t^a$. Its components defined by $\t^a|_{\q=0}$ and 
$(-\cD^b \t^a) |_{\q=0}$ describe the isometries of AdS$_3$. The other isometry transformations 
of the (2,0) AdS superspace are contained not only in $\t^a$ but also, e.g., 
in the real scalar $t$ subject to the following equations:
\bea
\cD^2 t = \bar \cD^2 t =0~, \qquad (\ri \cD^\a \bar \cD_\a -8S) t =0~, \qquad
\cD_a t=0~.
\eea
At the component level, $t$ contains the  real constant parameter $t |_{\q=0}$ 
and the complex Killing spinor $\cD_\a t|_{\q=0}$,
which generate the $R$-symmetry  and supersymmetry transformations of  
the (2,0) AdS superspace respectively.

Given a matter tensor superfield $\cV$ 
(with all indices suppressed), its (2,0) AdS transformation law is 
\bea
\d \cV = ( \t+\ri t\cJ+\hf t^{bc}\cM_{bc} )\cV~.
\eea

Supersymmetric actions can be constructed either by integrating a real scalar 
$\cL$ over the full (2,0) AdS superspace,\footnote{The component 
inverse vierbein is defined as usual, 
$e_a{}^m (x) = E_a{}^m |_{\q=0}$, with $e^{-1}=\det(e_a{}^m)$.}
\begin{align}\label{eq_22AdSDcomp}
\int \rd^3x\, \rd^4\q\, E\, \cL
	&= \int \rd^3x\, e\, \Big(
	\frac{1}{16} \cD^\alpha \bar \cD^2 \cD_\alpha 
	+ \ri S  \bar \cD^\alpha \cD_\alpha 
	\Big)    \cL\Big|_{\q=0} \non \\
	&= \int \rd^3x\, e\, \Big(
	\frac{1}{16} \bar \cD_\alpha  \cD^2 \bar \cD^\alpha 
	+ \ri S  \cD^\alpha  \bar\cD_\alpha
	\Big)    \cL   \Big|_{\q=0}
	~,
\end{align}
with $E^{-1}= {\rm Ber} (E_A{}^M)$,
or by integrating a covariantly chiral scalar $\cL_c$
over the chiral subspace of the (2,0) AdS superspace,
\begin{align}\label{chiralac}
\int \rd^3x\, \rd^2\q\, \cE\, \cL_c
	= -\frac{1}{4} \int \rd^3x\, e\, \cD^2 \cL_c\Big|_{\q=0}~, \qquad
\bar\cD^\alpha \cL_c =0~,
\end{align}
with $\cE$ being the chiral density.
The Lagrangians $\cL$ and $\cL_c$ are required to possess certain ${\rm U}(1)_R$ charges:
\begin{align}
\cJ \cL = 0~, \qquad \cJ \cL_c = -2 \cL_c~.
\end{align}
The two types of supersymmetric actions are related to each other
by the rule
\begin{align}
\int \rd^3x\, \rd^4\q\, E\, \cL &= -\frac{1}{4} \int \rd^3x\, \rd^2\q\, \cE\, \bar\cD^2 \cL~.
\end{align}
Using the AdS transformation law of the Lagrangian in \eqref{eq_22AdSDcomp}, 
$\d \cL = \t \cL$,  and the Killing equation 
\eqref{2_0-Killing_iso-def}, one may check explicitly that the component action 
defined by the right-hand side of \eqref{eq_22AdSDcomp} is invariant under 
the (2,0) AdS isometry group. A similar consideration may be given in the case of  
the chiral action \eqref{chiralac}, with the only difference that the AdS transformation of 
the chiral Lagrangian is $\d \cL_c = (\t - 2\ri \,t) \cL_c$.

For component reduction, it may be useful to choose a Wess-Zumino gauge such that
\bea
\cD_a |_{\q =0} = e_a{}^m(x) \pa_m + \hf \o_a{}^{bc} (x) \cM_{bc}~.
\label{stcd}
\eea
In this gauge, we will use the same symbol $\cD_a$ for the space-time covariant derivative 
in the right-hand side of \eqref{stcd}.

Given an Abelian vector multiplet, we can introduce gauge covariant derivatives
\bea
{\bm \cD}_A := \cD_A +\ri\, V_A \,\cZ~, \qquad [\cZ, \cD_A ]= [\cZ, {\bm \cD}_A ]=[ \cZ, \cJ ] =0~,
\label{gcd}
\eea
where $V_A$ is the gauge connection associated with  the U(1) generator $\cZ$.
The gauge covariant derivatives are required to obey the relations 
\bea
&\{ {\bm \cD}_\a, {\bm \cD}_\b\}
=0
~,~~~~~~
\{ {\bm \cD}_\a,\bar {\bm \cD}_\b\}
=
-2\ri {\bm \cD}_{\a\b}
-4\ri \, \ve_{\a\b} (S \cJ +G \cZ)
+4\ri \,S\, \cM_{\a\b} ~, 
\label{2.4}
\eea
and the other (anti) commutators can be restored by making use of the Bianchi identities
and complex conjugation. The gauge invariant field strength $G$ is real, 
$G = \bar G$, and covariantly linear, 
\bea
{\bm \cD}^2  G = \bar {\bm \cD}^2 G = 0~.
\eea

Suppose the vector multiplet under consideration is {\it intrinsic}, that is 
its field strength $G$ is a non-zero constant, which can be conveniently normalized as
\bea
G=S ~.
\eea
The name `intrinsic' is due to the fact that 
such a vector multiplet generates a unique super-Weyl transformation which
conformally relates the (2,0) AdS superspace to a flat one  \cite{KT-M11}.
In the case of the intrinsic vector multiplet, 
the second relation in \eqref{2.4} can be rewritten in the form 
\bea
\{ {\bm \cD}_\a,\bar {\bm \cD}_\b\}
=
-2\ri {\bm \cD}_{\a\b}
-4\ri \, \ve_{\a\b} S {\bm \cJ}
+4\ri \,S\, \cM_{\a\b} ~, \qquad {\bm \cJ} := \cJ +\cZ~.
\label{intrVM}
\eea
We see that the gauge covariant derivatives ${\bm \cD}_A$,
which are associated with the intrinsic vector 
multiplet, describe (2,0) AdS superspace  with a deformed U(1)$_R$ generator. 

The supersymmetric nonlinear $\s$-models with (2,0) AdS supersymmetry 
were studied in \cite{KT-M11}. In general, each $\s$-model can be described in terms
of chiral scalar superfields $\f^\ra $, $\bar \cD_\a \f^\ra =0$, 
taking their values in a K\"ahler manifold $\cM$. 
There are two different cases to consider. 

The first option corresponds to the situation where
the chiral variables $\f^\ra$ are neutral with respect to  
the $R$-symmetry group U(1)$_R$, 
\bea 
\cJ \f^\ra =0~.
\eea
In this case, {\it no superpotential} is allowed, and the most general  $\s$-model action is 
\bea
S= \int \rd^3x \,\rd^4\q \, E\,K(\f^{\ra}, \bar \f^{\bar \rb} )~,
\label{2.8}
\eea
where $K(\f, \bar \f )$ is the K\"ahler potential of  $\cM$. 
The target space, $\cM$, may be an arbitrary K\"ahler manifold. 
The action \eqref{2.8} is invariant under the (2,0) AdS isometry supergroup,
$\rm OSp(2|2;{\mathbb R}) \times Sp(2,{\mathbb R}) $. 
As follows from \eqref{eq_22AdSDcomp}, 
the above $\s$-model  possesses the K\"ahler symmetry 
\begin{align}
K (\f , \bar \f ) \rightarrow K (\f , \bar \f)+ F  (\f )  + \bar F  ( \bar \f ) ~, 
\end{align}
where $ F(\phi)$ is an arbitrary  holomorphic function. 

The second option is that  a superpotential 
is allowed,\footnote{We will show below that no superpotential 
is allowed for those $\s$-models 
in (2,0) AdS superspace which have additional supersymmetries, except the case
of critical (4,0) supersymmetry.}  
\bea
S= \int \rd^3x \,\rd^4\q \, E\,K(\f^{\ra}, \bar \f^{\bar \rb} ) 
~+ ~\Big(\int \rd^3x\, \rd^2\q\, \cE\, W (\f^\ra )+ \textrm{c.c.}\Big)~,
\label{2.8+W}
\eea
for some holomorphic function $W(\f)$. 
In this case, the target space must have a U(1) isometry group generated by a holomorphic 
Killing vector $J^\ra (\f)$ defined by 
\bea
J= J^\ra (\f) \pa_\ra +\bar J^{\bar \ra} (\bar \f) \pa_{\bar \ra} ~, \qquad
J^\ra(\f):= \ri \cJ \f^\ra~.
\label{2.11}
\eea
As is known, the Killing condition amounts to  
\bea
J^\ra K_\ra + \bar J^{\bar \ra} K_{\bar \ra} = F  + \bar F ~, 
\non
\eea
for some holomorphic function $F(\f)$. However, 
since the Lagrangian in  \eqref{eq_22AdSDcomp} 
has to be a scalar superfield, 
the K\"ahler potential must be neutral under ${\rm U}(1)_R$ and hence $F=0$, 
\bea
J^\ra K_\ra + \bar J^{\bar \ra} K_{\bar \ra} =0~. 
\eea
The infinitesimal  transformation  
of $\f^\ra$ under the (2,0) AdS isometry supergroup is
\bea
\d \f^\ra = \t \f^\ra + t J^\ra(\f)~.
\eea
The action \eqref{2.8+W} is invariant under this transformation 
if the superpotential $W(\f)$ obeys the condition \cite{KT-M11} 
\bea \label{eq_WU1}
J^\ra W_\ra = - 2 \ri\, W~.
\eea

Suppose the target space of the $\s$-model \eqref{2.8+W} possesses 
a holomorphic Killing vector field 
\bea
Z= Z^\ra (\f) \pa_\ra +\bar Z^{\bar \ra} (\bar \f) \pa_{\bar\ra} ~, \qquad
Z^\ra (\f):= \ri \cZ \f^\ra~,
\eea
which commutes with the Killing vector field \eqref{2.11}, 
\bea
[J, Z] =0~.
\eea
Without loss of generality, we can assume that\footnote{In the general case that  
$Z^\ra K_\ra + \bar Z^{\bar \ra} K_{\bar \ra} = H + \bar H$, for some holomorphic function $H(\f)$, 
we can introduce, following \cite{HKLR},  a new chiral superfield $\f^0$ and 
Lagrangian $K' = K -\f^0  -\bar \f^0 $, 
where $\f^0$ transforms as $\ri \cZ \f^0 =  H(\f^\ra)$. The Lagrangian $K'$
possesses the required property \eqref{2.17}. The field $\f^0$ is a purely gauge degree of freedom. }
\bea
Z^\ra K_\ra + \bar Z^{\bar \ra} K_{\bar \ra} =0~.
\label{2.17} 
\eea
We then can gauge the U(1) isometry generated by the Killing vector $Z^\ra (\f) $ by means of 
introducing gauge covariant derivatives ${\bm \cD}_A$,  eq. \eqref{gcd}, 
and replacing the chiral superfields $\f^\ra$ in \eqref{2.8} with gauge covariantly chiral ones 
${\bm \f}^\ra$, 
\bea
\bar {\bm \cD}_\a {\bm \f}^\ra =0~.
\eea
The algebra of covariant derivatives remains unchanged except that we replace
$\cJ$ with $\bm\cJ$ and identify ${\bm J}^\ra = J^\ra + Z^\ra$. In what follows,
we often will not distinguish between these cases.

Using the component reduction rule, one can show that
\begin{align}
S = \int \rd^3x\, \rd^4\q\, E\, K +
	\left(\int \rd^3x\, \rd^2\q\, \cE\, W + \textrm{c.c.}\right)
	= \int \rd^3x\, e\, L~,
\end{align}
where
\begin{align}
L &= -g_{\ra \bar \ra} \cD_m \varphi^\ra \cD^m \bar\varphi^{\bar \ra}
	- \ri g_{\ra \bar \ra} \bar\psi_\alpha^{\bar \ra} \hat \cD^{\alpha \beta} \psi_\beta^{\ra}
	+ F^\ra \bar F^{\bar \ra} g_{\ra \bar \ra}
	+ \frac{1}{4} R_{\ra \bar \ra \rb \bar \rb} (\psi^{\ra} \psi^\rb) 
	(\bar\psi^{\bar \ra} \bar\psi^{\bar \rb})
	\eol & \quad
	+ S\, (\psi^\ra \bar\psi^{\bar \ra}) (\ri g_{\ra \bar \ra} + \tsD_\ra J_{\bar \ra} 
	- \tsD_{\bar \ra} \bar J_{\ra})
	- 4 S^2 (J^\ra \bar J^{\bar \ra} g_{\ra \bar \ra} -D)
	\eol & \quad
	+ W_\ra F^\ra + \bar W_{\bar \ra} \bar F^{\bar \ra}
	- \frac{1}{2} \tsD_\ra W_\rb (\psi^\ra \psi^\rb)
	- \frac{1}{2} \tsD_{\bar \ra} \bar W_{\bar \rb} (\bar\psi^{\bar \ra} \bar\psi^{\bar \rb})~.
\end{align}
We use the shorthand
$(\psi^{\ra} \psi^\rb) := \psi^{\alpha\ra} \psi_\alpha^\rb$,
$(\psi^{\ra} \bar\psi^{\bar\rb}) := \psi^{\alpha\ra} \bar\psi_\alpha^{\bar\rb}$,
and $(\bar\psi^{\bar\ra} \bar\psi^{\bar\rb}) := \bar\psi_\alpha^{\bar\ra} \bar\psi^{\alpha\bar\rb}$.
The Killing potential $D$ is defined by $J^\ra K_\ra = -\ri D/2$.
Here and below, $\nabla_\ra$ 
and $\nabla_{\bar \ra}$ denote 
 the target-space covariant derivatives. 
 
We have defined the components of $\phi^\ra$ as
\bsubeq
\begin{align}
\varphi^\ra &:= \phi^\ra\vert~, \qquad
\psi_\alpha^\ra := \frac{1}{\sqrt 2} \cD_\alpha \phi^\ra \vert~, \\
F^\ra &:= -\frac{1}{4} g^{\ra \bar \rb} \cD^2 K_{\bar \rb} \vert
	= -\frac{1}{4} (\cD^2 \phi^\ra + \Gamma^\ra{}_{\rb\rc} \cD^\alpha \phi^\rb \cD_\alpha \phi^\rc)\vert
~.
\end{align}
\esubeq
In particular, the auxiliary field $F^\ra$ transforms covariantly under 
target space reparametrizations.
The vector derivative on the fermion is similarly reparametrization covariant,
\begin{align}
\hat \cD_m \psi_\alpha^\ra := \cD_m \psi_\alpha^\ra 
+ \Gamma^\ra{}_{\rb\rc} \cD_m \varphi^\rb \psi_\alpha^\rc~,
\end{align}
and the action of $\cJ$ on the physical fields is defined as  
\bea
\ri \cJ \vf^\ra = J^\ra(\vf), \qquad \ri \cJ \j^\ra_\a = \j^\rb_\a \pa_\rb J^\ra (\vf) + \ri \j^\ra_\a~.
\eea 
The fermion mass terms are given in terms of covariant field derivatives
\begin{align}
\tsD_\ra W_\rb := \pa_\ra \pa_\rb W - \Gamma^\rc{}_{\ra\rb} \pa_\rc W~.
\end{align}
Eliminating the auxiliary fields $F^\ra$ leads to the component Lagrangian
\begin{align}\label{eq_20compaction}
L &= -g_{\ra \bar \ra} \cD_m \varphi^\ra \cD^m \bar\varphi^{\bar \ra}
	- \ri g_{\ra \bar \ra} \bar\psi_\alpha^{\bar \ra} \hat \cD^{\alpha \beta} \psi_\beta^{\ra}
	+ \frac{1}{4} R_{\ra \bar \ra \rb \bar \rb} (\psi^{\ra} \psi^\rb) (\bar\psi^{\bar \ra} \bar\psi^{\bar \rb})
	\eol & \quad
	+ S\, (\psi^\ra \bar\psi^{\bar \ra}) (\ri g_{\ra \bar \ra} + \tsD_\ra J_{\bar \ra} 
	- \tsD_{\bar \ra} \bar J_{\ra})
	- 4 S^2 ( J^\ra \bar J^{\bar \ra} g_{\ra \bar \ra} -D)
	\eol & \quad
	- \frac{1}{2} \tsD_\ra W_\rb (\psi^\ra \psi^\rb)
	- \frac{1}{2} \tsD_{\bar \ra} \bar W_{\bar \rb} (\bar\psi^{\bar \ra} \bar\psi^{\bar \rb})
	- g^{\ra \bar \ra} W_\ra \bar W_{\bar \ra} ~.
\end{align}

One broad class of interest is when the target space is a K\"ahler cone,
see Appendix \ref{AppA}. Then the target space admits a homothetic
conformal Killing vector $\chi$, obeying the conditions \eqref{hcKv}
and \eqref{hcKv-pot}. For a superconformal $\s$-model,
the ${\rm U}(1)_R$ Killing vector fields $J$ should commute with $\chi$,
$[\chi, J]^\m = \chi^\n \nabla_\n J^\m - J^\n \nabla_\n \chi^\m = 0$,
(since these vector fields generate the ${\rm U}(1)_R$ and scale
transformations), which is equivalent to 
$\chi^\rb \nabla_\rb J^\ra = J^\ra$ \cite{deWKV}.
It follows that
\begin{align}
J^\ra \chi_\ra + J^{\bar \ra} \chi_{\bar \ra} = 0~.
\end{align}
The superpotential must obey \eqref{eq_WU1}; if the action is
additionally superconformal, it must also obey
\begin{align}\label{eq_Wconf}
W = \frac{1}{4} \chi^\ra W_\ra~.
\end{align}
It is natural to decompose $J^\ra$ as
\begin{align}\label{eq_Jconformal}
J^\ra = -\frac{\ri}{2} \chi^\ra + Z^\ra~,
\end{align}
where (by construction) $Z^\ra$ is a Killing vector which leaves the superpotential invariant,
$Z^\ra W_\ra = 0$, and commutes with $\chi$. 
In light of our discussion about gauged $\s$-models,
one may interpret the $\chi^\ra$ term in \eqref{eq_Jconformal} as the natural
part of the ${\rm U}(1)_R$ Killing vector and the $Z^\ra$ term as arising
from gauging the K\"ahler cone with a frozen vector multiplet.
Introducing a Killing potential $D^{(z)}$ for $Z$ using $Z^\ra \chi_\ra = -\ri D^{(z)} / 2$,
one finds that the component Lagrangian reduces to
\begin{align}
L &= -g_{\ra \bar \ra} \cD_a \varphi^\ra \cD^a \bar\varphi^{\bar \ra}
	- \ri g_{\ra \bar \ra} \bar\psi_\alpha^{\bar \ra} \hat \cD^{\alpha \beta} \psi_\beta^{\ra}
	+ \frac{1}{4} R_{\ra \bar \ra \rb \bar \rb} (\psi^{\ra} \psi^\rb) (\bar\psi^{\bar \ra} \bar\psi^{\bar \rb})
	\eol & \quad
	+ S\, (\psi^\ra \bar\psi^{\bar \ra}) (\tsD_\ra Z_{\bar \ra}  - \tsD_{\bar \ra} \bar Z_{\ra})
	+ 3 S^2 K
	- 4 S^2 \big( Z^\ra \bar Z^{\bar \ra} g_{\ra \bar \ra} - \frac{1}{2} D^{(z)}\big)
	\eol & \quad
	- \frac{1}{2} \tsD_\ra W_\rb (\psi^\ra \psi^\rb)
	- \frac{1}{2} \tsD_{\bar \ra} \bar W_{\bar \rb} (\bar\psi^{\bar \ra} \bar\psi^{\bar \rb})
	- g^{\ra \bar \ra} W_\ra \bar W_{\bar \ra} ~.
\end{align}
The potential-like term $3S^2 K$ may be combined with the scalar kinetic term to
give
\begin{align}
L &= g_{\ra \bar \rb} \chi^\ra (\hat \cD_a \hat \cD^a - \frac{1}{8} \cR) \bar\chi^{\bar \rb}
	- \ri g_{\ra \bar \rb} \bar\psi_\alpha^{\bar\rb} \hat \cD^{\alpha \beta} \psi_\beta^{\ra}
	+ \frac{1}{4} R_{\ra \bar \rb \rc \bar \rd} (\psi^{\ra} \psi^\rc) (\bar\psi^{\bar \rb} \bar\psi^{\bar \rd})
	\eol & \quad
	+ S\, (\psi^\ra \bar\psi^{\bar \ra}) (\tsD_\ra Z_{\bar \ra}  - \tsD_{\bar \ra} \bar Z_{\ra})
	- 4 S^2 \big( Z^\ra \bar Z^{\bar \ra} g_{\ra \bar \ra} - \frac{1}{2} D^{(z)}\big)
	\eol & \quad
	- \frac{1}{2} \tsD_\ra W_\rb (\psi^\ra \psi^\rb)
	- \frac{1}{2} \tsD_{\bar \ra} \bar W_{\bar \rb} (\bar\psi^{\bar \ra} \bar\psi^{\bar \rb})
	- g^{\ra \bar \ra} W_\ra \bar W_{\bar \ra} 
\end{align}
after identifying the scalar curvature of AdS as $\cR = -24 S^2$ and discarding
a total derivative. The scalar kinetic
operator in the first term is the conformal d'Alembertian in three dimensions.
The actual mass terms are confined to the second and third lines
which arise respectively from gauging a $\rm U(1)$ isometry with the intrinsic vector multiplet
and from introducing a superpotential.

\subsection{Nonlinear sigma models in (1,1) AdS superspace}
\label{11AdSderivatives} 

The geometry of (1,1) AdS superspace is described in terms of covariant derivatives
\bea
{\frak D}_A = ({\frak D}_a , {\frak D}_\a ,\bar {\frak D}^\a) =E_A{}^M \pa_M + \hf \O_A{}^{cd}\cM_{cd} 
\eea
which obey the following (anti) commutation relations:
\bsubeq \label{11AdSsuperspace}
\bea
&\{ {\frak D}_\a, {\frak D}_\b\}
=
-4\bar{\mu}\cM_{\a\b}
~,~~~
 \{\bar {\frak D}_\a,\bar {\frak D}_\b\}
=
4\mu\cM_{\a\b}
~,~~~
\{ {\frak D}_\a,\bar {\frak D}_\b\}
=
-2\ri {\frak D}_{\a\b}~, 
\label{AdS_(1,1)_algebra_1}
\\
&{[} {\frak D}_{a}, {\frak D}_\b{]}
=
\ri\bar{\mu}(\g_a)_\b{}^\g\bar {\frak D}_{\g}
~,~~~
{[}{\frak D}_{a},\bar {\frak D}_\b{]}
=
-\ri\mu(\g_a)_\b{}^\g {\frak D}_{\g}
~,~~~~
\label{AdS_(1,1)_algebra_2} \\
&{[} {\frak D}_a, {\frak D}_b]{}
= -4 |\mu|^2 \cM_{ab} ~.
\label{AdS_(1,1)_algebra_3}
\eea
\esubeq
Unlike \eqref{1.8}, here and below we use the complex basis for $\cN=2$ covariant derivatives
introduced in \cite{KLT-M11}.

The isometries of the (1,1) AdS superspace are  described by Killing vector fields, 
$l =l^a {\frak D}_a+l^\a {\frak D}_\a+{\bar l}_\a\bar {\frak D}^\a$,
which are defined to obey the Killing equation
\bea
\Big{[}l+\hf\l^{ab}\cM_{ab}, {\frak D}_C\Big{]}=0~,
\eea
for a certain Lorentz parameter $\l^{ab} = - \l^{ba}$.
This is equivalent  to the following set of equations:\bsubeq
\bea
{\frak D}_\a\bar{l}_\b&=&
-\ri\mub \,l_{\a\b}
~,
\\
{\frak D}_\a l_{\b\g}&=&
4\ri\ve_{\a(\b}\bar{l}_{\g)}
~,
\\
{\frak D}_\a\l_{\b\g}&=&
8\mub \,\ve_{\a(\b}l_{\g)}
~,
\\
{\frak D}_\a l_\b&=&
\hf\l_{\a\b}
~,
\eea
\esubeq
and
\bsubeq
\bea
{\frak D}_al_b&=&
\l_{ab}
~,
\\
{\frak D}_al^\b&=&
\ri\mu\,\bar{l}_\g(\g_a)^{\g\b}
~,
\\
{\frak D}_a\l^{bc}&=&
4|\mu|^2\big(
\d_a{}^{b}l^{c}
-\d_a{}^{c}l^{b}
\big)
~.
\eea
\esubeq
The above equations are actually equivalent to the following relations
\begin{subequations}\label{1,1-SK}
\bea
&
l_{\a} = \frac{\ri}{6}
\bar {\frak D}^\b l_{\a\b} ~,
~~~
{\frak D}_{(\a} l_{\b\g)}=0~,~~~
{\frak D}_\a l^\a=\bar {\frak D}^\a l_\a=0~,~~~
\bar {\frak D}_{(\a}l_{\b)}
+\ri\mu\, l_{\a\b}=0
~,~~
\label{1,1-SK_1}
\\
&
\l_{\a\b} = 2 {\frak D}_{(\a}l_{\b)}
~,~~~
{\frak D}_{(\a}\l_{\b\g)}=0~,
~~~
{\frak D}^\b\l_{\a\b}
-12\mub\, l_{\a}=0~.
\label{1,1-SK_2}
\eea
\end{subequations}
It can be seen that the parameters  $l^\a$ and $\l^{ab}$ 
are determined in terms of the vector parameter $l^a$ obeying several constraints including the 
ordinary Killing equation\footnote{It may be seen that only the equation
${\mathfrak D}_{(\alpha} l_{\beta \gamma)} = 0$
is critical (along with the definitions of $l_\alpha$ and $\lambda_{\alpha \beta}$
in terms of $l_{\alpha \beta}$). This is the ``master equation'' from which all
the other constraints can be derived, compare with the 4D $\cN=1$  case \cite{BK}.}  
\bea
{\frak D}_a l_b + {\frak D}_b l_a =0~.
\eea
Its bosonic components defined by $l^a|_{\q=0}$ and 
$\l^{ab}|_{\q=0} =  (-{\frak D}^b l^a) |_{\q=0}$ describe the isometries of AdS$_3$. 
The only independent complex fermionic parameters are 
$l_{\a}|_{\q=0}  = \frac{\ri}{6}\bar {\frak D}^\b l_{\a\b}|_{\q=0}$ and its conjugate.
The Killing vector fields introduced generate the supergroup 
$\rm OSp(1|2;{\mathbb R}) \times OSp(1|2;{\mathbb R})$, the isometry group 
of the (1,1) AdS superspace.

In the $(1,1)$ AdS superspace, supersymmetric actions can be constructed either by integrating
a real function $\cL$ over the full superspace,
\begin{align}\label{eq_11AdSDcomp}
\int \rd^3x\, \rd^4\q\, E\, \cL
	&= \int \rd^3x\, e\, \Big(
	\frac{1}{16} {\frak D}^\alpha (\bar {\frak D}^2 - 6 \mu) {\frak D}_\alpha 
	- \frac{\mu}{4} {\frak D}^2 
	- \frac{\bar\mu}{4} \bar {\frak D}^2 
	+ 4 \mu \bar \mu 
	\Big)\cL\Big|_{\q=0}~,
\end{align}
or by integrating a chiral function $\cL_c$ over the chiral superspace,
\begin{align}
\int \rd^3x\, \rd^2\q\, \cE\, \cL_c
	= -\frac{1}{4} \int \rd^3x\, e\, ({\frak D}^2 - 16 \bar\mu) \cL_c~, \qquad
\bar{\frak D}^\alpha \cL_c =0~.
\end{align}
These two types of superspace integrals are related to each other
by the chiral action rule
\begin{align}
\int \rd^3x\, \rd^4\q\, E\, \cL &= -\frac{1}{4} \int \rd^3x\, \rd^2\q\, \cE\, (\bar{\frak D}^2 - 4 \mu) \cL
\end{align}
and its inverse
\begin{align}\label{eq_11AdSFtoD}
\int \rd^3x\, \rd^2\q\, \cE\, \cL_c = \int \rd^3x\, \rd^4\q\, E\, \frac{\cL_c}{\mu}~.
\end{align}
Eq.  \eqref{eq_11AdSFtoD} has no analogue in the $(2,0)$ AdS superspace.

The general form of a $(1,1)$ supersymmetric $\s$-model in AdS${}_3$ is
the single term
\begin{align}\label{eq_11AdSSigma}
\int \rd^3x\, \rd^4\q\, E\, \cK (\f^\ra , \bar \f^{\bar \rb} ) ~, \qquad \bar {\frak D}_\a \f^\ra =0~,
\end{align}
where $ \cK(\phi, \bar\phi)$ is a real function of chiral superfields
$\phi^\ra$ and their conjugates $\bar\phi^{\bar\ra}$. Since
\begin{align}
\int \rd^3x\, \rd^4\q\, E\, F = \int \rd^3x\, \rd^2\q\, \cE\, \mu F~,
\end{align}
for a holomorphic function $F = F(\phi)$, the model \eqref{eq_11AdSSigma}
does not possess the usual K\"ahler symmetry. Because 
the Lagrangian in \eqref{eq_11AdSSigma} 
corresponds to the K\"ahler potential of some K\"ahler manifold $\cM$, we conclude that 
$\cK (\f , \bar \f )$ should be  a globally defined function 
on $\cM$.
This immediately implies that the K\"ahler two-form, 
 $ \O=2\ri \,g_{\ra \bar \rb} \, \rd \f^\ra \wedge \rd \bar \f^{\bar \rb}$,  which is associated with 
the K\"ahler metric $g_{\ra \bar \rb} := \partial_\ra \partial_{\bar \rb} \cK $, 
is exact and hence  $\cM$ is necessarily non-compact. 
We see that the $\s$-model couplings with (1,1) AdS supersymmetry
 are more restrictive than in the Minkowski case, which is 
completely analogous to the observations made in \cite{Adams:2011vw, FS, BKsigma1} 
regarding the four-dimensional $\s$-models with $\cN=1$ AdS supersymmetry. 
We also see that in three dimensions the $\s$-model couplings with (2,0) and (1,1) 
AdS supersymmetry types are rather different. In particular, compact target spaces are allowed 
in the (2,0) case, while the (1,1) AdS supersymmetry is consistent only 
with non-compact K\"ahler manifolds. 

It is worth noting that one may reintroduce K\"ahler symmetry by
separating the function $\cK$ into a K\"ahler potential $K$ and a
holomorphic superpotential $W$,
\begin{align}\label{eq_11cKfromK}
\cK = K + \frac{W}{\mu} + \frac{\bar W}{\bar \mu}~.
\end{align}
Then the action may be written in a familiar way
\begin{align}\label{eq_11AdSSigmaKW}
\int \rd^3x\, \rd^4\q\, E\, K + \left(\int \rd^3x\, \rd^2\q \, \cE\, W + \textrm{c.c.}\right)~.
\end{align}
However, in this case, the K\"ahler symmetry manifests itself as
\begin{align}
K \rightarrow K + F + \bar F~, \qquad W \rightarrow W - \mu F~,
\end{align}
where $F = F(\phi)$ is a holomorphic function. This implies that
$\cK$ is the only physically meaningful quantity.

Using the component reduction rule \eqref{eq_11AdSDcomp}, one can show that
\begin{align}
\int \rd^3x\, \rd^4\q\, E\, \cK = \int \rd^3x\, e\, L_\cK
\end{align}
where
\begin{align}
L_\cK &= -g_{\ra \bar \rb} {\frak D}_m \varphi^\ra {\frak D}^m \bar\varphi^{\bar \rb}
	- \ri g_{\ra \bar \rb} \bar\psi_\alpha^{\bar \rb} \hat {\frak D}^{\alpha \beta} \psi_\beta^{\ra}
	+ F^\ra \bar F^{\bar \rb} g_{\ra \bar \rb}
	+ \frac{1}{4} R_{\ra \bar \rb \rc \bar \rd} (\psi^{\ra} \psi^\rc) (\bar\psi^{\bar \rb} \bar\psi^{\bar \rd})
	\eol & \quad
	- \frac{\mu}{2} \tsD_\ra \cK_\rb \,(\psi^{\ra} \psi^\rb)
	- \frac{\bar\mu}{2} \tsD_{\bar \ra} \cK_{\bar \rb} \,(\bar \psi^\ra \bar\psi^{\rb})
	+ \mu \cK_\ra F^a + \bar \mu \cK_{\bar \ra} \bar F^{\bar \ra}
	+ 4 \mu \bar \mu \cK~.
\end{align}
We have defined the components of $\phi^\ra$ as
\bsubeq
\begin{align}
\varphi^\ra &:= \phi^\ra\vert~, \qquad
\psi_\alpha^\ra := \frac{1}{\sqrt 2} {\frak D}_\alpha \phi^\ra \vert~, \\
F^\ra &:= -\frac{1}{4} g^{\ra \bar \rb} {\frak D}^2 \cK_{\bar \rb} \vert
	= -\frac{1}{4} ({\frak D}^2 \phi^\ra 
	+ \Gamma^\ra{}_{\rb\rc} {\frak D}^\alpha \phi^\rb {\frak D}_\alpha \phi^\rc)\vert
	~.
\end{align}
\esubeq
In particular, the auxiliary field $F^\ra$ transforms covariantly under reparametrizations.
The vector derivative on the fermion is similarly reparametrization covariant,
\begin{align}
\hat {\frak D}_m \psi_\alpha^\ra := {\frak D}_m \psi_\alpha^\ra + \Gamma^\ra{}_{\rb\rc} {\frak D}_m 
\varphi^\rb \psi_\alpha^\rc~.
\end{align}
The fermion mass terms are given in terms of covariant field derivatives
of the K\"ahler potential,
\begin{align}
\tsD_\ra \cK_\rb := \pa_\ra \pa_\rb \cK - \Gamma^\rc{}_{\ra\rb} \pa_\rc \cK~.
\end{align}
Eliminating the auxiliary field $F^\ra$ and its conjugate $\bar F^{\bar \ra}$ leads to
\begin{align}\label{eq_11compaction}
L &= -g_{\ra \bar \rb} {\frak D}_m \varphi^\ra {\frak D}^m \bar\varphi^{\bar \rb}
	- \ri g_{\ra \bar \rb} \bar\psi_\alpha^{\bar \rb} \hat {\frak D}^{\alpha \beta} \psi_\beta^{\ra}
	+ \frac{1}{4} R_{\ra \bar \rb \rc \bar \rd} (\psi^{\ra} \psi^\rc) (\bar\psi^{\bar \rb} \bar\psi^{\bar \rd})
	\eol & \quad
	- \frac{\mu}{2} \tsD_\ra \cK_\rb \,(\psi^{\ra} \psi^\rb)
	- \frac{\bar\mu}{2} \tsD_{\bar \ra} \cK_{\bar \rb} \,(\bar \psi^\ra \bar\psi^{\rb})
	- \mu \bar \mu g^{\ra \bar \rb} \cK_\ra \cK_{\bar \rb} 
	+ 4 \mu \bar \mu \cK~.
\end{align}
Note that there is generally a scalar potential in AdS.

Let us again consider the case where the target space is a K\"ahler cone.
This implies that there exists a homothetic conformal Killing vector $\chi$
from which one may construct a K\"ahler potential as
$K = g_{\ra \bar\rb} \chi^\ra \bar\chi^{\bar\rb}$.
However, $\cK$ may differ from this choice of $K$ by the real part of
a holomorphic field, which, inspired by \eqref{eq_11cKfromK},
we can choose to parametrize as
\begin{align}
\cK = K + \frac{W}{\mu} + \frac{\bar W}{\bar \mu}~.
\end{align}
If the action is superconformal, the holomorphic function $W$ must obey
\eqref{eq_Wconf} and so $\cK$ consequently obeys
\begin{align}
\chi^\ra \cK_\ra  = \cK + \frac{3}{\mu} W - \frac{1}{\bar \mu} \bar W
\end{align}
and the component action takes the form
\begin{align}
L &= -g_{\ra \bar \rb} {\frak D}_m \varphi^\ra {\frak D}^m \bar\varphi^{\bar \rb}
	- \ri g_{\ra \bar \rb} \bar\psi_\alpha^{\bar \rb} \hat {\frak D}^{\alpha \beta} \psi_\beta^{\ra}
	+ \frac{1}{4} R_{\ra \bar \rb \rc \bar \rd} (\psi^{\ra} \psi^\rc) (\bar\psi^{\bar \rb} \bar\psi^{\bar \rd})
	\eol & \quad
	- \frac{1}{2} \tsD_\ra W_\rb \,(\psi^{\ra} \psi^\rb)
	- \frac{1}{2} \tsD_{\bar \ra} \bar W_{\bar \rb} \,(\bar \psi^\ra \bar\psi^{\rb})
	+ 3 \mu\bar\mu K
	- g^{\ra \bar\rb} W_\ra W_{\bar \rb}~,
\end{align}
which, as in the $(2,0)$ case, can be rewritten as
\begin{align}
L &= g_{\ra \bar \rb} \chi^\ra (\hat {\frak D}_m \hat {\frak D}^m - \frac{1}{8} \cR) \bar\chi^{\bar \rb}
	- \ri g_{\ra \bar \rb} \bar\psi_\alpha^{\bar \rb} \hat {\frak D}^{\alpha \beta} \psi_\beta^{\ra}
	+ \frac{1}{4} R_{\ra \bar \rb \rc \bar \rd} (\psi^{\ra} \psi^\rc) (\bar\psi^{\bar \rb} \bar\psi^{\bar \rd})
	\eol & \quad
	- \frac{1}{2} \tsD_\ra W_\rb \,(\psi^{\ra} \psi^\rb)
	- \frac{1}{2} \tsD_{\bar \ra} \bar W_{\bar \rb} \,(\bar \psi^\ra \bar\psi^{\rb})
	- g^{\ra \bar\rb} W_\ra W_{\bar \rb}
	~.
\end{align}
This reveals that the mass terms arise solely from the holomorphic function $W$.


\section{Sigma models with (3,0) AdS supersymmetry: Off-shell approach}\label{section3}
\setcounter{equation}{0}

In this and the next sections we provide a detailed study of the nonlinear $\s$-models 
with (3,0) AdS supersymmetry. 

The off-shell  (3,0) supersymmetric $\s$-model in the (2,0) AdS superspace 
considered in \cite{KLT-M11} has the form
\begin{subequations} \label{(3,0)sssm}
\bea
S &=& \oint_\g  \frac{\rd \zeta}{2\pi \ri \zeta}
\int \rd^3x\, \rd^4{ \q}  \, {E}\, 
 K (\U^{I} ,  \breve{\U}^{\bar J} ) ~,
\label{(3,0)ProjAction}
\eea
with the contour integral being evaluated along a closed path $\g$ around the origin in $\mathbb C$. 
Here 
$K(\F^I, \bar \F^{\bar J})$ is a {\it real analytic} function subject to the homogeneity conditions 
\bea
 \F^I \frac{\pa}{\pa \F^I} 
K(\F, \bar \F) =  K( \F,   \bar \F) ~, \qquad
 \bar \F^{\bar I} \frac{\pa}{\pa \bar \F^{\bar I}} 
K(\F, \bar \F) =  K( \F,   \bar \F) ~.
\label{Kkahler}
\eea
\end{subequations}
This condition means that $K(\F,  \bar \F) $ can be interpreted as the K\"ahler potential 
of a K\"ahler cone $\cX$, see Appendix \ref{AppA}. 
The dynamical variables in \eqref{(3,0)ProjAction} are
covariant  {\it weight-one arctic} multiplets 
\bea
{ \U}^I (\z) = \sum_{n=0}^{\infty}  \, \z^n \U_n^I  = \F^I + \z \S^I + \dots~, 
\label{3.3}
\eea
 and their smile-conjugate {\it weight-one antarctic} multiplets
\bea
\breve{ \U}^{\bar I} (\z) =\sum_{n=0}^{\infty}  \,  (-\z)^{-n}\,
{\bar \U}_n^{\bar I}~.
\label{3.4}
\eea
Here the component superfields $\F^I:=\U^I_0$ and $\S^I :=\U^I_1$ are chiral and complex linear
respectively, 
\bea
\bar \cD_\a \F^I =0~, \qquad \bar \cD^2 \S^I =0~,
\label{3.5}
\eea
while the remaining Taylor coefficients in \eqref{3.3},  $\U^I_2, \U^I_3, \dots, $ 
are complex unconstrained superfields. The latter superfields are auxiliary, 
since they are unconstrained and appear in the action without derivatives.  
The superfields $\F^I$ and $\S^I$ are physical. 

The theory defined by the action \eqref{(3,0)ProjAction} and 
the homogeneity condition \eqref{Kkahler}
is not the most general off-shell nonlinear $\s$-model  \cite{KLT-M11} with (3,0) AdS supersymmetry. 
The latter is given by an action of the form: 
\begin{subequations} \label{(3,0)sssm-general}
\bea
S &=& \oint_\g  \frac{\rd \zeta}{2\pi \ri \zeta}
\int \rd^3x\, \rd^4{ \q}  \, {E}\, \cL (\U^{I} ,  \breve{\U}^{\bar J};\z )~, \qquad  
\cL (\U^{I} ,  \breve{\U}^{\bar J};\z ):=\frac{1}{\z}  {\frak K} (\U^{I} , \z \breve{\U}^{\bar J} ) ~,
\label{(3,0)ProjAction_strange}
\eea
where ${\frak K} (\F , \bar \O )$ is a homogeneous function of $2n$ 
complex variables $\F^I $ and $\bar \O^{\bar J} $, 
\bea
\Big( \F^I \frac{\pa}{\pa \F^I} + \bar  \O^{\bar I} \frac{\pa}{\pa \bar \O^{\bar I} } \Big) {\frak K} (\F, \bar \O )
= 2 {\frak K} (\F , \bar \O )~,
\label{hom_cond_strange}
\eea
under the reality condition
\bea
\bar {\frak K} (\bar \F , -\O ) = -{\frak K} (\O , \bar \F )~,
\label{reality_condition_strange}
\eea
\end{subequations}
where $\bar {\frak K} (\bar \F ,  \O)$ denotes the complex conjugate of  ${\frak K} ( \F , \bar \O )$.  
In the case that ${\frak K} (\F , \bar \O )$ is also homogeneous with respect to $\F$ (or, equivalently, 
with respect to  $\bar \O$), 
\bea
 \F^I \frac{\pa}{\pa \F^I}  {\frak K} (\F, \bar \O )
=  {\frak K} (\F , \bar \O )~, \qquad
\bar \O^{\bar I} \frac{\pa}{\pa\bar  \O^{\bar I}}  {\frak K} (\F, \bar \O )
=  {\frak K} (\F , \bar \O )~,
\label{hom_cond_strange2}
\eea
the reality condition \eqref{reality_condition_strange} is equivalent to 
$\bar {\frak K} (\bar \F , \F ) = {\frak K} (\F , \bar \F )$, that is ${\frak K} (\F , \bar \F )$
is real.

The (3,0) supersymmetric $\s$-model \eqref{(3,0)sssm} has a simple geometric interpretation. 
This theory is associated with a K\"ahler cone $\cX$ for which $K(\F, \bar \F)$ is the 
preferred K\"ahler potential (see Appendix A). The $\s$-model target space turns out to be 
the cotangent bundle $T^*\cX$ which is a hyperk\"ahler cone \cite{K-duality}.
In the case of the most general (3,0) supersymmetric $\s$-model \eqref{(3,0)sssm-general}, 
a geometric interpretation of ${\frak K} (\F , \bar \O )$ is unclear to us. 
However, as will be shown below, the $\s$-model target space is a hyperk\"ahler cone
provided the off-shell theory \eqref{(3,0)sssm-general} leads to a non-degenerate  metric for the 
target space. We will come back to a general discussion of the $\s$-model \eqref{(3,0)sssm}  
at the end of this section. Right now, we turn to eliminating the auxiliary superfields 
in the theory  \eqref{(3,0)sssm-general} in a formal way  (that is, we  assume that 
the function ${\frak K} (\F , \bar \O )$ is properly chosen such that all  the auxiliary superfields 
can be eliminated  in a unique way).

The fact that $\U^I(\z)$ is a covariant weight-one arctic multiplet is encoded in its transformation law 
under the (3,0) AdS isometry supergroup, $\rm OSp(3|2;{\mathbb R}) \times Sp(2,{\mathbb R}) $
given in \cite{KLT-M12}. 
When realized in the (2,0) AdS superspace, the most general  (3,0) isometry transformation of any 
(3,0) supermultiplet splits into two different transformations: (i) a (2,0) AdS isometry transformation
generated  by superfield parameters specified
 in eqs. \eqref{2_0-Killing_iso-def} and \eqref{2_0-Killing_iso}; 
(ii) an extended supersymmetry transformation generated by 
a chiral superfield parameter\footnote{Ref.  \cite{KLT-M12} used a different parameter, denoted  
$\ve$, which is related to $\r$ as $\ve = -8S \r$. }
$\r$ and its conjugate $\bar \r$ which are subject to the constraints 
\bea\label{eq_rho30constraints}
\bar \cD_\a \r =0~, \qquad 
\cD_\a \r = \bar \cD_\a \bar \r \equiv \r_\a~, \qquad 
\cD^2 \r = -8\ri S \bar \r \quad \Longrightarrow \quad \cD_{\ab}\r=0~.
\eea
These constraints prove to imply  that the only independent components of $\r$ are the following:
(a) the constant complex parameter $\r|_{\q =\bar \q=0} $ which describes an $R$-symmetry 
transformation from the coset $\rm SU(2)/U(1)$; (b) the Killing spinor parameter 
$\cD_\a \r|_{\q=\bar \q=0}$ which describes a third supersymmetry transformation. 
The (2,0) AdS isometry transformation of $\U^I (\z)$ is 
\bea
\d_\t \U^I = (\t + \ri t \cJ) \U^I~, \qquad
\cJ = \Big( \,\z\frac{\pa}{\pa \z}
-\frac{1}{2} \Big) \quad \Longleftrightarrow \quad
\cJ \U^I_n = (n-\hf )\U^I_n ~. 
\label{shadow0}
\eea
In particular, the U$(1)_R$ charges of the dynamical superfields are
\bea
\cJ \F^I = -\hf \F^I~, \qquad \cJ \S^I = \hf \S^I~.
\label{charges}
\eea
A finite transformation generated by $\cJ$ is 
 \bea
\U(\z) \quad &\longrightarrow & \quad \U'(\z)=
{\rm e}^{-({\rm i} /2) \a}\, \U({\rm e}^{{\rm i} \a} \z)
~,\qquad
\a \in {\mathbb R}
~, 
\label{shadow4}
\eea
or in components
\begin{align}
\Phi \rightarrow \re^{-\ri \alpha/2} \Phi~, \qquad
\Sigma \rightarrow \re^{\ri \alpha/2} \Sigma~, \qquad
\U_2 \rightarrow \re^{3\ri \alpha/2} \U_2~, \quad \cdots~
\label{shadow4a}
\end{align}
This transformation coincides with the so-called shadow chiral rotation 
introduced in the context of 4D $ \cN=2$ superconformal $\s$-models
\cite{K-duality}. 
It is an instructive exercise to show that the transformation \eqref{shadow4}
is a symmetry of the theory \eqref{(3,0)sssm-general}.

The extended supersymmetry transformation of $\U^I$ is 
\bea
\d_\r \U^I = \Big\{  \ri\z\r^\a \cD_\a
+\frac{\ri}{\z}\r_\a \bar \cD^\a
- 4S \Big(\z \bar{\r}+ \frac{1}{\z} \r\Big) \z \frac{\pa}{\pa \z}
+4 S\,\z\bar{\r}
 \Big\} \U^I~.
\label{(3,0)SUSYtran}
\eea
For the physical superfields, this transformation law leads to 
\begin{subequations}\label{SUSY(3.10)}
\bea
\d_\r \F^I &=& (\ri \r_\a \bar \cD^\a - 4 S \r )\S^I
=  \frac{\ri}{2} \bar \cD^2 (\bar \r \S^I)~,   \label{SUSY(3.10)-a} \\
\d_\r \S^I &=& ( \ri \r^\a \cD_\a  +4S \bar \r ) \F^I 
+(\ri \r_\a \bar \cD^\a -8S\r)\U^I_2 
= \ri \bar \cD_\a (\r^\a \U^I_2 -\bar \r \cD^\a \F^I)~.~~~
\label{SUSY(3.10)-b}
\eea
\end{subequations}
It is seen that the variations $\d_\r \F^I$ and $\d_\r \S^I$ are chiral and complex linear respectively. 

As mentioned earlier, the complex unconstrained 
superfields   $\U^I_2, \U^I_3, \dots, $ and their conjugates 
appear in the action without derivatives, and therefore they are auxiliary. 
These superfields can in principle be eliminated using their algebraic nonlinear equations of motion, 
\begin{subequations}\label{AuxEOM(3,0)AdS}
\begin{alignat}{2}
\frac{\pa \mathbb L}{\pa \U_n^I} =  \oint_\g \frac{\rd\zeta}{2\pi \ri \zeta} 
\frac{\pa \cL}{\pa \U^I} \zeta^n  &=0~,
	&\qquad &n\geq 2~; \\
\frac{\pa \mathbb L}{\pa \bar\U_n^{\bar J}} =  \oint_\g \frac{\rd\zeta}{2\pi \ri \zeta} 
\frac{\pa \cL}{\pa \breve\U^{\bar J}} {(-\zeta)}^{-n} &=0~,
	&\qquad &n\geq 2~.
\end{alignat}
\end{subequations}
Here we have introduced the Lagrangian\footnote{In the case that ${\frak K}$ obeys 
the stronger homogeneity conditions 
\eqref{hom_cond_strange2}, the equations 
 \eqref{AuxEOM(3,0)AdS} coincide with the auxiliary superfield equations 
corresponding to the  4D $\cN=2$ supersymmetric $\s$-models 
on cotangent bundles of K\"ahler cones \cite{K-duality}, see also \cite{GK1,GK2}.} 
\bea
{\mathbb L} (\U_n , \bar \U_n) =  \oint_\g \frac{\rd \zeta}{2\pi \ri \zeta} \,  \cL(\U, \breve \U;\z)
= \oint_\g \frac{\rd \zeta}{2\pi \ri \zeta} \, 
\frac{1}{\z} {\frak K} (\U, \z \breve \U)
~.
\label{3.11}
\eea
If $\U^I(\z)$ and $\breve{\U}^{\bar I}(\z)$ give a solution of the auxiliary superfield equations 
\eqref{AuxEOM(3,0)AdS}, then $r \U^I(\z) $ 
and $r\breve{\U}^{\bar I}(\z)$ is also a solution for any $r \in {\mathbb R}-\{0\}$, 
as a consequence of the homogeneity condition \eqref{hom_cond_strange}.

Once the equations \eqref{AuxEOM(3,0)AdS} have been solved
and all the auxiliary superfields are expressed in terms of the physical ones,
 the Lagrangian 
 \eqref{3.11}  becomes a function of the physical superfields, 
 ${\mathbb L}(\F^I, \S^I , \bar \F^{\bar J}, \bar \S^{\bar J} )$, 
and the action reads
\bea
S= \int \rd^3x\, \rd^4\q\,  E\, {\mathbb L}(\F^I, \S^I , \bar \F^{\bar J}, \bar \S^{\bar J} )~.
\label{3.13}
\eea

The homogeneity condition \eqref{hom_cond_strange}
can be recast in the form
\bea
\sum_{k=0}^{\infty} \Big( 
\U_k^I \frac{\pa }{\pa \U_k^I} 
+ \bar \U^{\bar I}_k  \frac{\pa }{\pa \bar\U_k^{\bar I}} \Big) \, { \mathbb L}(\U_n , \bar \U_n) = 
2{ \mathbb L}(\U_n , \bar \U_n) ~.
 \eea
If the auxiliary superfield equations 
\eqref{AuxEOM(3,0)AdS} hold, then this condition simplifies 
\bea
\Big(  \Phi^I \frac{\pa }{\pa \Phi^I}+ \Sigma^I \frac{\pa }{\pa \Sigma^I}
	+ \bar\Phi^{\bar I} \frac{\pa }{\pa \bar\Phi^{\bar I}}
	+ \bar\Sigma^{\bar I} \frac{\pa }{\pa \bar\Sigma^{\bar I}} \Big) \mathbb L =2 \mathbb L ~, 
	\qquad {\mathbb L} = {\mathbb L}(\F, \S, \bar \F, \bar \S )~.
\eea
Under the  equations 
\eqref{AuxEOM(3,0)AdS},  the property that the theory has the U$(1)_R$ 
symmetry  \eqref{shadow4a} means
\bea
\Big(  \Phi^I \frac{\pa }{\pa \Phi^I}- \Sigma^I \frac{\pa }{\pa \Sigma^I}
	- \bar\Phi^{\bar I} \frac{\pa }{\pa \bar\Phi^{\bar I}}
	+ \bar\Sigma^{\bar I} \frac{\pa }{\pa \bar\Sigma^{\bar I}} \Big) \mathbb L =0 ~.
\eea
Combining these two results gives the homogeneity conditions 
\bea
\Big(  \Phi^I \frac{\pa }{\pa \Phi^I}
	+ \bar\Sigma^{\bar I} \frac{\pa }{\pa \bar\Sigma^{\bar I}} \Big) \mathbb L &=& \mathbb L ~, 
	\qquad
\Big(  \bar\Phi^{\bar I} \frac{\pa }{\pa \bar\Phi^{\bar I}}
+ \Sigma^I \frac{\pa }{\pa \Sigma^I}
	\Big) \mathbb L = \mathbb L ~.
\label{BlackboardL-hom}
\eea

By construction, the action \eqref{3.13} must be invariant under the extended 
supersymmetry transformation \eqref{SUSY(3.10)}
in which  $\U^I_2 = \U^I_2 (\F, \S, \bar \F,  \bar \S)$ is part of the solution of the auxiliary equations 
\eqref{AuxEOM(3,0)AdS}. In practice, the functions $\U^I_2$ are known only for special 
manifolds. However, supersymmetry considerations \cite{K-duality,K-comments}  allow one 
to develop a self-consistent scheme to determine both the Lagrangian 
${\mathbb L}(\F , \S , \bar \F , \bar \S )$ and the functions
$ \U^I_2 (\F, \S, \bar \F,  \bar \S)$ appearing in the supersymmetry transformation law 
\eqref{SUSY(3.10)}. Requiring the action \eqref{3.13} to be invariant under \eqref{SUSY(3.10)}
leads to several equations 
\begin{subequations}\label{eq_XiEqnsAdS}
\begin{align}
\frac{\partial \mathbb L}{\partial \Phi^I} +
	\frac{\partial \mathbb L}{\partial \Sigma^J} \frac{\partial \Upsilon_2^J}{\partial \Sigma^I}
	&= \frac{\partial \Xi}{\partial \Sigma^I}~,   \label{eq_XiEqnsAdS-a} \\
-\frac{\partial \mathbb L}{\partial \bar\Sigma^{\bar I}} +
	\frac{\partial \mathbb L}{\partial \Sigma^J} \frac{\partial \Upsilon_2^J}{\partial \bar \Phi^{\bar I}}
	&= \frac{\partial \Xi}{\partial \bar\Phi^{\bar I}}~,  \label{eq_XiEqnsAdS-b} \\
\frac{ \partial \mathbb L}{\partial \Sigma^J} \frac{\partial \Upsilon_2^J}{\partial \bar \Sigma^{\bar I}}
	&= \frac{\partial \Xi}{\partial \bar\Sigma^{\bar I}}~,
	\label{eq_XiEqnsAdS-c}
\end{align}
\end{subequations}
as well as 
\bea
2\X = - {\bar \F}^{\bar I} \frac{\pa \mathbb L}{\pa {\bar \S}^{\bar I} } 
+  {\S}^{ I} \frac{\pa \mathbb L}{\pa { \F}^{ I} } + 2 {\U}_2^{ I} \frac{\pa \mathbb L}{\pa { \S}^{ I} },
\label{X-first}
\eea
for some function $\X (\F, \S, \bar \F,  \bar \S)$.
The existence of $\X$ satisfying the three requirements  
\eqref{eq_XiEqnsAdS-a} -- \eqref{eq_XiEqnsAdS-c}
can be proved by using 
the contour integral definition of $\mathbb L$, eq. \eqref{3.11}. 
It is an instructive exercise to check that the function 
$\X$ defined as  (compare with \cite{BKLT-M})
\begin{align}
\Xi &:=  \oint_\g\frac{\rd\zeta}{2\pi \ri \zeta} 
\frac{1}{\z^2} {\frak K} (\U, \z \breve \U)
\label{3.20}
\end{align}
does obey the conditions \eqref{eq_XiEqnsAdS}. 

Eq. \eqref{X-first}  can actually be deduced from the conditions \eqref{eq_XiEqnsAdS} in conjunction 
with some additional observations. The first observation is that 
(i) $\U^I_2$ is 
a homogeneous function of $ \U^I(\z) $ and $\breve{\U}^{\bar I} (\z)$
of degree one; and (ii) if $\U^I(\z)$ is a solution of the auxiliary equations 
\eqref{AuxEOM(3,0)AdS}, then $r \U^I(\z) $ is also a solution for any $r \in {\mathbb R}-\{0\}$,
as a consequence of the homogeneity condition \eqref{hom_cond_strange}.
Upon elimination of the auxiliary superfields, this means
\begin{subequations}
\bea
\Big(  \Phi^I \frac{\pa }{\pa \Phi^I}+ \Sigma^I \frac{\pa }{\pa \Sigma^I}
	+ \bar\Phi^{\bar I} \frac{\pa }{\pa \bar\Phi^{\bar I}}
	+ \bar\Sigma^{\bar I} \frac{\pa }{\pa \bar\Sigma^{\bar I}} \Big) \U^J_2 = \U^J_2  ~,
	\label{U2-hom}
\eea
where $\U^J_2 = \U^J_2 (\F, \S, \bar \F,  \bar \S)$.
The second observation is that the symmetry transformation \eqref{shadow4a}
acts on  $ \U^J_2 (\F, \S, \bar \F,  \bar \S)$ exactly as in \eqref{shadow4a}, and thus
\bea
\Big(  \Phi^I \frac{\pa }{\pa \Phi^I}- \bar\Phi^{\bar I} \frac{\pa }{\pa \bar\Phi^{\bar I}}
- \Sigma^I \frac{\pa }{\pa \Sigma^I}	
	+ \bar\Sigma^{\bar I} \frac{\pa }{\pa \bar\Sigma^{\bar I}} \Big) \U^J_2 = -3\U^J_2~.
	\label{U2-sym}
\eea
\end{subequations}
Combining these two results gives
\bea
\Big(  \Phi^I \frac{\pa }{\pa \Phi^I}
	+ \bar\Sigma^{\bar I} \frac{\pa }{\pa \bar\Sigma^{\bar I}} \Big) \U^J_2 =- \U^J_2  ~, 
	\qquad
\Big(  \bar\Phi^{\bar I} \frac{\pa }{\pa \bar\Phi^{\bar I}}
+ \Sigma^I \frac{\pa }{\pa \Sigma^I}	
	 \Big) \U^J_2 = 2\U^J_2~.
	 \label{U2-hom+sym}
\eea
The third observation is that 
analogs of eqs. \eqref{U2-hom} and \eqref{U2-sym} may be derived for the function $\X$  
by making use of the contour-integral representation \eqref{3.20}. Specifically, one derives
\begin{subequations}
\bea
\Big(  \Phi^I \frac{\pa }{\pa \Phi^I}+ \Sigma^I \frac{\pa }{\pa \Sigma^I}
	+ \bar\Phi^{\bar I} \frac{\pa }{\pa \bar\Phi^{\bar I}}
	+ \bar\Sigma^{\bar I} \frac{\pa }{\pa \bar\Sigma^{\bar I}} \Big) \X &=& 2\X  ~, \\
	\Big(  \Phi^I \frac{\pa }{\pa \Phi^I}- \bar\Phi^{\bar I} \frac{\pa }{\pa \bar\Phi^{\bar I}}
- \Sigma^I \frac{\pa }{\pa \Sigma^I}	
	+ \bar\Sigma^{\bar I} \frac{\pa }{\pa \bar\Sigma^{\bar I}} \Big) \X &=& -2\X~.
\eea
\end{subequations}
Combining these two results gives
\bea
\Big(  \Phi^I \frac{\pa }{\pa \Phi^I}
	+ \bar\Sigma^{\bar I} \frac{\pa }{\pa \bar\Sigma^{\bar I}} \Big) \X =0  ~, 
	\qquad
\Big(  \bar\Phi^{\bar I} \frac{\pa }{\pa \bar\Phi^{\bar I}}
+ \Sigma^I \frac{\pa }{\pa \Sigma^I}	
	 \Big) \X = 2\X~.
 \label{U2-hom+sym-X}
\eea
Now, applying the relations \eqref{eq_XiEqnsAdS}, \eqref{U2-hom+sym} and \eqref{U2-hom+sym-X}
gives (with all indices suppressed)
\begin{align}
2\Xi &=  \bar\Phi \frac{\pa \Xi}{\pa \bar\Phi}
	+ \Sigma \frac{\pa \Xi}{\pa \Sigma} 
	= \bar\Phi \Big(-\frac{\pa\mathbb L}{\pa\bar\Sigma} 
	+ \frac{\pa\mathbb L}{\pa\Sigma} \frac{\pa\U_2}{\pa \bar\Phi}\Big)
	+ \Sigma \Big(\frac{\pa\mathbb L}{\pa\Phi} 
	+ \frac{\pa\mathbb L}{\pa\Sigma} \frac{\pa\U_2}{\pa \Sigma}\Big) \eol
	&= -\bar\Phi\frac{\pa\mathbb L}{\pa\bar\Sigma}
	+ \Sigma \frac{\pa\mathbb L}{\pa\Phi}
	+ \frac{\pa\mathbb L}{\pa\Sigma} \Big(\bar \Phi \frac{\pa\U_2}{\pa \bar\Phi}
		+ \Sigma\frac{\pa\U_2}{\pa \Sigma}\Big) 
	= -\bar\Phi\frac{\pa\mathbb L}{\pa\bar\Sigma}
	+ \Sigma \frac{\pa\mathbb L}{\pa\Phi}
	+ \frac{\pa\mathbb L}{\pa\Sigma} 2 \U_2~,
	\non
\end{align}
which is exactly the equation \eqref{X-first}.

In order to understand the target space geometry 
of the $\s$-model \eqref{3.13}, we have to construct a dual formulation for this theory
in which
the complex linear superfield
$\S^I$ is traded for a covariantly chiral superfield $\Psi_I$, $\bar \cD_\a \J_I =0$.
For this, we introduce the first-order action
\begin{align}\label{3.17FO}
S_{\text{first-order}} &= \int \rd^3x\, \rd^4\q\,  E\,
	\Big({\mathbb L} + \S^I \Psi_I + \bar \S^{\bar J} \bar \Psi_{\bar J}\Big)~, \qquad 
	\cJ \J_I = -\hf \J_I~,
\end{align}
in which the superfields $\S^I$ are complex unconstrained.  
The choice of the U$(1)_R$ charge of $\J_I$ follows from \eqref{charges}. 
The first-order model introduced is equivalent to the original $\s$-model \eqref{3.13}.
Indeed, varying \eqref{3.17FO} with respect to $\J_I$ enforces
complex linearity of $\S^I$,  
and then $S_{\text{first-order}} $ reduces to the original action. 
On the other hand,  if we apply the equation of motion for the unconstrained $\S^I$, we find
\begin{align}\label{eq_dualActionAdS}
S_{\rm dual} &= \int \rd^3x\, \rd^4\q\,  E\, {\mathbb K}(\F^I, \J_I , \bar \F^{\bar J}, \bar \J_{\bar J} )~,
\end{align}
where $\mathbb K$ is defined as
\begin{align}
{\mathbb K} := {\mathbb L} + \S^I \Psi_I + \bar \S^{\bar J} \bar \Psi_{\bar J}
\end{align}
and $\S^I$ is chosen to obey
\begin{align}
\frac{\pa \mathbb L}{\pa \S^I} = - \Psi_I~.
\end{align}
The homogeneity conditions \eqref{BlackboardL-hom} turn into 
\bea
\Big( \F^I \frac{\pa}{\pa \F^I}  +  { \J}_{ I}   \frac{\pa }{\pa { \J}_{ I} }
 \Big) {\mathbb K} = {\mathbb K}  ~, \qquad
 \Big( \bar \F^{\bar I} \frac{\pa}{\pa \bar \F^{\bar I}}  +  {\bar  \J}_{\bar I}   \frac{\pa }{\pa {\bar  \J}_{\bar I} }
 \Big) {\mathbb K} = {\mathbb K} ~.
\label{homo5}
\eea

It is easy to work out the extended supersymmetry transformation of the first-order action 
\eqref{3.17FO}
starting from the original transformation \eqref{SUSY(3.10)}.
Since $\S^I$ is complex unconstrained in  \eqref{SUSY(3.10)}, 
the supersymmetry transformation of $\F^I$ turns into
\bea
\d_\r \F^I 
=  \frac{\ri}{2} \bar \cD^2 (\bar \r \S^I)
= (\ri \r_\a \bar \cD^\a - 4 S \r )\S^I + \frac{\ri}{2}   \bar \r \bar \cD^2 \S^I~.
\label{3.31}
\eea
We keep  the transformation law of $\S^I$ unchanged, eq.  \eqref{SUSY(3.10)-b}. 
Then, it may be seen that $S_{\text{first-order}}$ is invariant provided $\J_I$ transforms 
as follows:
\bea
\d_\r \J_I 
=  -\frac{\ri}{2} \bar \cD^2 \Big(\bar \r \frac{\pa \mathbb L}{\pa \F^I}\Big)~.
\label{3.32}
\eea
${}$From \eqref{3.31} and \eqref{3.32} we read off the extended supersymmetry 
of the dual action \eqref{eq_dualActionAdS}: 
\bea
\d_\r \F^I 
=  \frac{\ri}{2} \bar \cD^2 \Big(\bar \r \frac{\pa \mathbb K}{\pa \J_I}\Big)~, \qquad
\d_\r \J_I 
=  -\frac{\ri}{2} \bar \cD^2 \Big(\bar \r \frac{\pa \mathbb K}{\pa \F^I}\Big)~.
\label{ESDA(3.0)}
\eea
If  we introduce the condensed notation 
\bea
\f^\ra := (\F^I\,, \J_I) ~, \qquad {\bar \f}^{\,\bar \ra} = ({\bar \F}^{\bar I}\,, {\bar \J}_{\bar I})~, 
\label{nott}
\eea
the above transformation law can be rewritten as 
\bea
\d_\r \f^\ra 
=  \frac{\ri}{2} \bar \cD^2 \Big(\bar \r \o^{\ra\rb} \frac{\pa \mathbb K}{\pa \f^\rb}\Big)~,
\label{SST(3,0)}
\eea
where 
\begin{align}
\omega^{\ra\rb} =
\begin{pmatrix}
0 & \delta^I{}_J \\
-\delta_I{}^J & 0
\end{pmatrix}~, \qquad
\omega_{\ra\rb} =
\begin{pmatrix}
0 & \delta_I{}^J \\
-\delta^I{}_J & 0
\end{pmatrix}~.
\end{align}
In accordance with the analysis in the next section, the $\s$-model target space 
is a hyperk\"ahler manifold. The structure of the supersymmetry transformation 
\eqref{SST(3,0)} shows that $\o^{(2,0)} = \o_{\ra\rb}\, \rd \f^\ra \wedge \rd \f^\rb $ 
is a covariantly constant holomorphic two-form
associated with the two complex structures orthogonal to the diagonal one. 
The explicit form of $\o_{\ra\rb}$ shows that the coordinates $\Phi^I$ and $\Psi_I$  
are holomorphic Darboux coordinates for the target space.

The relation \eqref{homo5} means that the target space possesses a homothetic conformal 
Killing vector $\c$ which looks like 
 \bea
\c = \f^\ra \frac{\pa}{\pa \f^\ra} + {\bar \f}^{\bar \ra}  \frac{\pa}{\pa {\bar \f}^{\bar \ra}}~.
\eea
The holomorphic Killing vector $J$ associated with the $R$-symmetry generator $\cJ$ 
is 
\bea
J^\ra:= \ri \cJ \f^\ra = -\frac{\ri}{2} \f^\ra~.
\eea 
It is obvious that the vector fields $\c$ and $J$ commute, $[\c, J]=0$.

Let us now discuss the family of $(3,0)$  supersymmetric $\s$-models described
by the action \eqref{(3,0)ProjAction} and  the homogeneity condition \eqref{Kkahler}.
In terms of the most general off-shell  action \eqref{(3,0)ProjAction_strange}, 
these theories are characterized by ${\frak K} (\U , \z \breve{\U} ) =\z{K} (\U,  \breve{\U} ) $.
The specific feature of such $\s$-models is the rigid U(1) symmetry
 \bea
\U(\z) \quad &\longrightarrow & \quad \U'(\z)=
{\rm e}^{-({\rm i}/2)  \a}\, \U( \z)
~,\qquad
\a \in {\mathbb R}
~, 
\label{internal_symmetry}
\eea
in addition to  the U$(1)_R$ symmetry \eqref{shadow4}. This internal symmetry has a number of 
nontrivial implications. To uncover them, it is useful to consider a combination 
of the symmetry transformations \eqref{shadow4} and \eqref{internal_symmetry} given by 
$\U(\z) \to \U'(\z)= \U({\rm e}^{{\rm i}  \a} \z)$.
This invariance implies that 
\bea
{ \S}^{ I} \frac{\pa \mathbb L}{\pa { \S}^{ I} }  &=&
{\bar \S}^{\bar I}
\frac{\pa \mathbb L}{\pa {\bar \S}^{\bar I} }  ~.
\label{semi-homo1} 
\eea
The same invariance enforces  
certain conditions on the functions $\U^I_2$ and $\X$ \cite{K-duality}: 
\begin{subequations}
\bea
{ \S}^{ J} \frac{\pa \U_2^I}{\pa { \S}^{ J} } &=&
{\bar \S}^{\bar J} \frac{\pa \U_2^I}{\pa {\bar \S}^{\bar J} }  +2\U_2^I~, 
 \label{semi-homo2}\\
   { \S}^{ J}
\frac{\pa \X}{\pa { \S}^{ J} } &=&{\bar \S}^{\bar J}
 \frac{\pa \X}{\pa {\bar \S}^{\bar J} }  +\X~.
\label{semi-homo3}
\eea
\end{subequations} 
These relations in conjunction with eqs. \eqref{eq_XiEqnsAdS-a}
and \eqref{eq_XiEqnsAdS-c} give
\bea
\X =   {\S}^{ I} \frac{\pa \cL}{\pa { \F}^{ I} } + 2 {\U}_2^{ I} \frac{\pa \cL}{\pa { \S}^{ I} }~.
\label{X-second}
\eea
This is compatible with \eqref{X-first} provided an alternative representation for $\X$ holds:
\bea
\X = - {\bar \F}^{\bar I} \frac{\pa \cL}{\pa {\bar \S}^{\bar I} } ~.
\label{X-third}
\eea

In the dual formulation, the condition of U(1) invariance, eq. \eqref{semi-homo1}, turns into
\bea
{ \J}_{ I} \frac{\pa \mathbb K}{\pa { \J}_{ I} }  &=& {\bar \J}_{\bar I}
\frac{\pa \mathbb K}{\pa {\bar \J}_{\bar I} }  ~.
\eea
Using this along with the hyperk\"ahler cone conditions \eqref{homo5}, we can show that
\bea
{ \F}^{ I} \frac{\pa \mathbb K}{\pa { \F}^{ I} }  - { \J}_{ I}
\frac{\pa \mathbb K}{\pa { \J}_{ I} } 
+ \textrm{c.c.} = 0
\eea
which, one can check, is a tri-holomorphic isometry,  
\bea
X^\ra = (\ri\Phi^I , -\ri \Psi_I)~, \qquad 
X^\ra  {\mathbb K}_\ra +  \bar X^{\bar \ra} {\mathbb K}_{\bar \ra} =0~,
\qquad \cL_X \omega_{\ra \rb}=0~,
\eea
of the hyperk\"ahler cone under consideration. 

In summary, the characteristic feature of all (3,0) supersymmetric $\s$-models, which are defined by 
the action \eqref{(3,0)ProjAction} and  the homogeneity condition \eqref{Kkahler}, 
is that the corresponding target space is a 
hyperk\"ahler cone  with a  U(1) tri-holomorphic isometry.
These models are a subclass of the general case
\eqref{(3,0)sssm-general}, whose target
space is required merely to be a hyperl\"ahler cone, but whether they
are a \emph{proper} subclass
remains unclear to us.

In the case of the $\s$-model \eqref{(3,0)sssm}, it is not difficult to see that the 
target space metric is non-degenerate in a neighborhood  of the zero section of $T^*\cX$. 
This follows from the explicit structure of the corresponding K\"ahler potential  \cite{K-duality}:
\bea
{\mathbb K} (\f^\ra , \bar \f^{\bar \rb} ) \equiv
{\mathbb K} (\F^I ,  \J_I, \bar \F^{\bar J},  \bar \J_{\bar J}) = K \big( \F, \bar{\F} \big)+    
\cH \big(\F,  \J ,\bar \F, \bar \J \big)~, 
\label{10.1}
\eea
where
\begin{align}
\cH \big(\F,  \J , \bar \F,  \bar \J \big)&= g^{I {\bar J}} \big( \F, \bar{\F} \big) \J_I \bar \J_{\bar J}
+\sum_{n=2}^{\infty} \cH^{I_1 \cdots I_n {\bar J}_1 \cdots {\bar 
J}_n }  \big( \F, \bar{\F} \big) \J_{I_1} \dots \J_{I_n} 
{\bar \J}_{ {\bar J}_1 } \dots {\bar \J}_{ {\bar J}_n } ~.~~~~~
\label{10.2}
\end{align} 
Here  the coefficients $\cH^{I_1 \cdots I_n {\bar J}_1 \cdots {\bar 
J}_n }$, with  $n=2,3,\dots$, 
are  real tensor functions of the K\"ahler metric
$g_{I \bar{J}} \big( \F, \bar{\F}  \big) 
= \pa_I 
\pa_ {\bar J}K ( \F , \bar{\F} )$ on $\cX$,  
the Riemann curvature $R_{I {\bar J} K {\bar L}} \big( \F, \bar{\F} \big) $ and its covariant 
derivatives. 
In accordance with \eqref{homo5}, $\cH^{I_1 \cdots I_n {\bar J}_1 \cdots {\bar 
J}_n }$ is a   homogeneous function of $\F^K$  of degree $(1-n)$.  

If we turn to the more general $\s$-model \eqref{(3,0)sssm-general}, 
the requirement that the target space metric be non-singular should be satisfied only 
under some restrictions on the function ${\frak K} (\F , \bar \O )$  appearing in the 
off-shell action. The explicit form of such restrictions remains unclear to us.
Even if such restrictions are met, different choices of  
${\frak K} (\F , \bar \O )$ 
may lead to the same target space geometry. 
As an illustration, we discuss the case of a single superconformal hypermultiplet.
It is known that any four-dimensional hyperk\"ahler cone 
is a {\it flat } space, for its Riemann curvature  identically vanishes \cite{deWKV}.  
To describe such a hyperk\"ahler cone, it suffices to use  
the off-shell $\s$-model  \eqref{(3,0)sssm} with $K(\U, \breve{ \U}) = \breve{ \U} \U$.
The corresponding hyperk\"ahler potential is 
$ {\mathbb K}(\F , \J , \bar \F , \bar \J ) = \bar \F \F + \bar \J \J$, and the hyperk\"ahler metric is 
{\it manifestly} flat, $g_{\ra\bar \rb} = \d_{\ra \bar \rb}$. 
On the other hand, for  a single hypermultiplet 
there exist infinitely many $\s$-models 
  \eqref{(3,0)sssm-general} of the form 
\bea 
\cL (\U ,  \breve{\U} ;\z )=\frac{1}{\z}  {\frak K} (\U, \z \breve{\U} ) 
=  \breve{ \U} \U\, f\Big( \frac{\z \breve{\U} }{\U} \Big)~,
 \eea
with $f (z) $ a function satisfying the  reality condition $\bar f (z) = f(-1/z)$. 
The corresponding target space is always flat, but the hyperk\"ahler metric generated 
in complex coordinates $\F$ and $\J$ 
is {\it not manifestly} flat.


\section{Sigma models with (3,0) AdS supersymmetry: On-shell approach}
\label{section4}
\setcounter{equation}{0}

In the previous section, it was shown that off-shell $(3,0)$ $\s$-models
lead to a formulation in $(2,0)$ superspace involving the  K\"ahler
potential ${\mathbb K}(\F , \J , \bar \F , \bar \J )$
of a hyperk\"ahler cone $\cM$, 
where the  chiral superfields $\Phi^I$ and $\Psi_I$  correspond to a choice
of complex Darboux coordinates on $\cM$.
We may proceed instead ``from the ground up'' and consider the most general $(2,0)$
models which exhibit a $(3,0)$ extended supersymmetry.

A $(2,0)$ supersymmetric $\s$-model without a superpotential is given by the action
\begin{align}
S = \int \rd^3x\, \rd^4\q\,E\, K
\end{align}
for a K\"ahler potential $K = K(\phi, \bar\phi)$.
The chiral superfield $\phi^\ra$ must transform under $(2,0)$ AdS supersymmetry as
\begin{align}
\delta \phi^\ra = \tau \phi^\ra + t J^\ra~, \qquad J^\ra := \ri \cJ \phi^\ra~,
\end{align}
where $J^\ra$ is a holomorphic Killing vector.

Suppose now that the action possesses a full $(3,0)$ supersymmetry, with $\phi^\ra$
transforming under the extended supersymmetry as
\begin{align}
\delta \phi^\ra = \frac{\ri}{2} \bar\cD^2 (\bar \rho \,\Omega^\ra)
\end{align}
for some function $\Omega^\ra(\phi,\bar\phi)$, where $\bar\rho$ obeys the
same conditions \eqref{eq_rho30constraints} as in the previous section.
The variation of the action must be zero,
\begin{align}\label{eq_varyS30}
0 = \delta S = \int \rd^3x\, \rd^4\q\,E\,\Big(
	\frac{\ri}{2} K_\ra \bar \cD^2 (\bar \rho \Omega^\ra)
	- \frac{\ri}{2} K_{\bar \ra} \cD^2 (\rho \Omega^{\bar \ra})
	\Big)~.
\end{align}
This condition, together with the requirement that the algebra should close,
imposes a number of conditions on the function $\Omega^\ra$ as well as the target space.
The analysis is somewhat technical, so we delay the discussion to appendix \ref{SUSY_Derivs}.
It turns out that the following properties must hold:
\begin{enumerate}

\item The target space possesses a homothetic conformal Killing vector $\chi^\ra$
(see Appendix A)
which is related to the $\rm U(1)$ Killing vector by $J^\ra = -\dfrac{\ri}{2} \chi^\ra$.

\item The quantity $\omega_{\ra \rb} := g_{\ra \bar \ra} \pa_\rb \Omega^{\bar \ra}$
is a covariantly constant holomorphic two-form, $\nabla_\rc \o_{\ra \rb} =  
\nabla_{\bar \rc} \o_{\ra \rb} =0$,
obeying the equation $\omega^{\ra\rb} \omega_{\rb\rc} = -\delta^\ra_\rc$, 
where $\omega^{\ra\rb} := g^{\ra \bar \rc} g^{\rb \bar \rd}\omega_{\bar \rc \bar \rd} $.

\end{enumerate}
One may also check that $\cL_J \omega_{\ra\rb} = -\ri\, \omega_{\ra\rb}$. 
In terms of these geometric objects, it holds that $\Omega^\ra = \omega^{\ra\rb} \chi_\rb$,
and the extended supersymmetry transformation may equivalently be written
\begin{align}\label{eq_dphi30}
\delta \phi^\ra
	= \ri \,\bar\rho_\alpha \,\omega^\ra{}_{\bar \rb} \bar \cD^\alpha \bar\phi^{\bar \rb}
	- 4 S \rho\, \omega^{\ra\rb} \chi_\rb
	+ \frac{\ri}{2} \bar \rho \,\omega^{\ra\rb} \bar \cD^2 \chi_\rb~,
\end{align}
with the extended supersymmetry algebra closing only on-shell.
Note that the last term in \eqref{eq_dphi30} vanishes on-shell as a consequence
of the absence of a superpotential.

Together these conditions imply that the target space geometry is a hyperk\"ahler
cone. The three covariantly constant complex structures can be taken as
\begin{align}\label{eq_ComplexStructures}
(J_1)^\mu{}_\nu =
\begin{pmatrix}
0 & \omega^\ra{}_{\bar \rb} \\
\omega^{\bar \ra}{}_\rb & 0                  
\end{pmatrix}~, \quad
(J_2)^\mu{}_\nu =
\begin{pmatrix}
0 & \ri \,\omega^\ra{}_{\bar \rb} \\
-\ri \,\omega^{\bar \ra}{}_\rb & 0                  
\end{pmatrix}~, \quad
(J_3)^\mu{}_\nu =
\begin{pmatrix}
\ri \,\delta^\ra{}_\rb & 0 \\
0 & - \ri \,\delta^{\bar \ra}{}_{\bar \rb}
\end{pmatrix}~.
\end{align}
Using $J_A$ and $\chi$, we may construct three $\rm SU(2)$ Killing vectors
\begin{align}
V_A^\mu := -\frac{1}{2} (J_A)^\mu{}_\nu \chi^\nu~,
\end{align}
with $V_3^\mu$ coinciding with the $\rm U(1)$ Killing vector $J^\mu$ required
by $(2,0)$ supersymmetry.
These vectors commute with the homothetic conformal
Killing vector, obey an $\rm SU(2)$ algebra amongst themselves,
\begin{align}
[V_A, \chi] = 0~, \qquad [V_A, V_B] = \epsilon_{ABC} V_C
\end{align}
and act as an $\rm SU(2)$ rotation on the complex structures
\begin{align}
\cL_{V_A} J_B = \epsilon_{ABC} J_C~.
\end{align}

The component action may be easily derived using \eqref{eq_20compaction}
and applying what we have learned of the target space geometry:
\begin{align}
L &= -g_{\ra \bar \rb} \cD_m \varphi^\ra \cD^m \bar\varphi^{\bar \rb}
	- \ri g_{\ra \bar \rb} \bar\psi_\alpha^{\bar \rb} \hat \cD^{\alpha \beta} \psi_\beta^{\ra}
	+ \frac{1}{4} R_{\ra \bar \rb \rc \bar \rd} (\psi^{\ra} \psi^\rc) (\bar\psi^{\bar \rb} \bar\psi^{\bar \rd})
	+ 3 S^2 K
\end{align}
where we have used $K = \chi^\ra \bar \chi^{\bar\rb} g_{\ra \bar \rb}$. The apparent scalar
potential actually corresponds to a massless model in AdS, confirmed by the absence
of fermionic mass terms. This can be made apparent by rewriting
the Lagrangian as
\begin{align}\label{eq_30compaction}
L &= g_{\ra \bar \rb} \chi^\ra (\hat\cD_a \hat\cD^a - \frac{1}{8} \cR) \bar\chi^{\bar \rb}
	- \ri g_{\ra \bar \rb} \bar\psi_\alpha^{\bar\rb} \hat \cD^{\alpha \beta} \psi_\beta^{\ra}
	+ \frac{1}{4} R_{\ra \bar \rb \rc \bar \rd} (\psi^{\ra} \psi^\rc) (\bar\psi^{\bar \rb} \bar\psi^{\bar \rd})
\end{align}
after identifying the scalar curvature of AdS as $\cR = -24 S^2$. The scalar kinetic
operator in the first term is indeed the conformal d'Alembertian in three dimensions.

We note that it is not possible to introduce any mass deformations, either via a
superpotential or by deforming the Killing vectors. We will see that the
(2,1) situation is quite different.


\section{Sigma models with (2,1) AdS supersymmetry: Off-shell approach}
\setcounter{equation}{0}

In the case of (2,1) AdS supersymmetry, general off-shell $\s$-models can be realized in 
terms of weight-zero polar hypermultiplets living in the (2,1) AdS superspace \cite{KLT-M12}. 
The geometry of this superspace is encoded in a covariantly constant real isotriplet 
$w^{ij} = w^{ji}$, conventionally normalized by $w^{ij}w_{ij}=2$,
which can be interpreted as the field strength of a frozen vector multiplet
(called the {\it intrinsic vector multiplet} in \cite{KLT-M12}). The local SU(2) gauge freedom 
can  partially be fixed by choosing a gauge $w^{ij} =\rm const$ and then mapping $w^{ij}$ 
to any particular position on the two-sphere  $w^{ij}w_{ij}=2$. Depending on the explicit choice 
of $w^{ij}$ made, the manifestly (2,1) supersymmetric $\s$-models can be 
reduced to either (2,0)  or (1,1) AdS superspace \cite{KLT-M12}.
Here we 
will use these off-shell realizations in order to reformulate the (2,1) supersymmetric $\s$-models 
in terms of covariantly chiral superfields on (2,0)  or (1,1) AdS superspace. 

\subsection{Formulation in (2,0) AdS superspace}

The off-shell  (2,1) supersymmetric $\s$-model in the (2,0) AdS superspace \cite{KLT-M12} is
\bea
S &=& \oint_\g  \frac{\rd \zeta}{2\pi \ri \zeta}
\int \rd^3x\, \rd^4{ \q}  \, {E}\, 
 K (\U^{I} ,  \breve{\U}^{\bar J} ) ~.
\label{(2,1--2,0)ProjAction}
\eea
Formally this coincides  with the (3,0) supersymmetric $\s$-model action \eqref{(3,0)ProjAction}. 
However, the conceptual difference between the two cases is that the 
Lagrangian in \eqref{(2,1--2,0)ProjAction}
is not required to obey any homogeneity condition like \eqref{Kkahler}. The only conditions 
on the Lagrangian in  \eqref{(2,1--2,0)ProjAction} are that (i)  
$ K (\F^{I} ,  \bar{\F}^{\bar J} )$ is a real analytic function of $n$ complex variables $\F^I$ and their 
conjugates; and (ii) the matrix $g_{I\bar J} := \pa_I \pa_{\bar J} K$ is nonsingular. 
One can consistently interpret $ K (\F ,  \bar{\F} )$ as the K\"ahler potential of a K\"ahler 
manifold $\cX$, since the action \eqref{(2,1--2,0)ProjAction} may be seen 
 to be invariant under K\"ahler transformations of the form
\be
{ K}({ \U}, \breve{\U})~\to ~{ K}({ \U}, \breve{ \U})
+{ \L}({\U}) +{\bar { \L}} (\breve{\U} )~,
\label{1.26}
\ee
with ${ \L}(\F^I)$ a holomorphic function.

The dynamical variables $\U^I(\z)$ and $\breve{\U}^{\bar J} (\z)$
in \eqref{(2,1--2,0)ProjAction} are {\it covariant weight-zero arctic} 
and {\it antarctic} multiplets. Considered as (2,0) AdS superfields, they are completely specified 
by eqs. \eqref{3.3} -- \eqref{3.5}. The information that  $\U^I(\z)$ is a covariant weight-zero arctic
multiplet is encoded in its transformation law under the (2,1) AdS isometry supergroup, 
$\rm OSp (2|2; {\mathbb R} ) \times   OSp(1|2, {\mathbb R})$, given in \cite{KLT-M12}.
Upon reduction to the (2,0) AdS superspace, 
the most general  (2,1) isometry transformation of any (2,1) supermultiplet splits into two different 
transformations: (i) a (2,0) AdS isometry transformation
generated  by superfield parameters specified
 in eqs. \eqref{2_0-Killing_iso-def} -- \eqref{2_0-Killing_iso}; 
(ii) a  third supersymmetry  transformation generated by 
a real spinor parameter $\r_\a$ obeying the constraints 
\bea
\cD_\b \r_\a = \bar \cD_\b\r_\a=0~.
\label{2_1-extra}
\eea
These conditions mean that $\r_\a$ is an ordinary  Killing spinor, 
\bea
\cD_{\b\g} \r_\a = S(\ve_{\a \b } \r_\g + \ve_{\a\g} \r_\b)~.
\eea
The (2,0) AdS isometry transformation of $\U^I (\z)$ is 
\bea
\d_\t \U^I = (\t + \ri t \cJ) \U^I~, \qquad
\cJ =  \z\frac{\pa}{\pa \z}
\quad \Longleftrightarrow \quad
\cJ \U^I_n = n \U^I_n ~. 
\eea
A finite transformation generated by $\cJ$,  
\bea
\U(\z) \quad &\longrightarrow & \quad \U'(\z)=
\U({\rm e}^{{\rm i} \a} \z)
~,\qquad
\a \in {\mathbb R}~, 
\label{5.6}
\eea
is a symmetry of the $\s$-model \eqref{(2,1--2,0)ProjAction}.
It coincides with the U(1) symmetry of
 the  4D $\cN=2$ supersymmetric $\s$-models 
on cotangent bundles of K\"ahler manifolds \cite{GK1,GK2}.

The third supersymmetry transformation of $\U^I$ is 
\bea
\d_\r \U^I = \Big\{  \ri\z\r^\a \cD_\a
+\frac{\ri}{\z}\r_\a \bar \cD^\a
 \Big\} \U^I~.
\label{5.7}
\eea
It is useful to represent 
\bea \label{eq_rho2120constraints1} 
\r_\a = \cD_\a \r = \bar \cD_\a \bar \r~, \qquad  \cJ \r = - \r~,
\eea
for some complex superfield parameter $\r $ defined modulo arbitrary antichiral shifts
\bea
\r ~\to ~\r+ \bar \l ~, \qquad \cD_\a \bar \l  = 0~.
\label{5.9}
\eea
Due to \eqref{2_1-extra}, this scalar parameter is subject to the constraints
\bea \label{eq_rho2120constraints2} 
\cD^2 \r = 0~, \qquad \bar \cD_\a \cD_\b \r =0~, 
\eea
in addition to the reality condition $ \cD_\a \r = \bar \cD_\a \bar \r$.
For the physical superfields, the transformation law \eqref{5.7}  leads to 
\begin{subequations}
\label{SUSY(5.11)}
\bea
\d_\r \F^I &=& \ri \r_\a \bar \cD^\a  \S^I
=  \frac{\ri}{2} \bar \cD^2 (\bar \r \S^I)~,   
\label{SUSY(5.11)-a} 
\\
\d_\r \S^I &=&  \ri \r^\a \cD_\a   \F^I 
+\ri \r_\a \bar \cD^\a  \U^I_2 
= \ri \bar \cD_\a (\r^\a \U^I_2 -\bar \r \cD^\a \F^I)~.~~~
\label{SUSY(5.11)-b}
\eea
\end{subequations}
The variations $\d_\r \F^I$ and $\d_\r \S^I$ are chiral and complex linear respectively. 
One should remember  that the U(1) charges of the physical superfields are
\bea
\cJ \F^I = 0~, \qquad \cJ \S^I = \S^I~.
\eea

Similar to the (3,0) case considered earlier, the (2,1) supersymmetric $\s$-model 
\eqref{(2,1--2,0)ProjAction} can be reformulated solely in terms of the physical superfields
and then, upon performing a duality transformation, in terms of (2,0) chiral superfields
(the procedure is almost identical to that described in \cite{K-duality,BKLT-M} for  4D $\cN=2$ 
supersymmetric $\s$-models).
Most aspects  of these procedures are completely analogous to the (3,0) case, 
but some differences  also occur.  
Upon elimination of the auxiliary superfields $\U^I_2, \U^I_3, \dots,$ the action 
takes the form 
\bea
S= \int \rd^3x\, \rd^4\q\,  E\, {\mathbb L}(\F^I, \S^I , \bar \F^{\bar J}, \bar \S^{\bar J} )~.
\label{5.13}
\eea
The complex variables $(\F^I, \S^J)$ parametrize the holomorphic tangent 
bundle $T\cX$ of the K\"ahler manifold. In complete analogy with the four-dimensional 
analysis in  \cite{K98},  
this follows from the observation that a holomorphic reparametrization of the K\"ahler manifold,
$ \F^I  \to \F'{}^I=f^I \big( \F \big) $,
has the following counterpart
\bea
\U^I (\z) \quad  \longrightarrow  \quad \U'{}^I(\z)=f^I \big (\U(\z) \big)
\label{kahl3}
\eea
for the (2,1) arctic multiplets.  Therefore, the physical superfields
\bea
 \U^I (\z)\Big|_{\z=0} ~=~ \F^I ~,\qquad  \quad \frac{ {\rm d} \U^I (\z) 
}{ {\rm d} \z} \Big|_{\z=0} ~=~ \S^I ~,
\label{kahl4} 
\eea
should be regarded, respectively, as  coordinates of a point in 
$\cX$ and a tangent vector at  the same point. 
The general form of the Lagrangian in \eqref{5.13} is (compare with  \cite{GK1,GK2})
\bea
{\mathbb L}  \big(\F, \S ,\bar \F,  \bar \S \big)
= K(\F, \bar \F) +\sum_{n=1}^{\infty}  {\mathbb L}_{I_1 \cdots I_n {\bar J}_1 \cdots {\bar 
J}_n }  \big( \F, \bar{\F} \big) \S^{I_1} \dots \S^{I_n} 
{\bar \S}^{ {\bar J}_1 } \dots {\bar \S}^{ {\bar J}_n }
~,~~~~~~~~
\label{act-tab}
\eea
where ${\mathbb L}_{I {\bar J} }=  - g_{I \bar{J}} \big( \F, \bar{\F}  \big) $ 
and the Taylor coefficients ${\mathbb L}_{I_1 \cdots I_n {\bar J}_1 \cdots {\bar 
J}_n }$, for  $n>1$, 
are tensor functions of the K\"ahler metric
$g_{I \bar{J}} \big( \F, \bar{\F}  \big) 
= \pa_I 
\pa_ {\bar J}K ( \F , \bar{\F} )$,  the Riemann curvature $R_{I {\bar 
J} K {\bar L}} \big( \F, \bar{\F} \big) $ and its covariant 
derivatives.  Each term in the action contains equal powers
of $\S$ and $\bar \S$, since the off-shell action \eqref{(2,1--2,0)ProjAction}
is invariant under the  U(1)  transformation \eqref{5.6}.

By construction, the tangent-bundle action \eqref{5.13} 
must be invariant under the third supersymmetry transformation \eqref{SUSY(5.11)} 
in which   $\U^I_2 $ has to be 
the function  $ \U^I_2 (\F, \S, \bar \F,  \bar \S)$ obtained by solving 
the equations of motion for the auxiliary superfields, eq.  \eqref{AuxEOM(3,0)AdS}. 
Requiring the action \eqref{5.13} to be invariant under  \eqref{SUSY(5.11)} 
leads to the consistency conditions \eqref{eq_XiEqnsAdS}, for some function $\X$
 given by eq. \eqref{3.20}. Using the contour integral representation \eqref{3.20}, 
 one may prove  that all conditions \eqref{eq_XiEqnsAdS} hold identically.  
Unlike the (3,0) $\s$-model, eq. \eqref{X-first} does not appear in the present case.

A novel feature occurs when we turn to the first-order formulation of the $\s$-model \eqref{5.13} 
given by 
\begin{align}
S_{\text{first-order}} &= \int \rd^3x\, \rd^4\q\,  E\,
	\Big({\mathbb L} + \S^I \Psi_I + \bar \S^{\bar J} \bar \Psi_{\bar J}\Big)~, \qquad 
	\cJ \J_I = - \J_I~,
\label{5.17}
\end{align}
in which the superfields $\S^I$ are complex unconstrained, while the Lagrange multipliers 
$\J_I$ are chiral, $\bar \cD_\a \J_I =0$.  Since the action must be invariant under holomorphic
reparametrizations of the K\"ahler manifold $\cX$, the variables $\J_I$ 
describe the components of a (1,0) form at the point $\F$ of $\cX$.
Using the invariance of  \eqref{5.13} under  \eqref{SUSY(5.11)},
this action may be seen to be invariant under the following supersymmetry transformation:
\begin{subequations}\label{5.18}
\bea
\d_\r \F^I 
&=&  \frac{\ri}{2} \bar \cD^2 (\bar \r \S^I)
= \ri \r_\a \bar \cD^\a \S^I + \frac{\ri}{2}   \bar \r \bar \cD^2 \S^I~, \\
\d_\r \J_I 
&=&  -\frac{\ri}{2} \bar \cD^2 \Big(\bar \r \frac{\pa \cL}{\pa \F^I}\Big)~.
\eea
\end{subequations}
It should be pointed out that the original supersymmetry transformation  \eqref{SUSY(5.11)}
involved the complex parameter $\bar \r$, defined modulo the gauge freedom \eqref{5.9},  
only via its spinor derivative $\r_\a = \bar \cD_\a \bar \r$. 
On the contrary, the supersymmetry transformation of $S_{\text{first-order}}$, 
\eqref{5.18} involves the naked parameter $\bar \r$. 
One may see that eq. \eqref{5.18} describes not only the third supersymmetry, but also a gauge
symmetry obtained by choosing $\bar \r$ to be a chiral superfield  $\l$, compare with \eqref{5.9}. 
It is not difficult to understand that this gauge invariance is a trivial gauge symmetry of the theory
with action \eqref{5.17}.

Starting from the first-order action \eqref{5.17} and integrating out the auxiliary superfields
$\S^I$ and $\bar \S^{\bar I}$ leads  to the dual action 
\begin{align}
S_{\rm dual} &= \int \rd^3x\, \rd^4\q\,  E\, {\mathbb K}(\F^I, \J_I , \bar \F^{\bar J}, \bar \J_{\bar J} )~.
\end{align}
Its target space is (an open domain of the zero section of) the cotangent bundle
$T^* \cX$ (compare with \cite{GK1,GK2}).
The extended supersymmetry of this $\s$-model is given by eq. \eqref{ESDA(3.0)} or equivalently, 
using the condensed notation \eqref{nott}, by eq. \eqref{SST(3,0)}.

\subsection{Formulation in (1,1) AdS superspace}

The off-shell  (2,1) supersymmetric $\s$-model in the (1,1) AdS superspace \cite{KLT-M12} is
\bea 
S =\hf  \oint_\g   \frac{\rd \zeta}{2\pi \ri \zeta}
 \int \rd^3x\, \rd^4\q \, {E}\, w^{ [2]} \,
 K({ \U}, \breve{\U})~, 
\qquad w^{[2]} : =  \frac{\ri}{| \m |} \Big( \frac{\m}{\z}+ \bar \m \z\Big)~.
\label{action(2,1)-->(1,1)}
\eea
The dynamical variables $\U^I (\z)$ and $\breve{\U}^{\bar I}(\z) $ have the functional form
\bea
{ \U}^I (\z) &=& \sum_{n=0}^{\infty}  \, \z^n \U_n^I  = \F^I + \z \S^I + \dots~, 
\qquad
\breve{ \U}^{\bar I} (\z) =\sum_{n=0}^{\infty}  \,  (-\z)^{-n}\,
{\bar \U}_n^{\bar I}~,
\eea
where $\F^I$ and $\S^I$ are chiral and complex linear superfields  respectively, 
\bea
\bar {\frak D}_\a \F^I = 0 ~, \qquad ( \bar {\frak D}^2 - 4\m ) \S^I =0~,
\label{constraints(2,1)-->(1,1)}
\eea
and the other components $\U^I_2, \U^I_3, \dots$, are complex unconstrained 
superfields.\footnote{The spinor covariant derivatives of (1,1) AdS superspace, 
$ {\frak D}_\a$ and $\bar {\frak D}_\a$, are related to those used in \cite{KLT-M12}, 
$\nabla_\a$ and $\bar \nabla_\a$, as follows: ${\frak D}_\a = \sqrt{\ri \m/|\m|} \nabla_\a$
and ${\bar {\frak D}}_\a = \sqrt{-\ri \bar \m/|\m|} \bar \nabla_\a$.}

The (1,1) AdS isometry transformation of $\U^I (\z)$ is very simple
\bea
\d_l \U^I = l \U^I~.
\eea
The third supersymmetry transformation of $\U^I$ is 
\bea
\d_\ve \U^I = - \Big\{   \z ({\frak D}^\a \ve)  {\frak D}_\a
- \frac{1}{\z} (\bar {\frak D}_\a \ve)  \bar {\frak D}^\a
+2S  \ve  \Big(\z  w+ \frac{1}{\z} \bar w\Big) \z \frac{\pa}{\pa \z}
 \Big\} \U^I~.
\eea
Here $\ve$ is a real parameter constrained by 
\bea \label{eq_rho21constraints1}
&
\bar {\frak D}_\a \ve = - \ri \dfrac{\m}{|\m|} {\frak D}_\a \ve~,\qquad
({\frak D}^2-4\mub ) \ve=
(\bar {\frak D}^2 - 4\mu )\ve=0~,
\eea
and hence 
\bea \label{eq_rho21constraints2}
{\frak D}_\a\bar {\frak D}_\b \ve  =  \bar {\frak D}_\a{\frak D}_\b \ve = -2\ri  |\m| \ve_{\a\b} \ve 
\quad  \longrightarrow \quad {\frak D}_{\a\b} \ve=0~.
\eea
For the physical fields this transformation law gives
\begin{subequations}
\bea
\d_\ve \F^I &=& \big( (\bar {\frak D}_\a \ve \big)  \bar {\frak D}^\a +2 \m \ve )\S^I
= \hf  (\bar {\frak D}^2 -4\m) (\ve \S^I)~, \\
\d_\ve \S^I&=& -( {\frak D}^\a \ve)   {\frak D}_\a \F^I + ((\bar {\frak D}_\a \ve)  \bar {\frak D}^\a
+4\m \ve )\U^I_2 
= 
\bar {\frak D}_\a \Big(
\ri \frac{\bar \m}{|\m|} \ve\,  {\frak D}^\a \F^I 
+ \U_2^I \bar {\frak D}^\a \ve \Big)~.~~~~~
\eea
\end{subequations}

The off-shell action \eqref{action(2,1)-->(1,1)} and the constraints obeyed by the physical superfields, 
eq. \eqref{constraints(2,1)-->(1,1)}, are similar to those describing the most general 4D $\cN=2$ 
supersymmetric $\s$-model in AdS${}_4$ \cite{BKLT-M}.
We therefore can apply the four-dimensional results obtained in \cite{BKLT-M} 
to the $\s$-model under consideration without any additional calculation.
We summarize the results and refer the interested reader to \cite{BKLT-M} for the technical details. 
Upon elimination of the auxiliary superfields  $\U^I_2, \U^I_3, \dots$, from the action 
\eqref{action(2,1)-->(1,1)} and subsequent dualization of the complex linear superfield
$\S^I$ and its conjugate $\bar \S^{\bar I}$ into a chiral scalar $\J_I$
 and its conjugate $\bar \J_{\bar I}$, $\bar {\frak D}_\a \J_I=0$,
 we end up with a $\s$-model action of the form
 \begin{align}
S_{\rm dual} &= \int \rd^3x\, \rd^4\q\, E\, \cK(\F^I, \J_I , \bar \F^{\bar J}, \bar \J_{\bar J} )~,
\end{align}
where the K\"ahler potential $\cK(\F , \J, \bar \F , \bar \J )$
is a globally defined function on the target space.
This action is invariant under supersymmetry transformations
\begin{align}\label{3.24Dar}
\delta \Phi^I = \frac{1}{2} (\bar {\frak D}^2 - 4 \mu) \left(\ve \frac{\pa \cK}{\pa \Psi_I}\right)~,\quad
\delta \Psi_I = -\frac{1}{2} (\bar {\frak D}^2 - 4 \mu) \left(\ve \frac{\pa \cK}{\pa \Phi^I}\right)~.
\end{align}
Using the concise notation  $\f^\ra = (\F^I, \J_I)$, this transformation can be rewritten as
\begin{align}
\delta \f^\ra = \frac{1}{2} (\bar {\frak D}^2 - 4 \mu) (\ve \omega^{\ra \rb} \cK_\rb)~,
\label{3.25SUSY}
\end{align}
where we have introduced the symplectic matrices 
\begin{align}
\omega^{\ra \rb} =
\begin{pmatrix}
0 & \delta^I{}_J \\
-\delta_I{}^J & 0
\end{pmatrix}~, \qquad
\omega_{\ra \rb} =
\begin{pmatrix}
0 & \delta_I{}^J \\
-\delta^I{}_J & 0
\end{pmatrix}~.
\end{align}

\subsection{Sigma model gaugings with a frozen vector multiplet}
\label{subsection5.3}

An important feature that distinguishes the (2,1) AdS superspace from the (3,0) one
is that the former allows the existence of a frozen vector multiplet with the property  
that its field strength is covariantly constant. Following the four-dimensional terminology
\cite{KT-M-4D-2008,BKLT-M}, such a vector  multiplet is called intrinsic since 
it is intimately connected to the geometry of the (2,1) AdS superspace.
It can be employed in the context of gauged supersymmetric $\s$-models 
in the (2,1) AdS superspace.

Let us consider a U(1) vector 
multiplet in the (2,1) AdS superspace. It can be described in terms of gauge
covariant derivatives 
\bea
{\bm \cD}_A = ({\bm \cD}_a , {\bm \cD}^{ij}_\a ) = 
\cD_A  + \ri \,V_A \cZ~, \qquad [\cZ,\cD_A] =[\cZ, {\bm \cD}_A ]=[\cZ, \cJ_{ij}]=0~,
\eea
where $\cD_A=(\cD_a , \cD^{ij}_\a)$ are the covariant derivatives of 
the (2,1) AdS superspace (see \cite{KLT-M12} for more details), $V_A$ and $\cZ$ are 
the U(1) gauge connection and generator respectively.
The anti-commutator of two spinor gauge covariant derivatives is
\bea
\{ {\bm \cD}_\a^{ij}, {\bm \cD}_\b^{kl}\}&=&
-2\ri\ve^{i(k}\ve^{l)j} {\bm \cD}_{\a\b}
+4\ri \,S\,(\ve^{i(k}\ve^{l)j}+w^{ij}w^{kl})\cM_{\a\b}
\non\\
&&
+\ri \,S\,\ve_{\a\b}\big(\ve^{i(k}w^{l)j}
+\ve^{j(k}w^{l)i}\big)
w^{pq}\cJ_{pq} - 2 \ve_{\a\b} \big(\ve^{i(k}W^{l)j}
+\ve^{j(k}W^{l)i}\big) \cZ
~,~~~~~
\label{5.26}
\eea
where the gauge invariant field strength $W^{ij} = W^{ji}$  is real,
$\overline{W^{ i j } } = W_{ij} = \ve_{ik} \ve_{jl} W^{kl}$, and obeys the Bianchi identity
\bea
\cD^{(ij}_\a W^{kl)} =0~.
\eea 
Suppose the field strength $W^{ij}$ is covariantly constant, 
$\cD^{ij}_\a W^{kl} =0$. In accordance with \eqref{5.26}, the integrability 
conditions for this constraint are $\cD_a W^{ij} =0$ and 
$W^{ij} = G \,w^{ij}$, with $G$ a real constant parameter. 
Without loss of generality, we can normalize $G=S$, and thus 
\bea
W^{ij} = S \,w^{ij}~.
\eea
The field strength is determined by the superspace curvature. 
This is why the vector multiplet under consideration is called intrinsic.
Now eq. \eqref{5.26} takes the form
\bea
\{ {\bm \cD}_\a^{ij}, {\bm \cD}_\b^{kl}\}&=&
-2\ri\ve^{i(k}\ve^{l)j} {\bm \cD}_{\a\b}
+4\ri \,S\,(\ve^{i(k}\ve^{l)j}+w^{ij}w^{kl})\cM_{\a\b}
\non\\
&&
-2S\ve_{\a\b}\big(\ve^{i(k}w^{l)j}
+\ve^{j(k}w^{l)i}\big) {\bm \cJ}~,
\label{5.36}
\eea
where
\bea
{\bm \cJ} :=  \cJ + \cZ~, \qquad \cJ :=  -\frac{\ri }{2} w^{pq}\cJ_{pq}~.
\eea

A universal procedure to construct gauged  supersymmetric 
$\s$-models with (2,1) AdS supersymmetry makes use of a real analytic
K\"ahler manifold $\cX$ possessing a U(1) isometry group.
Such a transformation group is generated by a holomorphic Killing vector field
\bea
Z= Z^I (\F) \pa_I + \bar Z^{\bar I} (\bar \F ) \pa_{\bar I}
\eea
defined by 
\bea
Z^I(\F) := \ri \cZ \F^I~,
\eea
where $\F^I$ are local complex coordinates for $\cX$. We assume that the K\"ahler potential 
$K(\F, \bar \F)$ is invariant under the U(1) isometry group, 
\bea
Z^I K_I + \bar Z^{\bar I} K_{\bar I}=0~.
\eea 
Associated with $\cX$  is an off-shell 
supersymmetric $\s$-model in the (2,1) AdS superspace described by 
the Lagrangian 
\bea
\cL^{(2)}= w^{(2)}K ({\bm \U}^{ I}, \breve{\bm \U}{}^{\bar J}) ~, \qquad 
w^{(2)}:=v_iv_j w^{ij}~,
\label{(2,1)gauge-cov-Lag}
\eea
with $v^i \in {\mathbb C}^2 -\{0\} $ the homogeneous coordinate for ${\mathbb C}P^1$.
The dynamical variables 
 ${\bm \U}^{ I}$ and $ \breve{\bm \U}{}^{\bar J}$ are gauge covariantly {\it arctic} and 
 {\it antarctic} multiplets, respectively. They obey the analyticity constraints  
\bea
{\bm \cD}^{(2)}_{\a} {\bm \U}^I  =0~, \qquad 
{\bm \cD}^{(2)}_{\a} \breve{\bm \U}^{\bar J}  =0~, \qquad 
{\bm \cD}_\a^{(2)}:=v_iv_j {\bm \cD}_\a^{ij}~
\label{anana}
\eea  
and have the following functional form: 
\begin{subequations}
\bea
{\bm \U}^I ( v) 
&=& \sum_{k=0}^{\infty} {\bm \U}^I_k \, \z^k = {\bm \F}^I + {\bm \S}^I \z + \dots~, \qquad 
\z= v^{\2}/v^{\1}~,\\
\breve{\bm \U}{}^{\bar I} (v) 
&=& \sum_{k=0}^{\infty}  {\bar {\bm \U}}^{\bar I}_k \,
(- \z)^{-k}
~.~~~
\eea
\end{subequations}
The  antarctic multiplet $\breve{\bm \U}{}^{\bar I} (v) $ 
is said to be the {\it smile-conjugate} of ${\bm \U}^I ( v) $.
If the background vector multiplet is switched off, the Lagrangian 
\eqref{(2,1)gauge-cov-Lag} reduces to that considered in \cite{KLT-M12}.

In accordance with \cite{KLT-M12}, the AdS superspace
reduction $ (2,1)  \to (2,0) $ can be performed by  
choosing $w^{ij}$ in the form:
$w^{\1 \1} = w^{\2\2}=0$ and $w^{\1\2} = -w_{\1\2}= -\ri$.
The spinor covariant derivatives for (2,0) AdS superspace can be chosen as  
${\bm \cD}_\a:= {\bm \cD}_\a^{\1\1} | $ and $\bar {\bm \cD}_\a := -{\bm \cD}_\a^{\2\2}|$. 
These operators obey the anti-commutation relation \eqref{intrVM}. 
When projected to (2,0) AdS superspace, the physical 
superfields $ {\bm \F}^I $ and $ {\bm \S}^I$ are constrained as
\bea
\bar {\bm \cD}_\a {\bm \F}^I =0~, \qquad 
\bar {\bm \cD}^2 {\bm \S}^I = 0~,
\eea
as a consequence of the analyticity constraints \eqref{anana}.
When projected to (2,0) AdS superspace,  
the $\s$-model action generated by the Lagrangian \eqref{(2,1)gauge-cov-Lag} proves to be
\bea
S &=& \oint_\g  \frac{\rd \zeta}{2\pi \ri \zeta}
\int \rd^3x\, \rd^4{ \q}  \, {E}\, 
 K ({\bm \U}^{I} ,  \breve{\bm \U}{}^{\bar J} ) ~.
\label{5.45}
\eea
If the background (2,0) vector multiplet is switched off, 
this action reduces to  \eqref{(2,1--2,0)ProjAction}. 
The only difference between the $\s$-model \eqref{5.45} 
and  the $\s$-model \eqref{(2,1--2,0)ProjAction} studied earlier 
is a different choice of the U$(1)_R$ generator; specifically, it is 
$\cJ$ for the  $\s$-model \eqref{(2,1--2,0)ProjAction}
and ${\bm \cJ} =\cJ +{ \cZ}$  the $\s$-model \eqref{5.45}. 
Therefore, all the results obtained for the  $\s$-model \eqref{(2,1--2,0)ProjAction}
can naturally be extended to the theory under consideration. 

As shown in \cite{KLT-M12}, the AdS superspace
reduction $ (2,1) \to (1,1)$ is carried out  by  
choosing $w^{ij}$ in the form: $w^{\1\2} = 0$, 
$w^{\1 \1} = - \bar \m /|\m|$  and $w^{\2\2}=-\m /|\m|$. 
The spinor covariant derivatives for (1,1) AdS superspace can be chosen as  
\bea
{\bm {\frak D}}_\a:= \sqrt{ \ri \frac{ \m}{|\m|} } {\bm \cD}_\a^{\1\1} | ~, \qquad 
\bar {\bm {\frak D}}_\a := -  \sqrt{ -\ri \frac{\bar \m}{|\m|} } {\bm \cD}_\a^{\2\2}|~.
\eea 
As follows from \eqref{5.36}, the operators ${\bm {\frak D}}_a \!\!:= {\bm \cD}_a|$, 
 ${\bm {\frak D}}_\a$ and $\bar {\bm {\frak D}}_\a$ obey the (1,1) AdS (anti) commutation relations, 
 eq. \eqref{11AdSsuperspace}, which do not involve any U$(1)_R$ curvature. 
 This means that the U$(1)_R$ connection associated with the covariant derivatives
 ${\bm {\frak D}}_A =({\bm {\frak D}}_a , {\bm {\frak D}}_\a,\bar {\bm {\frak D}}^\a)$
 can be completely gauged away, ending up with standard (1,1) AdS covariant derivatives
 ${ {\frak D}}_A =({ {\frak D}}_a , { {\frak D}}_\a,\bar { {\frak D}}^\a)$.
When projected to (1,1) AdS superspace, the physical 
superfields $ {\bm \F}^I $ and $ {\bm \S}^I$ can be seen to be constrained as
\bea
\bar {\frak D}_\a {\bm \F}^I=0~, \qquad
-\frac{1}{4} (\bar {\frak D}^2 - 4\mu ) {\bm \S}^I = 2 Z^I( {\bm \F})~.
\label{5.47}
\eea
Thus ${\bm \F}^I$ is an ordinary chiral superfield in (1,1) AdS superspace 
(and this can simply be denoted as $\F^I$), 
while $ {\bm \S}^I $ obeys a modified linear constraint. 
When projected to (1,1) AdS superspace,  
the $\s$-model action generated by the Lagrangian \eqref{(2,1)gauge-cov-Lag} proves to be
\bea 
S =\hf  \oint_\g   \frac{\rd \zeta}{2\pi \ri \zeta}
 \int \rd^3x\, \rd^4\q \, {E}\, w^{ [2]} \,
 K({\bm \U}, \breve{\bm \U})~, 
\qquad w^{[2]} : =  \frac{\ri}{| \m |} \Big( \frac{\m}{\z}+ \bar \m \z\Big)~.
\label{5.48}
\eea
Setting $Z^I =0$ in \eqref{5.47} reduces the $\s$-model action \eqref{5.48} to  
\eqref{action(2,1)-->(1,1)}. 
The off-shell action \eqref{5.48} and the constraints obeyed by the physical superfields, 
eq. \eqref{5.47}, are similar to those describing the gauged 4D $\cN=2$ 
supersymmetric $\s$-model in AdS${}_4$ \cite{BKLT-M}.
We therefore can apply the four-dimensional results obtained in \cite{BKLT-M} 
to the $\s$-model under consideration without any additional calculation.
Upon elimination of the auxiliary superfields  from the action 
\eqref{5.48} and subsequent dualization of the {\it deformed complex linear} superfield
${\bm \S}^I$ and its conjugate $\bar {\bm \S}^{\bar I}$ into a chiral scalar $\J_I$
 and its conjugate $\bar \J_{\bar I}$, $\bar {\frak D}_\a \J_I=0$,
 we end up with the $\s$-model
 \begin{align}
S_{\rm dual} = \int \rd^3x\, \rd^4\q \, E 
	\left( \cK(\F , \J, \bar \F , \bar \J ) + \frac{1}{\mu} W(\F , \J)
	+ \frac{1}{\bar \mu} \bar W ( \bar \F , \bar \J )\right)~,
\end{align}
where we have introduced the superpotential
\begin{align}
W (\F , \J )=  - 2\Psi_I X^I (\F)~.
\end{align}
The $\s$-model action 
is invariant under the extended supersymmetry transformation \eqref{3.24Dar}.


\section{Sigma models with (2,1) AdS supersymmetry: On-shell approach}\label{section6}
\setcounter{equation}{0}

In the previous section, off-shell $(2,1)$ $\s$-models
were shown to lead to formulations either in $(2,0)$ or $(1,1)$
superspace. In this section, we will again analyze the situation
in reverse and consider the most general $(2,0)$ and $(1,1)$
models possessing a full $(2,1)$ symmetry.

\subsection{Formulation in (2,0) AdS superspace}

We take the $(2,0)$ action involving a K\"ahler potential
\begin{align}
S = \int \rd^3x\, \rd^4\q\, E\, {\mathbb K}(\phi^\ra, \bar\phi^{\bar \rb})
\end{align}
and postulate the extended supersymmetry transformation law
\begin{align}
\delta \phi^\ra = \frac{\ri}{2} \bar \cD^2 (\bar \rho \Omega^\ra)
\end{align}
where $\Omega^\ra = \Omega^\ra(\phi, \bar\phi)$. The complex parameter
$\rho$ obeys the conditions \eqref{eq_rho2120constraints1} 
and \eqref{eq_rho2120constraints2}.
We require the action to be invariant. We delay the technical
analysis to Appendix \ref{SUSY_Derivs} and present the
result: in order for the action to be invariant,
the combination $\omega_{\ra\rb} := g_{\ra \bar \ra} \pa_\rb \Omega^{\bar\ra}$
must be a covariantly constant holomorphic two-form,
$\nabla_\rc \o_{\ra \rb} =  \nabla_{\bar \rc} \o_{\ra \rb} =0$.
A further analysis of the closure of the algebra dictates that
$\omega^{\ra\rb} \omega_{\rb\rc} = -\delta^\ra_\rc$, 
 where $\omega^{\ra\rb} := g^{\ra \bar \rc} g^{\rb \bar \rd}\omega_{\bar \rc \bar \rd} $,
and that the $\rm U(1)$ Killing
vector $J^\ra$ obeys $\cL_J \omega_{\ra\rb} = -\ri \, \omega_{\ra\rb}$.

In contrast to the $(3,0)$ situation, the target space is \emph{not}
in general a hyperk\"ahler cone. Rather, it is a
hyperk\"ahler manifold with the single
additional constraint that it possess a $\rm U(1)$ Killing vector $J^\ra$ which rotates
the complex structures; choosing these as in \eqref{eq_ComplexStructures} leads to
\begin{align}\label{eq_3020JRot}
\cL_J J_1 = J_2~, \qquad \cL_J J_2 = -J_1~, \qquad \cL_J J_3 = 0~.
\end{align}
An analogous result has recently been established for
$\cN=2$ $\s$-models in AdS$_4$ \cite{BKsigma1, BKsigma2, BKLT-M}
and for $\cN=1$ models in AdS$_5$ \cite{BaggerXiong, BaggerLi}.
The component Lagrangian (with auxiliaries eliminated)
can be calculated using \eqref{eq_20compaction}:
\begin{align}
\cL &= -g_{\ra \bar \ra} \cD_m \varphi^\ra \cD^m \bar\varphi^{\bar \ra}
	- \ri g_{\ra \bar \ra} \bar\psi_\alpha^{\bar \ra} \hat \cD^{\alpha \beta} \psi_\beta^{\ra}
	+ \frac{1}{4} R_{\ra \bar \ra \rb \bar \rb} (\psi^{\ra} \psi^\rb) (\bar\psi^{\bar \ra} \bar\psi^{\bar \rb})
	\eol & \quad
	+ S\, (\psi^\ra \bar\psi^{\bar \ra}) (\ri g_{\ra \bar \ra} + \tsD_\ra J_{\bar \ra} 
	- \tsD_{\bar \ra} \bar J_{\ra})
	- 4 S^2 ( J^\ra \bar J^{\bar \ra} g_{\ra \bar \ra} - D)~.
\end{align}

It is natural to ask what happens if we want to introduce additional
mass terms while maintaining the extended supersymmetry. Just as in the $(3,0)$
case, this cannot be done with a superpotential while maintaining the extended
supersymmetry. However, there is an alternative way to introduce masses
(or, more accurately, to deform the mass terms already present): this
is to deform the $\rm U(1)$ Killing vector.
We have already discussed in section \ref{NLSM20}
that the inclusion of a frozen vector multiplet deforms the $(2,0)$
supersymmetry algebra by replacing the $\rm U(1)$ generator $\cJ$ with
${\bm \cJ} = \cJ + \cZ$ where $\cZ$ is the generator associated with
the frozen vector multiplet. For a $\s$-model, this amounts to the
replacement of the $\rm U(1)$ Killing vector $J^\ra$ with ${\bm J}^\ra = J^\ra + Z^\ra$.
Since the effective $\rm U(1)$ Killing vector ${\bm J}^\ra$ generates
the masses in the component Lagrangian, this provides
an alternative way to introduce massive deformations.

When extended supersymmetry is taken into account, this procedure
still holds with one additional restriction: the Killing vector $Z^\ra$
associated with the gauging must be a tri-holomorphic isometry,
obeying $\cL_Z \omega_{\ra\rb} = 0$. Its addition to $J^\ra$ then leads
to a $\rm U(1)$ Killing vector ${\bm J}^a$ that still rotates
the complex structures as in \eqref{eq_3020JRot}. Deforming the masses
simply involves deforming the $\rm U(1)$ Killing vector.
Since the inclusion of a massive deformation only deforms the $\rm U(1)$
Killing vector, the form of the component action remains unchanged.
One simply replaces $J^\ra \rightarrow {\bm J}^\ra$ and $D \rightarrow {\bm D}$.

\subsection{Formulation in (1,1) AdS superspace}
Let us now consider the most general $(2,1)$ model written in $(1,1)$ superspace.
We begin with the most general $\s$-model in $(1,1)$ AdS,
\begin{align*}
S = \int \rd^3x\, \rd^4\q\, E \,\cK(\f^\ra , \bar \f^{\bar \rb})~.
\end{align*}
Inspired by the solution in projective superspace, we postulate
\begin{align}\label{eq_11extendedSusy}
\delta \phi^\ra = \frac{1}{2} (\bar {\frak D}^2 - 4 \mu) (\ve \Omega^\ra)
\end{align}
where the real parameter $\ve$ obeys \eqref{eq_rho21constraints1} and 
\eqref{eq_rho21constraints2}.
We find as usual that
\begin{align}
\omega_{\ra\rb} = g_{\ra \bar \ra} \pa_\rb \Omega^{\bar \ra}
\end{align}
is a covariantly constant holomorphic two-form; closure of the algebra
further imposes that $\omega^{\ra\rb} \omega_{\rb\rc} = - \delta^\ra_\rc$.
This demonstrates that the target space is hyperk\"ahler.
In addition, we discover that there exists a $\rm U(1)$ Killing vector
\begin{align}\label{eq_KillingV}
V^\ra = \frac{\mu}{2 S} \omega^{\ra\rb} \cK_\rb~.
\end{align}
Introducing the complex structures as in \eqref{eq_ComplexStructures},
one can check that $V^\ra$ acts as a rotation:
\begin{align}\label{eq_VRotJ}
\cL_V J_1 = \textrm{Im} \,\frac{\mu}{S} \,J_3~,\qquad
\cL_V J_2 = -\textrm{Re} \,\frac{\mu}{S} \,J_3~, \qquad
\cL_V J_3 = \textrm{Re} \, \frac{\mu}{S} J_2 - \textrm{Im}\, \frac{\mu}{S} J_1~,
\end{align}
and leaves invariant one particular combination of complex structure
\begin{align}
J_{\rm AdS} = -\textrm{Re} \, \frac{\mu}{S} J_1 - \textrm{Im}\, \frac{\mu}{S} J_2~.
\end{align}
In fact, we may rewrite $V^\mu$ as
\begin{align}
V^\mu = -\frac{1}{2} (J_{\rm AdS})^\mu{}_\nu \tsD^\nu \cK
\end{align}
which identities $\cK$ as the Killing potential for $V^\mu$, with
respect to the complex structure $J_{\rm AdS}$. The component Lagrangian
can be derived using \eqref{eq_11compaction}:
\begin{align}
\cL &= -g_{\ra \bar \rb} {\frak D}_m \varphi^\ra {\frak D}^m \bar\varphi^{\bar \rb}
	- \ri g_{\ra \bar \rb} \bar\psi_\alpha^{\bar \rb} \hat {\frak D}^{\alpha \beta} \psi_\beta^{\ra}
	+ \frac{1}{4} R_{\ra \bar \rb \rc \bar \r} (\psi^{\ra} \psi^\rc) (\bar\psi^{\bar \rb} \bar\psi^{\bar \rd})
	\eol & \quad
	- \frac{\mu}{2} \tsD_\ra \cK_\rb \,(\psi^{\ra} \psi^\rb)
	- \frac{\bar\mu}{2} \tsD_{\bar \ra} \cK_{\bar \rb} \,(\bar \psi^\ra \bar\psi^{\rb})
	- 4 S^2 (g^{\ra \bar \rb} V_\ra V_{\bar \rb} - \cK)~.
\end{align}

As in the $(2,0)$ formulation, we should inquire about introducing mass terms.
Because of the extremely close relationship between the presentation here
and that of AdS$_4$ \cite{BKsigma2, BKLT-M}, the answer is quite apparent.
The introduction of a superpotential amounts to the replacement in the action of
\begin{align}\label{eq_11Wdeformation}
\cK \rightarrow {\bm \cK} = \cK + \dfrac{W}{\mu} + \dfrac{\bar W}{\bar\mu}~.
\end{align}
In order for this to be invariant under the extended supersymmetry \eqref{eq_11extendedSusy},
the superpotential must obey
\begin{align}
\cK_\ra \omega^{\ra\rb} W_\rb + \cK_{\bar \ra} \omega^{\bar \ra \bar \rb} \bar W_{\bar \rb} = F + \bar F
\end{align}
where $F = F(\phi)$ is a holomorphic function.
This means that the superpotential $W$ is associated with
a holomorphic isometry, which we may denote
\begin{align}
Z^\ra := \omega^{\ra\rb} W_\rb~.
\end{align}
In fact, one may check that $Z^\ra$ is tri-holomorphic, obeying $\cL_Z \omega_{\ra\rb} = 0$.
Examining the definition \eqref{eq_KillingV} of the $\rm U(1)$ Killing vector $V^\ra$,
we see that the replacement \eqref{eq_11Wdeformation} leads to
\begin{align}
{\bm V}^\ra := V^\ra + Z^\ra~.
\end{align}
where the new $\rm U(1)$ Killing vector ${\bm V}^\ra$ still obeys the same
conditions \eqref{eq_VRotJ} as before. This is precisely the same physical situation
we observed in the $(2,0)$ formulation: the allowed deformation of the masses
corresponds to deforming the $\rm U(1)$ Killing vector by the addition of a
tri-holomorphic piece.

\section{$\cN=4$ AdS superspaces}
\setcounter{equation}{0}
\label{N4AdSsuperspaces}

In the remainder of this paper, we will focus on nonlinear $\s$-models with $\cN=4$ supersymmetry 
in $\rm AdS_3$.
We have previously described $\cN=3$ supersymmetric $\s$-models 
using only two manifest supersymmetries. 
In the  $\cN=4$ case, we will likewise consider formulations involving a smaller amount
of manifestly realized supersymmetry. Our first task, which is the focus of this section,
will be to analyze how  the various $\cN=4$ AdS superspaces (and their isometries) 
can be projected to AdS superspaces with a smaller number of Grassmann variables.

\subsection{Geometry of $\cN=4$ AdS superspaces}
\label{subsection7.1}

We begin by reviewing the geometry of the $\cN=4$ AdS superspaces constructed in
\cite{KLT-M12}. Consistent with the analysis of \cite{AT}, 
there are three types of $\cN=4$ AdS supersymmetry in three dimensions, 
and therefore three inequivalent superspaces. 
These three cases, which we call the (4,0), (3,1) and (2,2) AdS superspaces,
are special backgrounds allowed by the superspace geometry of
three-dimensional  $\cN=4$ conformal supergravity  of \cite{KLT-M11}.

All the three $\cN=4$ AdS supergeometries have covariant derivatives
of the form \cite{KLT-M12}
\bea
\cD_A=(\cD_a,\cD_\a^{i\bai})=E_A{}^M\pa_M+\hf\O_A{}^{cd}\cM_{cd}
+\F_A{}^{kl}\bL_{kl}
+\F_A{}^{\bak\bal}\bR_{\bak\bal}
~.
\label{N4-dev}
\eea
The operators $\bL_{kl}$ and $\bR_{\bak\bal}$ generate the $R$-symmetry group
SU(2)$_\rL\times$SU(2)$_\rR$ and act on the covariant derivatives as
\bea 
[\bL^{kl},\cD_\a^{i\bai}]=\ve^{i(k}\cD_\a^{l)\bai}
~,~~~~~~
[\bR^{\bak\bal},\cD_\a^{i\bai}]=\ve^{\bai(\bak}\cD_\a^{i\bal)}
~.
\eea
For each of the $\cN=4$ AdS superspaces, the algebra of covariant derivatives is
\bsubeq \label{N=4alg}
\bea
\{\cD_\a^{i\bai},\cD_\b^{j\baj }\}&=&\phantom{+}
2\ri\,\ve^{ij}\ve^{\bai \baj }\cD_{\a\b}
-\,4\ri(\cS^{ij}{}^{\bai \baj }+\ve^{ij}\ve^{\bai \baj }\cS)\cM_{\a\b}
\non\\
&&
+\,{2\ri}\ve_{\a\b}\ve^{\bai \baj }(2\cS+X)\bL^{ij}
-\,2\ri\ve_{\a\b}\ve^{ij}\cS^{kl}{}^{\bai \baj }\bL_{kl}
\non\\
&&
+\,2\ri\ve_{\a\b}\ve^{ij}(2\cS-X)\bR^{\bai\baj}
-\,2\ri\ve_{\a\b}\ve^{\bai \baj }\cS^{ij}{}^{\bak\bal}\bR_{\bak\bal}
~,
\label{N=4alg-1}
\\
{[}\cD_{\a\b},\cD_\g^{k\bak}{]}&=&
-\,2\Big(\d^k_l\d^\bak_\bal\cS+\cS^k{}_l{}^\bak{}_\bal\Big)\ve_{\g(\a}\cD_{\b)}^{l\bal}
~,
\label{N=4alg-3/2}
\\
{[}\cD_a,\cD_b{]}&=&-\,4\,S^2\cM_{ab}~,
\label{N=4alg-2}
\eea
\esubeq
where the real tensor  $\cS^{ij\bai\baj} = \cS^{(ij)(\bai\baj)}$ is covariantly constant, 
and the real scalars $\cS$, $X$ and $S$ are constant.
The parameter $S$ determines the curvature scale.
Depending on the different superspace geometries, the 
parameters  $\cS$, $\cS^{ij\bai\baj}$ and $X$ are
\cite{KLT-M12}
\bsubeq
\bea
&(4,0):&~~~\cS=S~,~~~\cS^{ij\bai\baj}=0~,~~~~~~
X{~\rm arbitrary}~;
\label{SSX40}
\\
&(3,1):&~~~\cS=\hf S~,~~
\cS^{ij\bai\baj}=\Big(\hf\ve^{ij}\ve^{\bai\baj}-w^{i\bai}w^{j\baj}\Big)S
~,~~~
X=0~;
\label{SSX31}
\\
&(2,2):&~~~\cS=0~,~~~
\cS^{ij\bai\baj}=l^{ij}r^{\bai\baj}\,S
~,~~~X=0~.
\label{SSX22}
\eea
\esubeq
In the (3,1) case, the covariantly constant tensor $w^{i\bai}$ is real, 
$\overline{w^{i\bai}} = w_{i\bai}= \ve_{ij}\ve_{\bai \baj} w^{j\baj}$, and normalized as 
\bea
w^{i\bak}w_{i\bak}=\d^i{}_j~, \qquad w^{k\bai}w_{k\baj}= \d^{\bai}{}_{\baj}~.
\eea
In the (2,2) case, the real iso-triplets $l^{ij} = l^{ji}$ and $r^{\bai\baj}=r^{\baj\bai}$ 
are covariantly constant and normalized as
\bea
l^{ik}l_{kj} = \d^i{}_j~, \qquad r^{\bai\bak}r_{\bak\baj}=\d^{\bai}{}_{\baj}~.
\eea

We emphasize that $X$ 
can appear in the algebra only in the (4,0) case. The (4,0) AdS superspace is conformally flat if and 
only if $X=0$ \cite{KLT-M12}.
For general values of $X$, the tangent space group of the (4,0) AdS supergeometry is the full 
$R$-symmetry group SU(2)$_\rL\times$SU(2)$_\rR$.
For the two critical values, $X=2S$ and $X=-2S$,
the SU(2)$_\rR$ or SU(2)$_\rL$ group  respectively can be gauged away.

For each of the (3,1) and (2,2) geometries, the $R$-symmetry sector of
the  superspace holonomy group is a subgroup of
SU(2)$_\rL\times$SU(2)$_\rR$ 
\cite{KLT-M12}.
For the (3,1) supergeometry, the relevant subgroup is SU(2)$_\cJ$ generated by
\bea
\cJ_{kl}=\bL_{kl}+w_k{}^\bak w_l{}^\bal\bR_{\bak\bal}
~,~~~{\rm or}~~~
\cJ_{\bak\bal}=w^k{}_\bak w^l{}_\bal\bL_{kl}+\bR_{\bak\bal}=w^k{}_\bak w^l{}_\bal\cJ_{kl}
~.
\eea
Indeed, the anti-commutator \eqref{N=4alg-1} in the (3,1) case can be rewritten 
in the following equivalent forms:
\begin{subequations}
\bea
\{\cD_\a^{i\bai},\cD_\b^{j\baj}\}&=&\phantom{+}
2\ri\ve^{ij}\ve^{\bai\baj}\cD_{\a\b}
+2\ri\ve_{\a\b}\,S\,\Big(
\ve^{\bai\baj}\d^i_k\d^j_l
+\ve^{ij}w_k{}^{\bai}w_{l}{}^{\baj}
\Big)\cJ^{kl}
\non\\
&&
-4\ri \,S\,\big(\ve^{ij}\ve^{\bai\baj}-w^{i\bai}w^{j\baj}\big)\cM_{\a\b}
\\
&=& \phantom{+}
2\ri\ve^{ij}\ve^{\bai\baj}\cD_{\a\b}
+2\ri\ve_{\a\b}\,S\,\Big(
\ve^{ij}\d^\bai_\bak\d^\baj_\bal
+\ve^{\bai\baj}w^i{}_{\bak}w^{j}{}_{\bal}
\Big)
\cJ^{\bak\bal}
\non\\
&&
-4\ri \,S\,\big(\ve^{ij}\ve^{\bai\baj}-w^{i\bai}w^{j\baj}\big)\cM_{\a\b}
~.
\eea
\end{subequations}
The generators $\cJ_{kl}$ and $\cJ_{\bak\bal}$ leave $w^{i\bai}$ invariant, 
 $\cJ_{kl}w^{i\bai}= \cJ_{\bak\bal} w^{i\bai}= 0$.
Since the $R$-symmetry curvature is spanned by the generators of  SU(2)$_\cJ$,  
it is possible to choose a gauge in which 
the $R$ symmetry connection takes its values in the Lie algebra of  SU(2)$_\cJ$;
in this gauge, the parameter $w^{i\bai}$ is constant.
This gauge choice is always assumed 
in the remainder of the paper.

In the (2,2) case,
 the $R$-symmetry sector of
the  superspace holonomy group is  
the Abelian subgroup $\rm U(1)_\rL \times U(1)_\rR $ of $\rm SU(2)_\rL \times SU(2)_\rR$    
generated by 
\bea
\bL:=l^{kl}\bL_{kl}
~,~~~~~~
\bR:=r^{\bak\bal}\bR_{\bak\bal}~.
\eea
This subgroup leaves invariant  the covariantly constant parameters $l^{kl}$ and  $r^{\bak\bal}$.
In the remainder of the paper, we choose a gauge in which only 
this subgroup 
appears in the (2,2) covariant derivatives; then the parameters
$l^{kl}$ and $r^{\bak\bal}$ are constant.

Given a particular  $\cN=4$ AdS superspace,
its isometry group  is generated by Killing vector 
fields, $\x=\x^a\cD_a+\x^\a_{i\bai}\cD_\a^{i\bai}$, obeying the equation
\bea
0&=&
\Big{[}
\x
+\hf \L^{\g\d}\cM_{\g\d}
+\L^{kl}\bL_{kl}
+\L^{\bak\bal}\bR_{\bak\bal}
,
\cD_A
\Big{]}
~.
\label{sK-1}
\eea
This Killing equation is equivalent to
\bsubeq
\bea
\cD_\a^{ i\bai}\x_{\b\g}&=&
4\ri\ve_{\a(\b}\x_{\g)}^{i\bai}
~,
\label{sK-2-1}
\\
\cD_\a^{i\bai}\x_{\b}^{j\baj}&=&
\x_{\a\b}\Big(\ve^{ij}\ve^{\bai\baj}\cS+\cS^{ij}{}^{\bai\baj}\Big)
+\hf\L_{\a\b}\ve^{ij}\ve^{\bai\baj}
+\L^{ij}\ve^{\bai\baj}\ve_{\a\b}
+\L^{\bai\baj}\ve^{ij}\ve_{\a\b}
~,
\label{sK-2-2}
\\
\cD_\a^{i\bai} \L_{\b\g}&=&
8\ri\ve_{\a(\b}\x_{\g) j\baj}(\cS^{ij}{}^{\bai \baj }+\ve^{ij}\ve^{\bai \baj }\cS)
~,
\label{sK-2-1b}
\\
\cD_\a^{i\bai}\L^{kl}&=&
-2\ri\ve^{i(k}\x_{\a}{}^{l)\bai}(2\cS+X)
-2\ri\x_{\a}{}^{i}{}_{\baj}\cS^{kl}{}^{\bai \baj}
~,
\label{sK-2-3}
\\
\cD_\a^{i\bai}\L^{\bak\bal}&=&
-2\ri\ve^{\bai(\bak}\x_{\a}{}^{i\bal)}(2\cS-X)
-2\ri\x_{\a j}{}^{\bai}\cS^{ij}{}^{\bak\bal}
~,
\label{sK-2-4}
\eea
\esubeq
and
\bsubeq \label{sK-2}
\bea
\cD_a\x_b&=&
\L_{ab}
~,
\label{sK-2-5}
\\
\cD_{a}\x^\b_{j\baj}
&=&
-\big(\cS\x^\g_{j\baj}
+\cS_{jk\baj\bak}\x^{\g k\bak}\big)(\g_a)_\g{}^\b
~,
\label{sK-2-6}
\\
\cD_a\L^{bc}
&=&
4S^2\big(\d_a^b\x^{c}
-\d_a^c\x^{b}
\big)
~,
\label{sK-2-7}
\\
\cD_a\L^{kl}&=&\cD_a\L^{\bak\bal}=0
~.
\label{sK-2-8}
\eea
\esubeq
Some useful implications of the above equations are
\bsubeq
\bea
&
\cD_{(\a}^{i\bai}\x_{\b\g)}
=\cD_{(\a}^{i\bai} \L_{\b\g)}=0
~,
\label{sK-2-1-b}
\\
&\cD^{\b i\bai}\x_{\a\b}=
6\ri\x_{\a}^{i\bai}~,~~~~~~
\cD^{\b i\bai} \L_{\a\b}=
12\ri\x_{\a j\baj}(\cS^{ij}{}^{\bai \baj }+\ve^{ij}\ve^{\bai \baj }\cS)
~,
\label{sK-2-1b-b}
\\
&\cD_{(\a}^{(i\bai}\x_{\b)}^{j)}{}_{\bai}=
\cD_{(\a}^{i(\bai}\x_{\b)}{}_i{}^{\baj)}=
0
~,~~~
\cD_{(\a}^{(i(\bai}\x_{\b)}^{j)\baj)}
=\x_{\a\b}\cS^{ij}{}^{\bai\baj}~,~~~
\cD_{(\a}^{i\bai}\x_{\b)}{}_{i\bai}=
4\x_{\a\b}\cS
+2\L_{\a\b}
~,~~~
\label{sK-2-5-ba}
\\
&\cD^{\a i\bai}\x_{\a i\bai}=\cD^{\a (i(\bai}\x_{\a}^{ j)\baj)}=0
~,~~~
\cD^{\a (i\bai}\x_{\a}^{j)}{}_{\bai}=
-4\L^{ij}
~,~~~
\cD^{\a i(\bai}\x_{\a}{}_i{}^{\baj)}=
-4\L^{\bai\baj}
~.
\label{sK-2-5-bb}
\eea
\esubeq
Here we have written the results in a form valid for the (4,0), (3,1) and (2,2) cases.
Depending on the $\cN=4$ AdS superspace under consideration, 
$\cS,\,X$ and $\cS^{ij\bai\baj}$ are constrained by 
(\ref{SSX40})--(\ref{SSX22}),
while the SU(2)$_\rL$ and SU(2)$_\rR$ parameters $\L^{kl}$ and $\L^{\bak\bal}$
are restricted by
\bsubeq
 \bea
&& (4,0)~~{\rm with}~~X=\phantom{+}2S:~~~\L^{\bak\bal}=0~;~~~~~~\\
&& (4,0)~~{\rm  with}~~X=-2S:~~~\L^{kl}=0~;
 ~~~~~~
 \\
&& (3,1):~~~\L^{\bak\bal}=w_k{}^{\bak}w_l{}^{\bal}\L^{kl}
 ~;
 \label{7.14c0}
 \\
&& (2,2):~~~
\L^{kl}=l^{kl}\L_\rL~,~~\overline{(\L_\rL)}=\L_\rL~,~~~~~~
\L^{\bak\bal}=r^{\bak\bal}\L_\rR~,~~\overline{(\L_\rR)}=\L_\rR~.
\label{7.14c}
 \eea
\esubeq


\subsection{From $\cN=4$ to $\cN=2$ AdS superspaces}
\label{fromN4toN2}

It was argued in the introduction that any $\cN=4$ AdS superspace can be reduced to the (2,0) 
AdS superspace. Here we elaborate on the details of such a reduction.

Let us fix a certain $\cN=4$ AdS superspace. 
We start by showing that its algebra of covariant derivatives possesses
 an $\cN=2$ AdS subalgebra associated with the covariant derivatives  $\cD_a$,
$\cD_\a^{1\bau}$ and $(-\cD_\a^{2\bad})$.\footnote{Given a tensor superfield $U$
of Grassmann parity $\e(U)$, 
the operation of complex conjugation maps $\cD_\a^{1\bau} U$ to
$\overline{\cD_\a^{1\bau} U}=- (-1)^{\e(U) } \cD_{\a 1\bau}\bar U=- (-1)^{\e(U)}\cD_\a^{2\bad}\bar U$.}
These operators obey the (anti) commutatation relations
\bsubeq \label{7.15}
\bea
\{\cD_\a^{1\bau},\cD_\b^{1\bau }\}&=&
-\,4\ri\cS^{11}{}^{\bau \bau }\cM_{\a\b}
~,
\label{7.15a}\\
\{\cD_\a^{1\bau},(-\cD_\b^{2\bad})\}&=&
-2\ri\,\cD_{\a\b}
+\,4\ri(\cS^{12}{}^{\bau \bad }+\cS)\cM_{\a\b}
\non\\
&&
-\,{2\ri}\ve_{\a\b}(2\cS+X)\bL^{12}
+\,2\ri\ve_{\a\b}\cS^{kl}{}^{\bau \bad }\bL_{kl}
\non\\
&&
-\,2\ri\ve_{\a\b}(2\cS-X)\bR^{\bau\bad}
+\,2\ri\ve_{\a\b}\cS^{12}{}^{\bak\bal}\bR_{\bak\bal}
~, \label{7.15b}
\\
{[}\cD_{\a\b},\cD_\g^{1\bau}{]}&=&
-\,2\Big(\d^1_l\d^\bau_\bal\cS+\cS^1{}_l{}^\bau{}_\bal\Big)\ve_{\g(\a}\cD_{\b)}^{l\bal}~,
\label{7.15c}\\
{[}\cD_a,\cD_b{]}&=&-\,4\,S^2\cM_{ab}~.
\eea
\esubeq
For the right-hand side of \eqref{7.15c} not to involve
$\cD_\a^{1\bad}$ and $\cD_\a^{2\bau}$, 
we must require
\bea
\cS^{11\bau\bad}=\cS^{12\bau\bau}=0
~,
\label{condONcS}
\eea
which leads to 
\bsubeq \label{1111}
\bea
\{\cD_\a^{1\bau},\cD_\b^{1\bau }\}&=&
-\,4\ri\cS^{11}{}^{\bau \bau }\cM_{\a\b}
~,
\label{1111-1}
\\
\{\cD_\a^{1\bau},(-\cD_\b^{2\bad})\}&=&
-2\ri\,\cD_{\a\b}
+\,4\ri(\cS+\cS^{12}{}^{\bau \bad })\cM_{\a\b}
-\,{2\ri}\ve_{\a\b}\big(2(\cS+\cS^{12}{}^{\bau \bad })\hat \cJ
+X \hat \cZ\big)
~,
~~~~~~~~~
\label{1111-2}
\\
{[}\cD_{\a\b},\cD_\g^{1\bau}{]}&=&
-\,2\Big(\cS+\cS^{12\bau\bad}\Big)\ve_{\g(\a}\cD_{\b)}^{1\bau}
-\,2\cS^{11\bau\bau}\ve_{\g(\a}\cD_{\b)}^{2\bad}~,
\label{1111-3}
\\
{[}\cD_a,\cD_b{]}&=&-\,4\,S^2\cM_{ab}~,
\label{1111-4}
\eea
\esubeq
where we have introduced two  U(1) generators 
\bsubeq
\bea
&\hat \cJ:=(\bL^{12}+\bR^{\bau\bad})~,~~~~~~
{[} \hat \cJ,\cD_\a^{1\bau}{]}=\cD_\a^{1\bau}
~,~~~
{[} \hat \cJ,(-\cD_\a^{2\bad}){]}=-(-\cD_\a^{2\bad})~,
\\
&\hat \cZ:=(\bL^{12}-\bR^{\bau\bad})~,~~~~~~
{[}\hat \cZ,\cD_\a^{1\bau}{]}=
{[}\hat \cZ,(-\cD_\a^{2\bad}){]}=0~,
\eea
\esubeq
such that 
\bea
[\hat \cJ, \hat \cZ]=0~.
\eea
As will be shown shortly,   the condition \eqref{condONcS} 
can always be satisfied.

In the (4,0) AdS superspace, $\cS^{ij\bai\baj}=0$ and  thus 
 the condition \eqref{condONcS} holds identically. 
Since in this case we also have $\cS=S$, 
the algebra \eqref{1111} can be seen to coincide with that defining the (2,0) AdS superspace, 
eq. \eqref{20AdSsuperspace}, if the U$(1)_R$ generator is identified with
\bea
{ \cJ}:=\hat \cJ + \frac{X}{2S}\hat \cZ ~.
\label{mod-cJ}
\eea
This identification is not unique.
Instead of  \eqref{mod-cJ}, 
one could have chosen $\cJ = \hat \cJ + \x \hat \cZ$ as the U$(1)_R$ 
generator and $\cZ = (\frac{X}{2S} -\x)\hat  \cZ $ as the central charge, 
for some real parameter $\x$. 
Such a choice would have led to the `gauged' realization of the (2,0)  AdS algebra, eq. \eqref{intrVM}. 
In what follows, we will use the identification   \eqref{mod-cJ}.

In the case of the  (3,1) and (2,2) AdS superspaces, it follows from 
the relations (\ref{SSX31}) and (\ref{SSX22})  that the condition (\ref{condONcS}) 
has only two solutions: either $\cS^{11\bau\bau}=0$ and then the resulting algebra (\ref{1111})
is  of the  (2,0) AdS type; or $\cS^{12\bau\bad}=0$ and then the (anti) commutation relations
(\ref{1111}) are of the (1,1) AdS type.
We will mainly be concerned with reductions to (2,0) AdS superspace and consider that
case in the remainder of this subsection. The reductions to (1,1) AdS superspace are included
for completeness in Appendix \ref{AppD}.

For any $\cN=4$ tensor superfield  $U(x,\q_{\imath\bar{\jmath}})$,
we define its $\cN=2$ projection by
\bea
U|:=U(x,\q_{\imath \bar{\jmath}})|_{\q_{1\bad}=\q_{2\bau}=0}~.
\label{N2red-1}
\eea
By definition,  $U|$  depends on the Grassmann coordinates
$\q^\mu:=\q^\mu_{1\bau}$ and their complex conjugates,  $\qb^\mu=\q^\mu_{2\bad}$.
For the $\cN=4$ AdS covariant derivatives\footnote{The reader should 
bear in mind that, depending on the choice of parameters $\cS,\,\cS^{ij\bai\baj},\,X$,
the $R$-symmetry connection may take values 
only in a subgroup of SU(2)$_\rL\times$SU(2)$_\rR$.}
\be
\cD_{{A}}=E_{{A}}{}^{{M}}\pa_{{M}}
+\hf\O_{{A}}{}^{bc}\cM_{bc}
+\F_{{A}}{}^{kl}\bL_{kl}
+\F_{{A}}{}^{\bak\bal}\bR_{\bak\bal}~,
\label{N2red-2}
\ee
the projection is defined as
\bea
\cD_{{A}}|=E_{{A}}{}^{{M}}|\pa_{{M}}
+\hf\O_{{A}}{}^{bc}|\cM_{bc}
+\F_{{A}}{}^{kl}|\bL_{kl}
+\F_{{A}}{}^{\bak\bal}|\bR_{\bak\bal}~.
\label{N2red-3}
\eea
Since the operators $\big(\cD_a,\,\cD_\a^{1\bau},\,-\cD_\a^{2\bad} \big)$ 
form a closed  algebra isomorphic to that of the (2,0) AdS superspace,
one can use the freedom to perform general coordinate, local Lorentz and SU(2) transformations
to choose a gauge in which
\bea
\cD_\a^{1\bau}|=\cD_\a
~,~~~~~~
-\cD_\a^{2\bad}|=\cDB_\a
~,
\label{7.24}
\eea
where 
\bea
\cD_A = (\cD_a , \cD_\a, \bar \cD^\a)= {E}_A{}^M \pa_M +\hf {\O}_A{}^{cd}\cM_{cd}
+ \ri \,{ \F}_A \cJ
\label{1.4}
\eea 
denote the covariant derivatives of the (2,0) AdS superspace.\footnote{We hope the reader will not 
be confused by the use of the same notation $\cD_A$ in \eqref{N2red-2} and \eqref{1.4}
for the covariant derivative of  the $\cN=4$ and (2,0) AdS superspaces respectively.} 
Note that  the  U$(1)_R$ generator $\cJ$ is defined by (\ref{mod-cJ})
and coincides with $\hat \cJ$ in the cases (3,1), (2,2) and (4,0) with $X=0$. 
We recall that the covariant derivatives of the (2,0) AdS superspace 
obey the (anti) commutation relations \eqref{20AdSsuperspace}.

In the coordinate system defined by \eqref{7.24}, 
the operators $\cD_\a^{1\bau}|$ and $\cD_{\a}^{2\bad}|$  involve no
partial derivative with respect to $\q_{1\bad},\,\q_{2\bau}$, 
and therefore, for any positive integer $k$,  
it holds that $\big( \cD_{\hat{\a}_1} \cdots  \cD_{\hat{\a}_k} U \big)\big|
= \cD_{\hat{\a}_1}| \cdots  \cD_{\hat{\a}_k}| U|$, 
where $ \cD_{\hat{\a}} :=\big( \cD_\a^{1\bau}, -{\cD}_\a^{2\bad} \big)$ 
and $U$ is a tensor superfield. This implies that $\cD_a| = \cD_a$ in (\ref{1.4}).

Let us now consider a Killing vector field $\x$ of one of the $\cN=4$ AdS superspaces, 
as specified by  eqs. \eqref{sK-1} -- \eqref{sK-2}.
We introduce $\cN=2$ projections of the $\cN=4$ Killing parameters:
\bsubeq
\begin{gather}
\t^a:=\x^a|
~,~~~
\t^\a:=\x^\a_{\1\1}|~,~~~
\bar{\t}^\a=\x^\a_{\2\2}|~,~~~
t:=\ri(\L^{12}+\L^{\bau\bad})|=\overline{t}
~, ~~~t^{ab}:= \L^{ab}|~
;
\label{N2proj-20-1}
\\
\ve^\a:=-\x^\a_{1\bad}|~,~~~
\bar{\ve}^\a=\x^\a_{2\bau}|
~,~~~
\s:= \ri (\L^{12}-\L^{\bau\bad})|=\bar{\s}
~;
\label{N2proj-20-2-0}
\\
\bar{\ve}_\rL:=-\frac{1}{4S}\L^{11}|~,~~~
\ve_\rL=-\frac{1}{4S}\L^{22}|
~,~~~
\bar{\ve}_\rR=-\frac{1}{4S}\L^{\bau\bau}|~,~~~
\ve_\rR=-\frac{1}{4S}\L^{\bad\bad}|
~.~~~~~~
\label{N2proj-20-2}
\end{gather}
\esubeq
The parameters $(\t^a, \, \t^\a,\,\bar{\t}_\a,\, t^{\a\b} ,\, t )$
describe the infinitesimal isometries of the (2,0) AdS superspace.
This can be easily proven  by $\cN=2$ projection of the equations (\ref{sK-1})--(\ref{sK-2-4}).
The remaining parameters
$(\veps^\alpha, \,\bar\veps_\alpha, \,\sigma, \,\veps_L, \,\bar\veps_L, \,\veps_R, \,\bar\veps_R)$ are 
associated with the remaining two supersymmetries
and the residual $R$-symmetry. Depending on the initial $\cN=4$ AdS superspace,
these parameters are constrained in different ways.

\subsubsection{AdS superspace reduction (4,0) $\to$ (2,0)}

For the (2,0) reduction of the (4,0) Killing vectors, we find
a set of differential relations between $\veps_\alpha$, $\veps_L$, $\veps_R$
and their complex conjugates:
\bsubeq \label{extra4020}
\begin{gather}
\cD_\a\bar{\ve}_\b =
4S\ve_{\a\b}\,\bar{\ve}_\rL
~,~~~
\cDB_\a\ve_\b=
-4S\ve_{\a\b}\,{\ve}_\rL
~,
\label{extra4020-1}
\\
\cD_\a\ve_\b
=-4S\ve_{\a\b}\,\bar{\ve}_\rR
~,~~
\cDB_\a\bar{\ve}_\b
=4S\ve_{\a\b}\,{\ve}_\rR
~,~~~~~~
\label{extra4020-1b}
\\
\cDB_\a\ve_\rL =\cDB_\a{\ve}_\rR=0
~,~~~~~~
\cD_\a{\ve}_\rL=
\ri \,\ve_{\a}\Big(1+\frac{X}{2S}\Big)
~,~~~
\cD_\a{\ve}_\rR=-\ri\,\bar{\ve}_{\a}\Big(1-\frac{X}{2S}\Big)
~.~~~~~~
\label{extra4020-2}
\end{gather}
\esubeq
The action of the U$(1)_R$ generator  \eqref{mod-cJ}  on these parameters is
\bea
{ \cJ}\ve_\a=-\frac{X}{2S}\ve_\a~,\qquad
{ \cJ}{\ve}_\rL=-\Big(1+\frac{X}{2S}\Big){\ve}_\rL~,\qquad
{ \cJ}{\ve}_\rR=-\Big(1-\frac{X}{2S}\Big){\ve}_\rR~.
~~~~~~
\label{7.28}
\eea
The real parameter $\sigma$, corresponding to one of the residual $R$-symmetries,
can be shown to obey 
\bea
\sigma - \dfrac{X}{2S} t = \textrm{const}~.
\label{7.30}
\eea
A finite U(1)  transformation generated by the constant parameter $(\sigma - t{X}/{2S} )$
does not act on the (2,0) AdS superspace, and thus it 
is analogous to the so-called shadow chiral rotation 
introduced in the context of 4D $ \cN=2$ superconformal $\s$-models
\cite{K-duality}.

In the critical cases the parameters are further constrained to satisfy
\bsubeq
\bea
X=2S&:&~~~~~~
\L^{\bak\bal}=\ve_\rR=0~,~~~~~~
\cD_\a\ve_\b=
\cDB_\a\bar{\ve}_\b
=0~;
\\
X=-2S&:&~~~~~~
\L^{kl}=\ve_\rL=0~,~~~~~~
\cDB_\a\ve_\b=
\cD_\a\bar{\ve}_\b
=0~.
\eea
\esubeq

\subsubsection{AdS superspace reduction (3,1) $\to$ (2,0)}
\label{subsubsection7.2.2}

For the reduction from (3,1) to (2,0) superspace,
a local $R$-symmetry transformation can be applied to bring 
$w^{i\bai}$ to look like
\bea
w^{1\bau}=w^{2\bad}=0~,~~~
w^{1\bad}=1~,~~~
w^{2\bau}=-(w^{1\bad})^*=-1
~,
\eea
and then the condition \eqref{condONcS}  holds.
This implies
\bea
&\L^{\bak\bal}=\d_{k}^{\bak}\d_{l}^{\bal}\L^{kl}
~,~~~
\ve_\rL=\ve_\rR:=\ve~.
\label{3120uuu}
\eea
The (2,0) projection of (\ref{sK-1})--(\ref{sK-2-4}) gives
\bsubeq \label{extra3120}
\begin{gather}
\cD_\a\bar{\ve}_\b=-\cD_\a\ve_\b=
4S\ve_{\a\b}\,\bar{\ve}
~,~~~
\cDB_\a\ve_\b=-\cDB_\a\bar{\ve}_\b=
-4S\ve_{\a\b}\,\ve
~,
\label{extra3120-1}
\\
\cDB_\a\ve=0
~,~~~
\cD_\a\ve=
\frac{\ri}{2} \big(\ve_{\a}-\bar{\ve}_{\a}\big)
~.
\label{extra3120-2}
\end{gather}
\esubeq
These imply
\bea
\cD_\a(\ve_\b+\bar{\ve}_\b)=\cDB_\a(\ve_\b+\bar{\ve}_\b)=0
~.
\label{extra3120-1b}
\eea
The real parameter $\sigma$ turns out to vanish.

\subsubsection{AdS superspace reduction (2,2) $\to$ (2,0)}

For the final case of the (2,2) to (2,0) reduction, a local $R$-symmetry transformation
can be applied to bring 
$l^{ij}$ and $r^{\bai\baj}$ to the form:
\bsubeq
\bea
&l^{11}=l^{22}=r^{\bau\bau}=r^{\bad\bad}=0~,~~~
l^{12}=-\ri~,~~~r^{\bau\bad}=\ri
~,
\\
&\ve_\rL=\L^{22}=0~,~~~
\ve_\rR=\L^{\bad\bad}=0~,
\label{2220uuu}
\eea
\esubeq
and then the condition \eqref{condONcS} holds.
The (2,0) projection of (\ref{sK-1})--(\ref{sK-2-4}) gives
\bea
&\cD_\a\ve_\b=\cDB_\a\ve_\b=0
~.
\label{extra2220-1}
\eea
The real parameter $\sigma$ must be constant.


\subsection{From $\cN=4$ to $\cN=3$ AdS superspaces}
\label{N4-3}

Rather than reduce the $\cN=4$ AdS superspaces to $\cN=2$, we can choose
instead to reduce to $\cN=3$. This will turn out to have a very interesting
consequence for $\cN=4$ supersymmetric  $\s$-models: it will be possible to directly
relate an $\cN=4$ $\s$-model in projective superspace to one in $\cN=3$
(and vice-versa) with very little work. However, in contrast to the
$\cN=2$ reductions considered above, a restriction must be imposed on the $\cN=4$
geometry: 
$X=0$. The point is that all $\cN=3$ AdS superspaces are conformally flat \cite{KLT-M12}.
The conformally flat $\cN=4$ AdS superspaces are those for which  $X=0$ \cite{KLT-M12}.

To start with, let us  look for an $\cN=3$ subalgebra of the $\cN=4$ algebra.
For this we break the $R$-symmetry group SU(2)$_\rL\times$SU(2)$_\rR$
to its central subgroup SU(2)$_{\cJ}$
generated by
\bea\label{eq_defJij}
\cJ^{ij}:= \bL^{ij} + \delta^i{}_{\bar i} \delta^j{}_{\bar j} \bR^{\bar i\bar j}~.
\eea
Here we identify the $\bai,\baj$ indices with the $i,j$ ones.
By choosing $\cJ^{ij}$ to be the SU(2) generator of the $\cN=3$ structure group
we naturally split the $\cN=4$ derivatives as
\bea
\cD_\a^{i\bai}=\cD_\a^{(i\bai)}-\hf\ve^{i\bai}\ve_{j\baj}\cD_\a^{j\baj}
~.
\label{decomp-4-3}
\eea
The second term is invariant under the action of $\cJ^{ij}$ while the first one, symmetric in $i$ and
$\bai$, transforms as
\bea
[\cJ^{ij},\cD_\a^{(k\bak)}]=
\hf\ve^{ki}\cD_\a^{(j\bak)}
+\hf\ve^{kj}\cD_\a^{(i\bak)}
+\hf\ve^{\bak i}\cD_\a^{(kj)}
+\hf\ve^{\bak j}\cD_\a^{(ki)}
~.
\eea
The remainder of the $\cN=4$ $R$-symmetry group is generated by
\begin{align}
\D^{ij}:=\bL^{ij} - \delta^i{}_{\bar i} \delta^j{}_{\bar j} \bR^{\bar i\bar j}~.
\end{align}

${}$From the
$\cN=4$ AdS algebra \eqref{N=4alg}, we derive the following 
vector-spinor commutator
\bea
{[}\cD_{\a\b},\cD_\g^{(k\bak)}{]}&=&
-2\ve_{\g(\a}\d_{\b)}^{\d}\Big(\d^{(k}_l\d^{\bak)}_\bal\cS
+\cS^{(k}{}_l{}^{\bak)}{}_\bal\Big)\cD_\d^{l\bal}
~.
\eea
By imposing the closure of the algebra of the operators 
$\big(\cD_\a^{(i\bai)},\cD_a,\cJ^{kl}\big)$,
 the following necessary  condition arises
\bea
\ve_{i\baj}\cS^{i(j\bai)\baj}=0
~,
\label{N=4N=3consistency-1}
\eea
together with the requirement that $X\equiv 0$.
The equation \eqref{N=4N=3consistency-1} implies
\bea
\cS^{ij\bai\baj}=\cS^{(ij\bai\baj)}
-\frac{1}{3}\ve^{i(\bai}\ve^{\baj) j}\ve_{k\bak}\ve_{l\bal}\cS^{kl\bak\bal}
\label{N=4N=3consistency-2}
~,
\eea
as well as $\cS^{ij\bai\baj}=\cS^{\bai\baj ij}$.
Then the algebra of the  covariant  derivatives $\cD_a $ and $ \cD_\a^{(i\bai)}$ becomes
\bsubeq
\bea
\{\cD_\a^{(i\bai)},\cD_\b^{(j\baj )}\}&=&
-2\ri\ve^{i(j}\ve^{\baj)\bai}\cD_{\a\b}
-4\ri\Big(\cS^{(ij\bai \baj) }
-\ve^{i(j}\ve^{\baj) \bai }\big(\cS-\frac{1}{6}\ve_{k\bak}\ve_{l\bal}\cS^{kl\bak\bal}\big)\Big)\cM_{\a\b}
\non\\
&&
-\ri\ve_{\a\b}\Big(\ve^{ij}\cS^{(\bai \baj kl)}\cJ_{kl}
+\ve^{\bai\baj}\cS^{(i j kl)}\cJ_{kl}\Big)
\non\\
&&
+2\ri\ve_{\a\b}\big(\cS-\frac{1}{6}\ve_{k\bak}\ve_{l\bal}\cS^{kl\bak \bal}\big)
\Big(
\ve^{i j }\cJ^{\bai\baj}
+\ve^{\bai \baj }\cJ^{ij}
\Big)
~,~~~~~~
\label{albgN4-N3-1}
\\
{[}\cD_{\a\b},\cD_\g^{(k\bak)}{]}&=&
-2\ve_{\g(\a}\Big(
\big(\cS-\frac{1}{6}\ve_{k\bak}\ve_{l\bal}\cS^{kl\bak\bal}\big)\cD_{\b)}^{(k\bak)}
+\cS^{(kl\bak\bal)}\cD_{\b) (l\bal)}\Big)
~,
\label{albgN4-N3-3/2}
\\
{[}\cD_a,\cD_b{]}&=&-4S^2\cM_{ab}~.
\label{albgN4-N3-2}
\eea
\esubeq
This has the form of a general $\cN=3$ AdS algebra \cite{KLT-M12} once we
identify the SU(2)$_\rL$ and SU(2)$_\rR$ indices.

There are two possibilities for the $\cN=3$ AdS algebra:
\bsubeq
\bea
(3,0):&&~~~~~~
\cS=S~,~~
\cS^{(ijkl)}=0
\\
(2,1):&&~~~~~~
\cS=\frac{1}{3}S~,~~
\cS^{(ijkl)}=-Sw^{(ij}w^{kl)}
~,~~
w^{ij}w_{ij}=2
~.~~~~~~~~~
\eea
\esubeq
Here $w^{ij}=w^{ji}$ is a real covariantly constant isotriplet which can be identified with the 
field strength of a frozen vector multiplet on the (2,1) AdS superspace, see subsection
\ref{subsection5.3} for more details.
We immediately conclude that if we begin with the  (4,0) algebra, then $\cS^{ij\bai\baj}=0$ and
only a (3,0) reduction is possible. Similarly, for the (2,2) case,
one finds $\cS^{(ij\bai\baj)}\ne0$, which always leads
to a (2,1) truncation. For the (3,1) case, we find two
distinct classes of solution to \eqref{N=4N=3consistency-1}:
one with $\cS^{(ij\bai\baj)}=0$ and a (3,0) geometry;
and another with $\cS^{(ij\bai\baj)}\ne0$ and a (2,1) geometry.

Now we sketch some of the technical details of the reduction.
Given an $\cN=4$ tensor superfield  $U(x,\q_{\imath\bar{\jmath}})$,
we define its $\cN=3$ projection as
\bea
U|:=U(x,\q_{\imath \bar{\jmath}})|_{\ve^{i\bai}\q_{i\bai}=0}~.
\label{N3red-1}
\eea
The superfield  $U|$  depends on only the six Grassmann coordinates
$\q^\mu_{(i\bai)}$.
The projection of the $\cN=4$ covariant derivatives 
to the $\cN=3$ subspace formally goes along the same lines of the
reduction from $\cN=4$ to $\cN=2$ described in the previous subsection
and here  we skip the technicalities.
The only important point is this:
since in the different consistent $\cN=3$ truncations 
the derivatives $\cD_a $ and $\cD_\a^{(i\bai)}$ form a closed  algebra,
one can use the freedom to perform general coordinate, local Lorentz and 
SU(2)$_\rL\times$SU(2)$_\rR$ transformations
to chose a gauge where
\bea
\cD_A| = (\cD_a| , \cD_\a^{(i\bai)}|)= {E}_A{}^M \pa_M +\hf {\O}_A{}^{cd}\cM_{cd}
+ \hf{ \F}_A{}^{kl} \cJ_{kl}
\label{N=3-1.4}
\eea 
denote the covariant derivatives of either (3,0) or (2,1) AdS.
In the (2,1) case ${ \F}_A{}^{kl}={ \F}_Aw^{kl}$ since the SU(2)$_\cJ$ is broken to a U(1) subgroup.

The reduction of an $\cN=4$ Killing vector $\x$ is more interesting.
The natural decomposition of $\xi$ is
\bea
\x = \x^a \cD_a + \x^\a_{i\bai} \cD^{i\bai}_\a
	= \x^a \cD_a + \x^\a_{(i\bai)} \cD^{(i\bai)}_\a
	+ \X^\a\ve_{i\bai}\cD^{i\bai}_\a~,\qquad
\X^\a:=-\hf\ve^{i\bai}\x^{\a}_{i\bai}~.
\eea
The parameters $\x^a\vert$ and $\xi^\a_{(i \bai)}\vert$ will play the
role of the $\cN=3$ Killing vectors while $\X^\a\vert$ will yield the
extra supersymmetry.

For the SU(2)$_\rL\times$SU(2)$_\rR$ symmetry parameters, we must
be a little careful. Because we are identifying the two types of
SU(2) indices, we will affix an additional label to the parameters,
denoting them $\L_\rL^{ij}$ and $\L_\rR^{\bar i \bar j}$. Then we
may decompose as
\bea
\L^{ij}:=\L_\rL^{ij}+\L_\rR^{\bar i \bar j}~,~~~
\tilde{\L}^{ij}:=\L_\rL^{ij}-\L_\rR^{\bar i \bar j}~.
\eea
The superfields $\L^{ij}$ and $\tilde{\L}^{ij}$ are 
associated respectively with the $\cJ^{ij}$ and $\D^{ij}$ generators.

By consistently reducing to $\cN=3$ AdS 
the $\cN=4$ Killing equations \eqref{sK-2-1}--\eqref{sK-2-8},
it can be proven that the superfields
$(\x^a\vert,\x^\a_{(i\bai)}\vert,\L^{ab}\vert,\L^{ij}\vert)$ parametrize a general isometry of 
(3,0) or (2,1) AdS
(see \cite{KLT-M12} for the $\cN=3$ AdS Killing equations).
Similarly, one can derive the differential constraints satisfied by the  
extra superfields $\X^\a\vert$ and $\tilde{\L}^{ij}\vert$ that parametrize
the extra supersymmetry and the remaining $R$-symmetry.
The most important equations are
\bsubeq
\bea
\cD_\a^{(i\bai)}\X_\b&=&-\hf\ve_{\a\b}\tilde{\L}^{i\bai}
~,~~~
\cD_a\X^\b=\Big(\hf\ve_{i\bai}\ve_{j\baj}\cS^{ij\bai\baj}-\cS\Big)\X^\g(\g_a)_\g{}^\b
~,
\label{N3uuuseful-2}
\\
\cD_\a^{(i\bai)}\tilde{\L}^{kl}&=&
\ri\Big(
8\ve^{k(i}\ve^{\bai)l}\big(\cS
+\frac{1}{6}\ve_{k\bak}\ve_{l\bal}\cS^{kl\bak\bal}\big)
-4\cS^{(i\bai kl)}\Big)\X_\a
~,~~~
\cD_a\tilde{\L}^{kl}=0
~.
\label{N3uuuseful-3}
\eea
\esubeq
Let us now address each case specifically.

\paragraph{(4,0) $\rightarrow$ (3,0)}
This is the easiest case since $\cS^{ij\bai\baj}\equiv 0$ and $\cS=S$.
There is no reduction of the structure group SU(2)$_\rL\times$SU(2)$_\rR$:
half of it is manifest in the (3,0) structure group SU(2)$_\cJ$ and
parametrized by $\L^{ij}$, while the non-manifest half is generated
by $\Delta^{ij}$ and parametrized by $\tilde \L^{ij}$.

\paragraph{(3,1) $\rightarrow$ (3,0)}
In the (3,1) case, we have a covariantly constant isovector $w^{i\bai}$, which
reduces the structure group to an SU(2) generated by
$\bL^{ij} + w^i{}_{\bar k} w^j{}_{\bar l} \bR^{\bar k \bar l}$.
A (3,0) reduction arises when this generator can be
identified with the (3,0) SU(2) generator $\cJ^{ij}$ \eqref{eq_defJij}.
This occurs when $w^{i \bar k}$ is antisymmetric; without
loss of generality, we can choose
$w_{i \bar k} = \veps_{i \bar k}$ and $w^{i \bar k} = -\veps^{i \bar k}$.
Using the expressions for $\cS$ and $\cS^{ij\bai\baj}$ in the (3,1) case,
one easily finds
\bea
\cS^{(ij\bai\baj)}=0~, \qquad \cS-\frac{1}{6}\ve_{k\bak}\ve_{l\bal}\cS^{kl\bak\bal}=S
~,
\eea
which is a necessary condition for the (3,1)$\to$(3,0) reduction to be consistent.

\paragraph{(3,1) $\rightarrow$ (2,1)}
The other possible reduction of (3,1) is to (2,1).
Recall that (2,1) is equipped with a symmetric $w_{ij}$ tensor.
The obvious thing to do here is to choose $w_{i \bar k}$ to be symmetric
and identify it with $w_{ij}$.
One can check that
\bea
\cS-\frac{1}{6}\ve_{k\bak}\ve_{l\bal}\cS^{kl\bak\bal}=\frac{1}{3}S~,~~~
\cS^{(ij\bai\baj)}=-Sw^{(ij}w^{\bai\baj)}
~,
\eea
which is the appropriate condition for a (2,1) AdS superspace.
To understand what happens to the $R$-symmetry group,
we observe that the (3,1) $R$-symmetry parameter is constrained by
$\L_\rL^{kl}=w^{k}{}_{\bak}w^{l}{}_{\bal}\L_\rR^{\bak\bal}$.
By choosing $w^{i\bai}$ to be symmetric, one can show that
\begin{align}
\L^{ij} = \L_L^{ij} + \L_R^{\bai \baj} = w^{ij} \L
\end{align}
for some parameter $\L$.
This is what we expect since the SU(2)$_\cJ$ in the (2,1) case is reduced to a U(1) subgroup.
The other parameter $\tilde \L^{ij}$ obeys
\begin{align}
\tilde \L^{ij} w_{ij} = 0~.
\end{align}
The unbroken SU(2) of the $(3,1)$ algebra has decomposed into a U(1) appearing in
the $(2,1)$ algebra, with the rest relegated to the extended supersymmetry.

\paragraph{(2,2) $\rightarrow$ (2,1)}
For the (2,2) geometry, there is a single possibility: reduction to (2,1).
We recall that the (2,2) geometry is characterized by 
two covariantly constant symmetric tensors: $l^{ij}$ and $r^{\bar k \bar l}$.
In reducing to $(2,1)$, however, we should end up with a single tensor $w^{ij}$.
We can always choose $l^{ij} = -r^{\bai \baj} = \pm w^{i j}$ and check that
(keeping in mind the fact that  $\cS=0$ in the (2,2) case)
\bea
-\frac{1}{6}\ve_{k\bak}\ve_{l\bal}\cS^{kl\bak\bal}=\frac{1}{3}S~,\qquad
\cS^{(ij\bai\baj)}=-Sw^{(ij}w^{\bai\baj)}
\eea
as expected.
Then it can be easily proven that the $R$-symmetry parameter $\L^{ij}$ becomes
\bea
\L^{ij}=w^{ij}\L~,
\eea
for some real parameter $\L$. 
We also find $\tilde\L^{ij}=w^{ij}\tilde\L$, for some real parameter $\tilde\L$.
These results are completely natural,  since the $R$-symmetry part of the (2,2) structure group 
is U(1)$_\rL \times$U(1)$_\rR$, in accordance with  eq. \eqref{7.14c}.
One U(1) subgroup turns into the U(1) $R$-symmetry of the (2,1) AdS superspace, 
and the other becomes part of the extended supersymmetry.


\section{Rigid $\cN=4$  supersymmetric field theories in AdS: Off-shell multiplets and invariant 
actions}
\setcounter{equation}{0}

In this and subsequent sections, our goal is to apply the supergravity
techniques of \cite{KLT-M11} to describe general 
nonlinear $\s$-models in $\rm AdS_3$ possessing $\cN=4$ supersymmetry. 
Our discussion here is a generalization of the $\cN=3$ analysis given in \cite{KLT-M12}.

\subsection{Covariant projective supermultiplets}

In complete analogy with matter couplings in $\cN=4$ supergravity \cite{KLT-M11}, 
a large class of rigid supersymmetric theories in (4,0), (3,1) and (2,2) AdS superspaces can be 
formulated in terms of covariant projective supermultiplets.
As described in \cite{KLT-M11},
because the supergravity structure group includes the factor SU(2)$_\rL\times$SU(2)$_\rR$, 
it is natural to introduce left and right isotwistors, $v_\rL= v^i$, $v_\rR=v^{\bai}$,
and to define inequivalent left and right projective multiplets. 
Here we will focus on the properties of left projective multiplets;
analogous results for right projective multiplets can be obtained by applying a mirror map 
\cite{KLT-M11}.

A {\em covariant left projective supermultiplet} of weight $n$,
$Q_\rL^{(n)}(z^M,v^i)$, is defined to be a Lorentz scalar and SU(2)$_\rR$ invariant superfield that 
lives on the appropriate $\cN=4$ AdS superspace,
is holomorphic with respect to isospinor variables $v^i $ on an open domain of 
${\mathbb C}^2 \setminus  \{0\}$, 
and is characterised by the following conditions:
\begin{itemize}

\item[(i)] $Q_\rL^{(n)}$  is  a homogeneous function of $v$ 
of degree $n$,
\be
Q_\rL^{(n)}(z,c\,v_\rL)\,=\,c^n\,Q_\rL^{(n)}(z,v_\rL)~, 
\qquad c\in \mathbb{C}^* \equiv {\mathbb C} \setminus  \{0\}~;
\label{weight}
\ee

\item[(ii)]  
$Q_\rL^{(n)}$ transforms under the AdS isometry supergroup as
\bea
\d_\x Q_\rL^{(n)} 
&=& \Big( \x + \L^{ij} \bL_{ij} \Big) Q_\rL^{(n)} ~,  
\non \\ 
\L^{ij} \bL_{ij}  Q_\rL^{(n)}&=& -\Big(\L^{(2)} {\bm \pa}_\rL^{(-2)} 
-n \, \L^{(0)}\Big) Q_\rL^{(n)} ~, \qquad 
{\bm \pa}_\rL^{(-2)} :=\frac{1}{(v_\rL,u_\rL)}u^{i}\frac{\pa}{\pa v^{i}}~,
\label{harmult1}   
\eea 
where $\x$ denotes an arbitrary  AdS Killing vector field, eq.  (\ref{sK-1}),
 and $\L^{ij}$ the associated SU(2)$_\rL$ 
 parameter;

\item[(iii)] $Q_\rL^{(n)}$  obeys the analyticity constraint
\be
\cD^{(1)\bai}_{\a} Q_\rL^{(n)} = 0~, \qquad \cD_\a^{(1)\bai}:=v_i\cD_\a^{i\bai}
~.
\label{ana}
\ee  
\end{itemize}

Some comments are necessary.
The homogeneity condition (\ref{weight}) and the analyticity constraint (\ref{ana}) 
are consistent with the interpretation that the isospinor
$ v^{i} \in {\mathbb C}^2 \setminus\{0\}$ is   defined modulo the equivalence relation
$ v^{i} \sim c\,v^{i}$,  with $c\in {\mathbb C}^*$.
Therefore, the projective multiplets live in ${\cM}^{3|6} \times {\mathbb C}P^1$.
On the other hand, the transformation law \eqref{harmult1} and the parameters
\bea
\L^{(2)} :=\L^{ij}\, v_i v_j 
~,\qquad
\L^{(0)} :=\frac{v_i u_j }{(v_\rL,u_\rL)}\L^{ij}~,
\qquad (v_\rL,u_\rL):=v^iu_i
\label{W2t3}
\eea
depend on an additional  isotwistor  $u_{i}$, which is 
subject only to the condition $(v_\rL,u_\rL)\ne0$ and otherwise is completely arbitrary.
Nevertheless, both $Q_\rL^{(n)}$ and $\d_\x Q_\rL^{(n)}$ are independent of $u_i$.

The analyticity condition \eqref{ana} is quite powerful. Requiring
the consistency condition $\{\cD_\a^{(1)\bai},\cD_\b^{(1)\baj}\} Q_\rL^{(n)} = 0$,
we conclude that all left projective multiplets must be Lorentz and SU(2)$_\rR$ scalars.
If instead we require only the conditions (i) and (ii) to be satisfied,
we find the so-called {\it left isotwistor} supermultiplets, which may
belong to nontrivial representations of the Lorentz and SU(2)$_\rR$ groups.
These isotwistor superfields can be used to construct projective ones
with the aid  of the left analytic projection operator $\D_\rL^{(4)}$ \cite{KLT-M11},
which in AdS is given by
\bea
\D_\rL^{(4)}&=&\frac{1}{48}
(\cD^{(2)\bak\bal}-4\ri \cS^{(2)\bak\bal})\cD^{(2)}_{\bak\bal}
~,~~~~~~
\cD^{(2)}_{\bai\baj}:=\cD^{(1)\g}_{(\bai}\cD^{(1)}_{\g \baj)}~.
\label{6.54a} 
\eea
In fact, given a Lorentz and SU(2)$_\rR$ invariant weight-$(n-4)$
left isotwistor superfield, 
$T_\rL^{(n-4)}$, 
a covariant left projective multiplet $Q_\rL^{(n)} (v_\rL) $ of weight $n$
can be constructed as
\bea
Q_\rL^{(n)}=\D_\rL^{(4)} T_\rL^{(n-4)}~.
\label{6.53}
\eea

General off-shell $\cN=4$ $\sigma$-models can be described by {\it left arctic}
weight-$n$ projective multiplets $\U_\rL^{(n)} (v)$. These are defined to be 
holomorphic in the north chart
of $\mathbb CP^1$, and so can be represented as
\bea
\U_\rL^{(n)} ( v_\rL) &=&  (v^{1})^n\, \U_\rL^{[n]} ( \z) ~, \qquad 
\U_\rL^{ [n] } ( \z) = \sum_{k=0}^{\infty} \U_k  \z^k 
~.
\label{arctic1}
\eea
Their smile-conjugates are {\it left antarctic} multiplets $\breve{\U}_\rL^{(n)} (v_\rL) $,
 \bea
\breve{\U}_\rL^{(n)} (v_\rL) &=& 
(v^{2}  \big)^{n}\, \breve{\U}_\rL^{[n]}(\z) =
(v^{1} \,\z \big)^{n}\, \breve{\U}_\rL^{[n]}(\z) ~, \qquad
\breve{\U}_\rL^{[n]}( \z) = \sum_{k=0}^{\infty}  {\bar \U}_k \,
\frac{(-1)^k}{\z^k}~.~~~
\label{antarctic1}
\eea
In complete analogy to the $\cN=3$ case,
we have introduced the inhomogeneous complex coordinate 
$\z= v^{2}/v^{1}$ on the north chart of  ${\mathbb C}P^1$.
The pair of fields $\U_\rL^{[n]} ( \z)$ and $\breve{\U}_\rL^{[n]}(\z)$ 
constitute the so-called left polar weight-$n$ multiplet.

For further details on $\cN=4$ multiplets, the reader can refer to
the full supergravity treatment in \cite{KLT-M11} and restrict to the
appropriate $\cN=4$ AdS background.


\subsection{Reduction to $\cN=2$ AdS superspaces}

It will be useful to reduce $\cN=4$ AdS projective multiplets to 
$\cN=2$ AdS superfields.
According to the analysis of section \ref{N4AdSsuperspaces}, one can choose to
reduce $\cN=4$ AdS either to (1,1) or (2,0) AdS. 
The (2,0) reduction is more interesting because it is more general --
it is available for \emph{all} the $\cN=4$ AdS geometries --
so we will restrict our attention to it.

Actions involving the polar multiplets
${\U}_\rL^{[n]}(z,\z)$ and ${\breve\U}_\rL^{[n]}(z,\z)$ 
yield the most general $\sigma$-models, so it is sufficient to discuss these alone.
We work in the north chart where the analyticity condition 
is equivalent to
\bea
\cD_\a^{1\bad}\U_\rL^{[n]}=\frac{1}{\z} \cD_\a^{2\bad}\U_\rL^{[n]}~,~~~~~~
\cD_\a^{2\bau}\U_\rL^{[n]}=\z\cD_\a^{1\bau}\U_\rL^{[n]}
~.
\label{anaN2}
\eea
These equations ensure that the dependence of $\U^{[n]} (x,\q_{\imath {\bar{\jmath}}} , \z)$ on 
the Grassmann coordinates $\q^\m_{1\bar{2}}$ and $\q^\m_{2\bar{1}}$
is entirely determined 
by its dependence on the other coordinates  $\q^\m_{1\bar{1}} $ and  $\q^\m_{2\bar{2}}$.
In other words, all the information about $\U_\rL^{[n]}(\z)$ 
is encoded in its $\cN=2$ projection $\U_\rL^{[n]}(\z)|$.
The same holds true for all projective multiplets written in the north chart.

The isometry transformations for the polar multiplet, reduced to (2,0) AdS, are
\bsubeq
\bea \label{eq_dUpL}
\d_\x \U_\rL^{[n]}|
&=& \Big{[} 
\t
+\ri t\Big(\z\frac{\pa}{\pa \z}-\frac{n}{2}\Big)
+\z\bar{\ve}^\a \cD_\a
-\frac{1}{\z}\ve_\a \cDB^\a
+\ri\s\Big(\z\frac{\pa}{\pa \z}-\frac{n}{2}\Big)
\non\\
&&~
-4S\Big( \bar{\ve}_\rL\z 
+\frac{1}{\z}{\ve}_\rL\Big)\z\frac{\pa}{\pa \z}
+4Sn\,\z\bar{\ve}_\rL
 \Big{]} \U_\rL^{[n]}| 
 ~,~~~~~~
 \\  
\d_\x \breve{\U}_\rL^{[n]} |
&=& \Big{[} 
\t
+\ri t\Big(
\z\frac{\pa}{\pa \z}
+\frac{n}{2}\Big)
+\z\bar{\ve}^\a \cD_\a
-\frac{1}{\z}\ve_\a \cDB^\a
+\ri \s\Big(
\z\frac{\pa}{\pa \z}
+\frac{n}{2}\Big)
\non\\
&&~
-4S\Big( \bar{\ve}_\rL\z 
+\frac{1}{\z}{\ve}_\rL\Big)\z\frac{\pa}{\pa \z}
-4Sn \frac{1}{\z}{\ve}_\rL
 \Big{]}
 \breve{\U}_\rL^{[n]}  |
 ~.
\eea 
\esubeq
Note that $\t=\t^a\cD_a+\t^\a\cD_\a+\bar{\t}_\a\cDB^\a$ is a (2,0) Killing vector.
The additional parameters appearing above are defined in \eqref{N2proj-20-1}--\eqref{N2proj-20-2}.

We will also need the (2,0) transformation for the projective superspace Lagrangian,
which is a real weight-two left projective superfield $\cL_\rL^{(2)}$.
In the north chart, we represent it as
\bea
\cL_\rL^{(2)}(v_\rL)=\ri (v^1)^{2}\z\cL_\rL^{[2]}(\z)~,
\label{8.11}
\eea
and we find
\bea
\d_\x \cL_\rL^{[2]} |
&=&\Big{[}
\t
+\ri t\z\frac{\pa}{\pa \z}
+\z\bar{\ve}^\a \cD_\a
-\frac{1}{\z}\ve_\a \cDB^\a
+\ri \s\z\frac{\pa}{\pa \z}
\non\\
&&~
-4S \bar{\ve}_\rL\z\Big(\z\frac{\pa}{\pa \z}
-1\,\Big)
-4S \ve_\rL\frac{1}{\z}\Big( \z\frac{\pa}{\pa \z}
+1\Big)
 \Big{]} \cL_\rL^{[2]} |
 ~.
 \label{L2n}
\eea


\subsection{Reduction to $\cN=3$ AdS superspaces}
\label{projective-4-3}

An interesting alternative reduction procedure is to take $\cN=4$ projective 
superfields to $\cN=3$.\footnote{Within the 3D harmonic superspace approach, 
the reduction  $\cN=4 \to \cN=3$ was worked out by Zupnik for the case 
of  Poincar\'e supersymmetry \cite{Zupnik:2010av}.} 
As discussed already in section \ref{N4AdSsuperspaces}, there are
several distinct cases to consider since neither (3,0) nor (2,1) AdS is universal
in the same sense that (2,0) is. Fortunately, the general features of the reduction
are independent of which exact reduction is being considered.

We have previously described, in subsection \ref{N4-3}, how the $\cN=3$ reduction
of a general $\cN=4$ superfield proceeds.  In the case of a left projective multiplet,
$Q_\rL^{(n)}$, a slight elaboration is needed since the multiplet also depends
on the isotwistor $v_\rL^i:= v^i$ parametrizing the manifold $\mathbb CP^1$.
Recall that such a multiplet must satisfy the analyticity condition $\cD_\a^{(1)\baj}Q^{(n)}_\rL=0$.
Decomposing the $\cN=4$ covariant derivative as in \eqref{decomp-4-3}, one finds
\bea
&&v_i\cD_\a^{i\baj}Q^{(n)}_\rL=0~~~\Longleftrightarrow~~~
\cD_\a^{(2)}Q^{(n)}_\rL=0~,~~~
\ve_{i\baj}\cD_\a^{i\baj}Q^{(n)}_\rL=-2\cD_\a^{(0)}Q^{(n)}_\rL
~,
\label{proj-4-3_1}
\eea
where we have defined 
\bea
\cD_\a^{(2)}:=v_iv_j \delta^j{}_{\baj} \cD_\a^{(i\baj)}~,~~~~~~
\cD_\a^{(0)}:=\frac{v_i u_j}{(v_\rL,u_\rL)} \delta^j{}_{\baj} \cD_\a^{(i\baj)}
~.
\eea
We have introduced an auxiliary isotwistor $u_{i}$ above, but
the multiplet $Q_\rL^{(n)}$ doesn't depend on it.

In the $\cN=3$ reduction introduced in subsection \ref{N4-3}, we identified
the left and right $\rm SU(2)$ indices. It follows that we should also identify
the left and right isotwistors. In other words, we take $v_\rL^i := v^i$
and $v_\rR^{\bai} := v^{\bai}$ to be equivalent, $v^i = \delta^i{}_{\bar i} v^{\bai}$. Similarly,
we take the auxiliary isotwistors to obey $u_i = \delta_i{}^{\bai} u_{\bai}$.
The meaning of the two constraints in \eqref{proj-4-3_1} then becomes clear. 
The first constraint $\cD_\a^{(2)}Q^{(n)}_\rL=0$, when projected to $\cN=3$
superspace, is simply the analyticity condition of an $\cN=3$  projective superfield \cite{KLT-M12}.
The second constraint
then fixes the dependence of $Q_\rL^{(n)} (x,\q_{\imath {\bar{\jmath}}} , v)$ on 
the Grassmann coordinate $\ve^{{\imath {\bar{\jmath}}}}\q^\a_{{\imath {\bar{\jmath}}}}$ 
in terms of its dependence on the other coordinates  $\q^\a_{(\imath {\bar{\jmath}})}$.
In other words, all the information about $Q_\rL^{(n)}$ 
is encoded in its $\cN=3$ projection 
$Q_\rL^{(n)}|_{\ve^{{\imath {\bar{\jmath}}}}\q^\a_{{\imath {\bar{\jmath}}}}=0}$.

We can now rewrite the isometry transformation \eqref{harmult1} for
a weight-$n$ left projective superfield projected to $\cN=3$:
\bea
\d_\x Q_\rL^{(n)} \vert
&=& \Big( \x^a\cD_a+\x^\a_{(i\baj)}\cD_\a^{(i\baj)}
	-\L^{(2)} {\bm \pa}^{(-2)} 
	+ n \, \L^{(0)}
\non\\
&&~~~
-2\X^\a\cD_\a^{(0)} 
-\tilde\L^{(2)} {\bm \pa}^{(-2)} 
+ n \, \tilde\L^{(0)} \Big) Q_\rL^{(n)} \vert
~.
\label{left-4-3}
\eea
The first line corresponds to an $\cN=3$ Killing isometry, with
parameters $(\xi^a\vert, \,\xi^\a_{(i\bar j)}\vert, \, \L^{ij}\vert)$, as described
in \cite{KLT-M12}. The second line, involving the parameters
$(\Xi^\alpha\vert,\,\tilde\L^{ij}\vert)$ corresponds to an extended supersymmetry
and $R$-symmetry transformation associated with the rest of the $\cN=4$
AdS isometry group. Naturally, the precise relations these parameters
satisfy depends on the case in question.

In order for the transformation law \eqref{left-4-3} to be sensible, $\delta_\xi Q_L^{(n)}\vert$
must remain a weight-$n$ projective multiplet in $\cN=3$ superspace.
The first line of \eqref{left-4-3} satisfies this condition automatically since
it is an $\cN=3$ Killing transformation. To verify that the second line
of \eqref{left-4-3} is sensible, we must check two conditions: it must be
annihilated by $\cD_\a^{(2)}$ and it must be independent of the auxiliary isotwistor
$u_i$.

Both conditions may be checked by exploiting an alternative parametrization
of the parameters $\Xi_\alpha$ and $\tilde \Lambda^{ij}$. Using 
eqs. \eqref{N3uuuseful-2} and \eqref{N3uuuseful-3}, one can show that
\begin{align}
\Xi_\alpha = -\frac{\ri}{2}\cD_\alpha^{(2)} \Omega^{(-2)}
~,~~~~~~
\tilde{\L}^{(2)}:=v_iv_j\tilde\L^{ij}=\frac{\ri}{2}\cD^{(4)}\Omega^{(-2)}~,~~~
\cD^{(4)}:=\cD^{\a(2)}\cD_\a^{(2)}~,~~~~~~
\label{useful-X-O}
\end{align}
where $\Omega^{(-2)}$ is some weight-$(-2)$ isotwistor superfield depending on $v^i$
but not $u_i$. Naturally, it must be chosen so that $\Xi_\alpha$ is independent of
$v^i$, ${\bm \pa}^{(-2)} \Xi_\alpha = 0$; in addition, a number of other conditions
are encoded in \eqref{N3uuuseful-2}--\eqref{N3uuuseful-3}. We will not attempt a comprehensive
analysis here, since the representation \eqref{useful-X-O} is sufficient
for our needs.

In terms of the parameter $\Omega^{(-2)}$,
the transformation \eqref{left-4-3} can be rewritten
\bea
\delta_\x Q_\rL^{(n)}\vert &=& \delta_\x^{\cN=3} Q_\rL^{(n)}\vert
	+ \D^{(4)}\Big(\Omega^{(-2)} {\bm\pa}^{(-2)}  
	+ \frac{n}{2} ({\bm\pa}^{(-2)} \Omega^{(-2)})\Big)Q_\rL^{(n)}\vert
	~,
	\label{transf-4-3_2}
\eea
where we have introduced 
the $\cN=3$ analytic projector operator \cite{KLT-M12}
\bea
\D^{(4)}:=\frac{\ri}{4}\big(\cD^{(4)}-4\ri\cS^{(4)}\big) 
~,~~~~~~
\cS^{(4)}:=v_i v_j v_k v_l\cS^{ijkl}
~.
\eea
Its presence guarantees that the second term in \eqref{transf-4-3_2} is analytic.
It is an instructive exercise to check that this expression is also independent
of $u_i$. In the form given above, this is straightforward.

Although our analysis has focused on left projective superfields,
it turns out that the only difference for the case of an $\cN=4$ \emph{right}
projective superfield is the overall sign of the extended supersymmetry
transformation:
\bea
\delta_\x Q_\rR^{(n)}\vert &=& \delta_\x^{\cN=3} Q_\rR^{(n)}\vert
	- \D^{(4)}\Big(\Omega^{(-2)} {\bm\pa}^{(-2)}  
	+ \frac{n}{2} ({\bm\pa}^{(-2)} \Omega^{(-2)})\Big)Q_\rR^{(n)}\vert
	~.
	\label{transf-4-3_2right}
\eea


\subsection{$\cN=4$ supersymmetric actions}
\label{N4action-full}

In order to formulate the dynamics of rigid $\cN=4$ supersymmetric field theories in 
$\rm AdS_3$,  a manifestly supersymmetric action principle is required. 
We can readily construct such an action principle
by restricting to AdS the locally supersymmetric action introduced in \cite{KLT-M11}.
The action is generated by a real Lagrangian  $\cL_\rL^{(2)} (z,v_\rL)$, 
which is a covariant weight-two left projective multiplet, and has the form: 
\bea
S[\cL_\rL^{(2)}]&=&
\frac{1}{2\pi\ri} \oint_\g  (v_\rL, \rd v_\rL)
\int \rd^3 x \,{\rm d}^8\q\,E\, \cC_\rL^{(-4)}\cL_\rL^{(2)}~, 
\qquad E^{-1}= {\rm Ber}(E_A{}^M)~.
\label{InvarAc}
\eea
Here the line integral  is carried out over a closed contour
$\g =\{v^i(t)\}$ in ${\mathbb C}P^1$.
The action involves a left isotwistor superfield $\cC_\rL^{(-4)} (z,v_\rL)$ which can be defined as
\bea
\cC_\rL^{(-4)}:=\frac{\cU_\rL^{(n)}}{\D_\rL^{(4)}\cU_\rL^{(n)}}
~,
\eea
for some left isotwistor multiplet $\cU_\rL^{(n)}$ such that 
$1/ \D_\rL^{(4) } \cU_\rL^{(n)} $ is well defined.  
The superfield $\cC_\rL^{(-4)}$ is 
required to write the action as an integral over the full AdS  superspace.
It can be proven that \eqref{InvarAc} is independent of the explicit choice of
 $\cU_\rL^{(n)}$. For a proof the reader may consult \cite{KLT-M11} or follow 
 the analogous derivation given for the $\cN=3$ case in \cite{KLT-M12}.

The crucial feature of the action \eqref{InvarAc}  is that it is
manifestly invariant under
isometry transformations of the appropriate $\cN=4$ AdS superspace.
On the other hand the action involves 
 the superfield $\cC^{(-4)}$, which is a purely gauge degree of freedom,
and is given by  an integral over all the eight Grassmann variables of the AdS superspaces
even though, due to the analyticity constraint,
 the Lagrangian $\cL_\rL^{(2)}$ depends only on four fermionic variables.
 For this reason, \eqref{InvarAc} is not the most practical version of the projective action principle.
 
By integrating out {\it four}, {\it two} or {\it all} the fermionic directions it is possible to 
write \eqref{InvarAc} in more useful
ways: as (i) an integral in $\cN=2$ AdS, (ii) an integral in $\cN=3$ AdS,
or (iii) as a component expression. The
price is that we lose
manifest invariance under the full $\cN=4$ isometry group.
Let us demonstrate each case in turn.

\subsubsection{$\cN=4$ supersymmetric action in $\cN=2$ AdS}
\label{subsubsection8.4.1}

Here we present the $\cN=4$ supersymmetric action reduced to (2,0) superspace.
We focus on (2,0) only to simplify the presentation, but the reader should keep
in mind that the same results hold for any consistent (1,1) reduction.

Recall that the Lagrangian $\cL_\rL^{(2)}(v_\rL)$ is a real weight-two left projective superfield,
and in the north chart it is associated with the superfield  $\cL_\rL^{[2]}(\z)$ defined by
$\cL_\rL^{(2)}(v) = \ri (v^{1})^2 \z \cL_\rL^{[2]}(\z)$.
We shall prove that for any $\cN=4$ AdS superspace
the supersymmetric action (\ref{InvarAc})
takes the following form in (2,0) AdS superspace:
\bea
S(\cL_\rL^{(2)}) &=& \int \rd^3x\, \rd^2{\q} \rd^2{\qb}\, { E}\, 
\oint_C \frac{\rd \zeta}{2\pi \ri \zeta}\, \cL^{[2]}_\rL|
~,~~~~~~
{E}^{-1}:={\rm Ber}({E}_A{}^M)~.
\label{AdS-N4-2_0_action}
\eea
To do this, we must check explicitly that \eqref{AdS-N4-2_0_action}
is invariant under the full isometry group.
By making use of the transformation rule (\ref{L2n}),
the variation of \eqref{AdS-N4-2_0_action} is
\bea
\d_\x S(\cL_\rL^{(2)}) &=& \int \rd^3x\, \rd^2{\q} \rd^2{\qb}\, { E}\, 
\oint_C \frac{\rd \zeta}{2\pi \ri \zeta}
 \Big{[} 
\t
+\ri t\z\frac{\pa}{\pa \z}
+\z\bar{\ve}^\a \cD_\a
-\frac{1}{\z}\ve_\a \cDB^\a
+\ri \s\z\frac{\pa}{\pa \z}
\non\\
&&
-4S\z\Big(\z\bar{\ve}_\rL
+ \frac{1}{\z}{\ve}_\rL\Big) \frac{\pa}{\pa \z}
-4S\Big(\frac{1}{\z}{\ve}_\rL
-\z\bar{\ve}_\rL\Big)
 \Big{]} \cL^{[2]}_\rL |
~.
\eea
The first term corresponds to the variation
under  an infinitesimal isometry transformation of the (2,0) AdS superspace. Since the action is 
manifestly invariant under the (2,0) AdS isometry group, this
variation vanishes. The second and fifth terms are
total derivatives in $\z$ and vanish under the contour integral.
Integrating by parts the remaining terms, we find
\bea
\d_\x S(\cL_\rL^{(2)}) &=& \int \rd^3x\, \rd^2{\q} \rd^2{\qb}\, {E}\, 
\oint_C \frac{\rd \zeta}{2\pi \ri\z }
 \Big{[} 
\frac{1}{\z}(\cDB_\a\ve^\a  -8S\ve_\rL)
+\z(\cD_\a\bar{\ve}^\a+8S\bar{\ve}_\rL)
 \Big{]} \cL_\rL|
 ~.~~~~~~
\eea
As a consequence of the equations (\ref{extra4020-1}),  \eqref{3120uuu}--\eqref{extra3120-1} and 
\eqref{2220uuu}--\eqref{extra2220-1}, the terms in parentheses vanish
and the action is invariant for all the $\cN=4$ geometries and consistent (2,0) reductions.

\subsubsection{$\cN=4$ supersymmetric action in $\cN=3$ AdS}

Next we consider the reduction of the $\cN=4$ action to $\cN=3$ AdS.
Begin by recalling the form of the $\cN=3$ action principle found by
restricting the locally supersymmetric $\cN=3$ action introduced in
\cite{KLT-M11} to the appropriate $\cN=3$ AdS superspace.
The action is generated by a Lagrangian  $\cL^{(2)} (z,v)$, 
which is a real weight-two $\cN=3$ projective multiplet, and has the form
\bea
S[\cL^{(2)}]&=&
\frac{1}{2\pi\ri} \oint_\g  (v, \rd v)
\int \rd^3 x \,{\rm d}^6\q\,E\, \cC^{(-4)}\cL^{(2)}~, 
\qquad E^{-1}= {\rm Ber}(E_A{}^M)~,
\label{InvarAc-N3}
\eea
where as usual the contour integral  is carried out over a closed contour
$\g =\{v^i(t)\}$ in ${\mathbb C}P^1$.
The isotwistor superfield $\cC^{(-4)} (z,v)$ appearing above is given by
\bea
\cC^{(-4)}:=\frac{\cU^{(n)}}{\D^{(4)}\cU^{(n)}}
~.
\eea
The action looks the same for both (3,0) and (2,1) AdS
except for the precise definition of the $\cN=3$ analytic projector $\D^{(4)}$.
Just as in the $\cN=4$ case, the superfield $\cC^{(-4)}$  is actually a
pure gauge degree of freedom in the sense that \eqref{InvarAc-N3} 
is independent of the explicit choice of $\cU^{(n)}$.

We must now relate the $\cN=4$ action \eqref{InvarAc} to an $\cN=3$
action \eqref{InvarAc-N3} for some choice of $\cL^{(2)}$. It turns
out that simply reducing the Lagrangian in the obvious way
\begin{align}
\cL^{(2)} = \cL_\rL^{(2)}\vert_{\veps^{i \bar j} \q_{i \bar j} = 0}
\end{align}
gives the correct answer. To prove this, we must check that the action
\eqref{InvarAc-N3} is invariant under the full $\cN=4$ AdS isometry
for this choice. Using \eqref{transf-4-3_2} for the case $n=2$, we have
\bea
\d_\x\cL_{\rL}^{(2)}\vert
=
 \delta_\x^{\cN=3} \cL_{\rL}^{(2)}\vert
	+ \D^{(4)}{\bm\pa}^{(-2)} \Omega^{(-2)}\cL_{\rL}^{(2)}\vert
~.
\label{varLagrang-4-3}
\eea
The first term automatically leaves the action \eqref{InvarAc-N3} invariant,
while the second term leads to
\bea
\d_\x S[\cL^{(2)}]&=&
\frac{1}{2\pi\ri} \oint_\g  (v, \rd v)
\int \rd^3 x \,{\rm d}^6\q\,E\, \cC^{(-4)}\D^{(4)}{\bm\pa}^{(-2)} \Omega^{(-2)}\cL_{\rL}^{(2)}~.
\eea
Integrating the analytic projector by parts, we find
\bea
\d_\x S[\cL^{(2)}]&=&
\frac{1}{2\pi\ri} \oint_\g  (v, \rd v){\bm\pa}^{(-2)}
\int \rd^3 x \,{\rm d}^6\q\,E\,  \Omega^{(-2)}\cL_{\rL}^{(2)}
~,
\eea
which vanishes identically upon integrating around the contour.

\subsubsection{$\cN=4$ supersymmetric action in components}

Finally, we present the evaluation of \eqref{InvarAc} upon integration of all
the Grassmann coordinates. We will apply the technique
first used in \cite{KT-M} to derive the $\cN=1$ supersymmetric action in  $\rm AdS_5$
and similarly used in \cite{KLT-M12} for the $\cN=3$
supersymmetric action in  $\rm AdS_3$.

We start with the  $\cN=4$ left projective superspace action 
in three-dimensional Minkowski space, 
which was introduced in
\cite{KPT-MvU}.
It has the form
\bea
S[L_\rL^{(2)}]
=\frac{1}{2\p} \oint_{\g}  { v_i {\rm d} v^i }
\int {\rm d}^3x \, D_\rL^{(-4)}L_\rL^{(2)} (z,v_\rL) \Big|_{\q =0}
~,
\label{flat-Ac}
\eea
where the Lagrangian $L_\rL^{(2)}(z,v_\rL)$ is a real left weight-two projective multiplet, 
and  the operator $D_\rL^{(-4)}$ is defined in terms of the {\it flat} spinor 
covariant derivatives $D_\a^{i\bai} $ as
\bea
D_\rL^{(-4)}&:=&\frac{1}{48}D^{(-2)\bak\bal}D^{(-2)}_{\bak\bal}~,
~~~
D^{(-2)}_{\bak\bal}:=D^{(-1)\g}_\bak D_{\g\bal}^{(-1)}
~,~~~
D_\a^{(-1)\bai}:=\frac{1}{(v_\rL,u_\rL)}u_iD_\a^{i\bai}~.
~~~~~~
\label{N4D-defs}
\eea
The $D_\rL^{(-4)}$ operator depends on the isotwistor
$v^i (t) $, which varies along the integration contour,  
but  also on a constant ($t$-independent) isotwistor $u_i$ which is 
chosen such that 
$v_i(t)$ and $u_i$  
are everywhere linearly independent along the contour $\g$,
that is $\big(v_\rL(t),u_\rL\big) \neq 0$. 
The flat action \eqref{flat-Ac} is actually independent of $u_i$, since it 
proves to be invariant under arbitrary
projective transformations of the form
\be
\big(u_i{}\,,\,v_i (t){}\big)~\to~\big(u_i{}\,,\, v_i (t){} \big)\,R_\rL(t)~,~~~~~~
R_\rL(t)\,=\,
\left(\begin{array}{cc}a_\rL(t)~&0\\ b_\rL(t)~&c_\rL(t)\end{array}\right)\,\in\, 
{\rm GL}(2,\mathbb{C})~,
\label{proj-inv}
\ee
where $a_\rL(t)$ and $b_\rL(t)$ obey the first-order differential equations
\bea
\dt{a}_\rL=b_\rL\,{(\dt{v}_\rL, v_\rL)\over (v_\rL,u_\rL)}~, \qquad 
\dt{b}_\rL=-b_\rL\,{(\dt{v}_\rL, u_\rL)\over (v_\rL,u_\rL)}~,
\label{ode}
\eea
with $\dt{f}:= \rd f(t)/ \rd t$ for any function $f(t)$. This invariance follows from the 
fact that the Lagrangian $L_\rL^{(2)} (v_\rL)$ has the following properties:
(i) it is a homogeneous function of $v^i$ of degree two; and (ii) it
satisfies the analyticity condition 
$D^{(1)\bai}_\a L_\rL^{(2)} (v_\rL)=0$.
Due to the property (ii) it turns out that  the action  \eqref{flat-Ac} is 
invariant under the standard $\cN=4$ super-Poincar\'e transformations in three dimensions 
\cite{KPT-MvU}.

Our goal is  to generalize the above construction to the $\cN=4$ AdS case. 
Let $z^M = (x^m, \q^\m_{\imath \bar{\jmath}})$
be local coordinates of the AdS superspace.
Given a  tensor superfield $U(x,\q)$,  
we define its restriction to the body of the superspace, $ \q^\m_{\imath \bar{\jmath}}=0$, 
according to the rule
\be
U||:= U(x,\q )|_{\q_{\imath \bar \jmath}=0}~.
\label{projection-0}
\ee
We also define the double-bar projection of the covariant derivatives 
\be
\cD_{{A}} ||:= 
E_{A}{}^M||\pa_M
+\hf\O_A{}^{bc}||\cM_{bc}
+\F_A{}^{kl}||\bL_{kl}
+\F_A{}^{\bak\bal}||\bR_{\bak\bal}
~.
\ee
Recall that for all the AdS geometries,
${[}\cD_{a},\cD_b{]} =-4\,S^2\,\cM_{ab}$.
This allows us to use general coordinate and local structure group transformations 
to choose a (Wess-Zumino) gauge in which
\be
\cD_{{a}} || = \nabla_{ a} = e_{a}{}^{m} (x) \, \pa_{ m} 
+ \hf \o_{a}{}^{ b c} (x) \,\cM_{ b c}~,
\label{WZ}
\ee
where $\nabla_{a}$ stands for  the covariant derivative
of anti-de Sitter space $\rm AdS_3$, 
\bea
{[} \nabla_{a},\nabla_{b} {]}&=&-4S^2 \cM_{ab}~.
\eea

We would like to construct
an AdS  generalization of the action (\ref{flat-Ac}) 
that describes \eqref{InvarAc} in a form where all the Grassmann
variables have been integrated out. 
We expect
\bea
S [ \cL_\rL^{(2)}] =S_0+\cdots~,~~~~~~
S_0=\frac{1}{2\p} \oint_{\g}  { v_i {\rm d} v^i }\int {\rm d}^3x\, e \,
\cD_\rL^{(-4)}
\cL_\rL^{(2)}  ||
~,
\label{comp-Ac-000}
\eea
where $S_0$ is the covariantization of the Minkowski result \eqref{flat-Ac},
with $e:={\rm det}{(e_m{}^a)}$.
The ellipsis stands for curvature-dependent corrections that are necessary for the
action to be  invariant under the symmetries of its parent action (\ref{InvarAc}).
Note that both the flat action (\ref{flat-Ac})
and the parent curved full superspace action (\ref{InvarAc})
are manifestly projective invariant (\ref{proj-inv}).
On the other hand, $S_0$  is not.
Projective invariance
can be used as a tool to iteratively
complete $S_0$ to $S[\cL_\rL^{(2)}]$.
We describe this procedure in Appendix \ref{AppB}, which the interested
reader may consult.
The final form is
\bea
S[\cL_\rL^{(2)}]&=&
\oint_{\g}  \frac{({\rm d} v_\rL,v_\rL)}{2\p} 
\int {\rm d}^3x \,e\, 
\Big{[}
\cD_\rL^{(-4)}
-\frac{1}{3}\cS^{(-2)}{}^{\bak \bal }\cD^{(-2)}_{\bak \bal }
-2\cS^{(-2)}{}^{\bak \bal }\cS^{(-2)}_{\bak \bal }
\Big{]}
\cL_\rL^{(2)}||
~.~~~~~~~~
\label{components-Ac}
\eea
Here we have used
\bea
\cS^{(-2)\bai\baj}:=\frac{u_iu_j }{(v_\rL,u_\rL)^2}\cS^{ij\bai\baj}~,
\eea
and presented the action in a form valid for all $\cN=4$ AdS superspaces. 
To pick up a specific AdS background, one should 
choose the appropriate value of the curvature $\cS^{ij\bai\baj}$ according to 
\eqref{SSX40}--\eqref{SSX22}.

One can obtain the action principle for right projective 
Lagrangians by applying the mirror map of \cite{KLT-M11} to \eqref{components-Ac}.


\section{Relating $\cN=3$ and $\cN=4$ supersymmetric sigma models}
\label{section9}
In three-dimensional Minkowski space,
 $\cN=3$ supersymmetry for $\s$-models is known to imply $\cN=4$ supersymmetry \cite{A-GF}.
The standard argument (see, e.g., \cite{Bagger}) goes
as follows: $\cN$-extended supersymmetry requires $\cN-1$ anti-commuting complex structures.
In the case $\cN=3$, the target space has two such complex structures, $I$ and $J$. 
Their product $K:=I \,J$ is a third complex structure which anticommutes with $I$ and $J$, 
and therefore the $\s$-model is $\cN=4$ supersymmetric. 
This argument tells us nothing about off-shell supersymmetry.
There have been developed alternative proofs \cite{KPT-MvU} which 
are intrinsically off-shell and make use 
of the formulations for general $\cN=3$ supersymmetric $\s$-models in terms of
(i) $\cN=2$ chiral superfields; and (ii) $\cN=3$ polar supermultiplets.  

The goal of this section is to understand whether a $(p,q)$ supersymmetric 
$\sigma$-model in AdS${}_3$ with $\cN= p+q =3$ actually possesses 
an enhanced $(p', q')$ supersymmetry with $\cN' = p' +q'=4$.  
We will see that this is always the case, and the necessary (but not sufficient) conditions
are $p' \geq p$ and $q' \geq q$.
To prove that, in this section we will use both $\cN=2$ and $\cN=3$ superfields.

\subsection{(2,0) AdS superspace approach: Formulation in terms of $\cN=3$ polar supermultiplets} 
The starting point of our analysis is the transformation law of the $\cN=4$ left arctic supermultiplet
of  weight-$n$ given in eq. \eqref{eq_dUpL}. We split it into two different transformations:
\begin{subequations}
\bea
\d_\t \U_\rL^{[n]}
&=& \Big\{ 
\t
+\ri t\Big(\z\frac{\pa}{\pa \z}-\frac{n}{2}\Big)\Big\}\U_\rL^{[n]}~;  \label{9.1a}\\
\d_\ve \U_\rL^{[n]}
&=& \Big\{ 
\z\bar{\ve}^\a \cD_\a
-\frac{1}{\z}\ve_\a \cDB^\a
-4S\Big( \bar{\ve}_\rL\z 
+\frac{1}{\z}{\ve}_\rL\Big)\z\frac{\pa}{\pa \z}
+4Sn\,\z\bar{\ve}_\rL
 \Big\} \U_\rL^{[n]} \non \\
 &&\qquad +\ri\s\Big(\z\frac{\pa}{\pa \z}-\frac{n}{2}\Big)\U_\rL^{[n]} ~.
 \label{9.1b}
\eea
\end{subequations}
We recall that the parameters $\t = \t^a \cD_a + \t^\a \cD_\a + \bar \t_\a \bar \cD^\a$ and $t$
describe the  isometry transformations of the (2,0) AdS superspace.
The other parameters
$(\veps^\alpha, \,\bar\veps_\alpha, \,\sigma, \,\veps_\rL, \,\bar\veps_\rL) $ 
as well as $(\veps_\rR, \,\bar\veps_\rR)$, which appear in the transformation laws of right projective 
supermultiplets, are associated with the remaining two supersymmetries
and the residual $R$-symmetry. 

\subsubsection{(3,0) AdS supersymmetry implies (4,0) AdS supersymmetry}  

In the case of (4,0) AdS supersymmetry, the equations obeyed by the parameters 
$(\veps^\alpha, \,\bar\veps_\alpha, \,\sigma, \,\veps_\rL, \,\bar\veps_\rL, \,\veps_\rR, \,\bar\veps_\rR)$
 follow from 
\eqref{extra4020} by setting $X=0$:  \bsubeq \label{9.2}
\bea
\cD_\a\bar{\ve}_\b &=&\phantom{+}
4S\ve_{\a\b}\,\bar{\ve}_\rL
~,\qquad
\cDB_\a\ve_\b=
-4S\ve_{\a\b}\,{\ve}_\rL
~,
\\
\cD_\a\ve_\b
&=&-4S\ve_{\a\b}\,\bar{\ve}_\rR
~,\qquad
\cDB_\a\bar{\ve}_\b
=\phantom{-}4S\ve_{\a\b}\,{\ve}_\rR
~,~~~~~~
\\
\cDB_\a\ve_\rL &=&\cDB_\a{\ve}_\rR=0
~,~~~~~~~~
\cD_\a{\ve}_\rL=
\ri \,\ve_{\a}
~,\qquad
\cD_\a{\ve}_\rR=-\ri\,\bar{\ve}_{\a}
~.~~~~~~
\eea
\esubeq
In accordance with eq. \eqref{7.28}, the action of the U$(1)_R$ generator 
 \eqref{mod-cJ}  on these parameters is as follows:
\bea
{ \cJ}\ve_\a=0~,\qquad
{ \cJ}{\ve}_\rL=-{\ve}_\rL~,\qquad
{ \cJ}{\ve}_\rR=-{\ve}_\rR~.
\eea
Finally, the relation \eqref{7.30} tells us that the parameter $\s$  is constant. 

In general, the supersymmetry parameters $\ve_\a$ and $\bar \ve_\a$ are linearly independent. 
Let us now consider a special choice 
\bea
\bar \ve_\a = -\ve_\a = \ri \r_\a~, \qquad \bar \r_\a = \r_\a~.
\label{9.4}
\eea
For this specific case we also denote 
\bea
\r\equiv {\ve}_\rL~, \qquad \bar \r \equiv \bar {\ve}_\rL~.
\label{9.5}
\eea
Then it follows from the relations \eqref{9.2} that 
the (3,0) AdS identities \eqref{eq_rho30constraints} hold. 
If we also choose
\bea
\s=0~,
\eea
then the (4,0) AdS supersymmetry transformation law \eqref{9.1b} reduces to the (3,0) one, eq. 
\eqref{(3,0)SUSYtran}, if the hypermultiplet weight is $n=1$. 
Moreover, for $n=1$ it is easy to see that a finite U(1) transformation
generated by the constant parameter $\s$ in  \eqref{9.1b} coincides with the U(1) symmetry 
\eqref{shadow4} of the off-shell (3,0) supersymmetric $\s$-model 
\eqref{(3,0)sssm-general}.
Now, it is an instructive exercise to check explicitly that the (3,0) supersymmetric $\s$-model 
\eqref{(3,0)sssm-general} is invariant under the (4,0) supersymmetry transformation \eqref{9.1b}
with $n=1$. 
The solution to this problem is actually given in section \ref{subsubsection8.4.1}.
We conclude that any off-shell (3,0) supersymmetric $\s$-model in AdS${}_3$ 
possesses (4,0) off-shell supersymmetry. 
This theory is actually $\cN=4$ superconformal. 

\subsubsection{(3,0) AdS supersymmetry implies (3,1) AdS supersymmetry}  

In the case of (3,1) AdS supersymmetry,  the equations obeyed by the parameters 
$(\veps^\alpha, \,\bar\veps_\alpha, \, \veps_\rL, \,\bar\veps_\rL, \,\veps_\rR, \,\bar\veps_\rR)$
are given in section \ref{subsubsection7.2.2}. We recall that the parameter $\s$ 
is absent in this case.

We can consider a special (3,1) supersymmetry transformations defined by the conditions
\eqref{9.4}. Using the identification \eqref{9.5}, we then deduce from the relations \eqref{extra3120}
that the (3,0) AdS identities \eqref{eq_rho30constraints} hold. 
As a result, the (3,1) AdS supersymmetry transformation law \eqref{9.1b} reduces to the (3,0) one, eq. 
\eqref{(3,0)SUSYtran}, if the hypermultiplet weight is $n=1$. 
The (3,0) supersymmetric $\s$-model 
\eqref{(3,0)sssm-general} is invariant under the (3,1) supersymmetry transformation \eqref{9.1b}
with $n=1$.  Any off-shell (3,0) supersymmetric $\s$-model in AdS${}_3$ 
possesses (3,1) off-shell supersymmetry. We point out again that this theory is 
in fact  $\cN=4$ superconformal. 

\subsubsection{(2,1) AdS supersymmetry implies (2,2) AdS supersymmetry}  

In the case of (2,2) AdS supersymmetry, the parameters $\ve_\rL$ and $\ve_\rR$ vanish, 
while the complex supersymmetry parameter $\ve_\a$ is almost constant, 
that is, it obeys the constraints \eqref{extra2220-1}. Finally, the $R$-symmetry parameter $\s$ 
is constant.

We consider a special  (2,2) supersymmetry transformation defined by the conditions
\eqref{9.4}.  In accordance with  \eqref{extra2220-1}, the real spinor parameter $\r_\a$ 
obeys the (2,1) constraints \eqref{2_1-extra}. Let us also choose $\s=0$.  
Then the (2,2) AdS supersymmetry transformation law \eqref{9.1b} reduces to the (2,1) one, 
eq. \eqref{5.7},  in the case of weight-zero hypermultiplets, $n=0$. 
For $n=0$, the finite U(1) transformation of $\U(\z)$ 
generated by the parameter $\s$ in  \eqref{9.1b} coincides with the U(1) symmetry 
 \eqref{5.6} of the most general off-shell  (2,1) supersymmetric $\s$-model 
 \eqref{(2,1--2,0)ProjAction}. 
 It is a simple exercise to show that this $\s$-model action is invariant under the 
 (2,2) AdS supersymmetry transformation  \eqref{9.1b}. 
 We conclude that any off-shell (2,1) supersymmetric $\s$-model in AdS${}_3$ 
possesses (2,2) off-shell supersymmetry. 

We recall that the  most general off-shell $\s$-model with (2,1) AdS supersymmetry is 
described by the action \eqref{(2,1--2,0)ProjAction}, with $K(\F, \bar \F)$ being a general real 
analytic function. An important subclass of these theories corresponds to the case that $K(\F, \bar \F)$  
obeys the homogeneity conditions
\bea
 \F^I \frac{\pa}{\pa \F^I} 
K(\F, \bar \F) =  K( \F,   \bar \F) ~, \qquad
 \bar \F^{\bar I} \frac{\pa}{\pa \bar \F^{\bar I}} 
K(\F, \bar \F) =  K( \F,   \bar \F) ~.
\eea
The resulting $\s$-model is $\cN=3$ superconformal.
In fact, it possesses $\cN=4$ superconformal invariance. 
The latter observation means that such a $\s$-model can support each of the three $\cN=4$
AdS supersymmetries: (4,0), (3,1) and (2,2).

\subsection{(2,0) AdS superspace approach: Formulation in terms of $\cN=2$  chiral superfields}

Next we turn to the most general  $\sigma$-models in (2,0) AdS superspace,
eq. \eqref{2.8+W}, 
possessing two additional supersymmetries.  We have already described 
in  section \ref{N4AdSsuperspaces} how these two supersymmetries and the residual 
$R$-symmetry transformations are described in terms of
the parameters $(\veps^\alpha, \,\bar\veps_\alpha, \,\sigma, \,\veps_\rL, \,\bar\veps_\rL)$. 
Similar to the discussion in the previous subsection, 
 the parameter $\sigma$ can be neglected in our discussion.
It is useful to decompose the complex spinor parameter $\veps_\alpha$ as
\begin{align}
\veps_\alpha = -\ri \rho_\alpha + \rho_{\alpha}'~, \qquad
\bar\veps_\alpha = \ri \rho_\alpha + \rho_{\alpha}'~,
\end{align}
where $\rho_\alpha$ and $\rho'_{\alpha}$ are both real spinors.
Due to the constraints on $\rho_\alpha$ and $\rho'_\alpha$,
we may introduce complex parameters $\rho$ and $\rho'$ which obey
\begin{align}
\rho_\alpha = \cD_\alpha \rho = \bar \cD_\alpha \bar \rho~, \qquad
\rho'_\alpha = \cD_\alpha \rho' = \bar \cD_\alpha \bar \rho'~.
\end{align}
When $\rho'_\alpha$ vanishes, an $\cN=4$ Killing transformation
reduces to an $\cN=3$ transformation with spinor parameter $\rho_\alpha$.

We have already shown that any (2,0) $\sigma$-model in AdS${}_3$
with the $\cN=3$ supersymmetry
transformation
\begin{align}\label{eq_3susy}
\delta \phi^\ra = \frac{\ri}{2}  \bar\cD^2 (\bar\rho \Omega^\ra)
\end{align}
imposes the requirement that the target space be hyperk\"ahler.
In the (3,0) case, the target space must be a hyperk\"ahler cone.
In the (2,1) case, the target space must possess a U(1) isometry
which acts as a rotation on the complex structures.
In both cases, we may choose $\Omega^\ra = \omega^{\ra \rb} K_\rb$.

It turns out that, just as in the off-shell models,
it is always possible to impose $\cN=4$ supersymmetry without
any further restrictions on the target space geometry. In particular, we may lift
(3,0) to either (4,0) or (3,1), while (2,1) may be lifted only to (2,2). In each case,
the full extended supersymmetry transformation is simply
\begin{align}\label{eq_4susy}
\delta \phi^\ra = \frac{\ri}{2}  \bar\cD^2 \Big((\bar\rho \pm \ri \bar\rho') \Omega^\ra\Big)~, \qquad
\Omega^\ra = \omega^{\ra \rb} K_\rb~,
\end{align}
where a plus sign is associated with a left $\cN=4$ hypermultiplet and a minus
sign with a right $\cN=4$ hypermultiplet. Let us now prove this for each case.

\subsubsection{(3,0) AdS supersymmetry implies (4,0) AdS supersymmetry}  

We begin with the lift of (3,0) AdS supersymmetry to (4,0).
The parameters $\rho$ and $\rho'$ for a (4,0) isometry transformation can be taken to obey
\begin{align}
\rho = \frac{1}{2} (\veps_\rL + \veps_\rR)~, \qquad \rho' = -\frac{\ri}{2} (\veps_\rL - \veps_\rR)~.
\end{align}
One can then check that $\rho$ obeys the conditions \eqref{eq_rho30constraints} consistent with
an extended (3,0) parameter. Remarkably, $\rho'$ obeys the same conditions.
To prove invariance of the action under \eqref{eq_4susy}, we need only recycle the proof
already given for the (3,0) case. The only change is the phase of the extended
supersymmetry transformation, which corresponds to a phase rotation of the complex
structure. We immediately conclude that the action must remain invariant
under the full (4,0) AdS supersymmetry.

\subsubsection{(3,0) AdS supersymmetry implies (3,1) AdS supersymmetry}  
Next, we address the (3,1) lift of (3,0) AdS supersymmetry.
The (3,1) Killing parameters obey $\veps_\rL = \veps_\rR = \veps$, and we may choose
\begin{align}
\rho = \frac{1}{2} \veps~.
\end{align}
Once again, $\rho$ obeys the (3,0) conditions \eqref{eq_rho30constraints}.
However, $\rho'_\alpha$ is nearly constant, so $\rho'$ only needs to obey the conditions
\begin{align}\label{eq_rhoprime31}
\cD^2\rho' = 0~, \qquad \bar\cD_\alpha \cD_\beta \rho' = 0~.
\end{align}
Since $\rho'$ is no longer quite the same sort of parameter as $\rho$,
it is a minor task to check that the action is indeed invariant
under \eqref{eq_4susy}, which we leave as an exercise to the reader.
We conclude any on-shell $\sigma$-model with (3,0) supersymmetry
can be lifted to (3,1).

\subsubsection{(2,1) AdS supersymmetry implies (2,2) AdS supersymmetry}  
Our last case to consider is the lift of (2,1) to (2,2). For a (2,2)
isometry transformation, the parameters $\veps_\rL$ and $\veps_\rR$ both vanish with
$\rho$ and $\rho'$ obeying the identical conditions
\eqref{eq_rho2120constraints2} and \eqref{eq_rhoprime31}. To prove invariance
of the action, we again need only recycle the proof given for the (2,1)
case since the only modification is a change in phase of the complex structure.
We conclude that an on-shell $\sigma$-model with (2,1) AdS supersymmetry
may always be lifted to (2,2).

\subsection{$\cN=3$ AdS superspace approach}

Let us now consider a theory with manifest off-shell $\cN=3$ AdS supersymmetry
written in projective superspace. Depending on whether the action possesses
(3,0) or (2,1) supersymmetry, the natural weight for the $\cN=3$ multiplets and the
conditions on the Lagrangian will be different. For the moment, we will remain
with a general case: a weight-two projective Lagrangian $\cL^{(2)}$ depending on
weight-$n$ projective multiplets $Q^{(n)}$, their smile-conjugates, and
possibly the isotwistor $v^i$ as well.

We have already seen in section \ref{projective-4-3}
that an $\cN=4$ theory constructed out of a Lagrangian $\cL_\rL^{(2)}$
built from left projective multiplets $Q_L^{(n)}$ can be reduced to
$\cN=3$ via a projection operation
\begin{align}
Q^{(n)} = Q_L^{(n)}\vert_{\veps^{\imath \bar{\jmath }} \q_{\imath \bar{\jmath }} = 0}~, \qquad
\cL^{(2)} = \cL^{(2)}\vert_{\veps^{\imath \bar{\jmath }} \q_{\imath \bar{\jmath }} = 0}~.
\end{align}
The $\cN=4$ isometry transformation on $Q^{(n)}$ is decomposed into the sum
of an $\cN=3$ Killing transformation and an extended supersymmetry transformation,
the latter being
\begin{align}\label{eq_deltaQext}
\delta_\Omega Q^{(n)}
	&= \D^{(4)}\Big(\Omega^{(-2)} {\bm\pa}^{(-2)} Q^{(n)}
	+ \frac{n}{2} ({\bm\pa}^{(-2)} \Omega^{(-2)}) Q^{(n)}\Big)~.
\end{align}
The $v^i$-dependent parameter $\Omega^{(-2)}$ was required to obey a number
of conditions encoded in \eqref{useful-X-O} and \eqref{N3uuuseful-2}--\eqref{N3uuuseful-3}.

It is a simple exercise to reverse this procedure. Given any $\cN=3$ theory,
we may simply \emph{impose} \eqref{eq_deltaQext} as the extended
supersymmetry transformation for the multiplet $Q^{(n)}$. This constitutes
its $\cN=4$ lift and completely determines the left projective multiplet
$Q_L^{(n)}$ into which $Q^{(n)}$ may be encoded.\footnote{One can also perform
a lift to a \emph{right} projective multiplet. The only difference is
the overall sign of the transformation law \eqref{eq_deltaQext}.} Naturally, we expect only
specific lifts to be possible: (3,0) certainly cannot be lifted to (2,2), nor
can (2,1) be lifted to (4,0), but most other possibilities are allowed.
We discuss (and demonstrate) these possibilities below.

\subsubsection{(3,0) AdS supersymmetry implies (4,0) and (3,1) AdS supersymmetry}  
A $\sigma$-model with off-shell (3,0) supersymmetry is described in projective
superspace by a weight-two Lagrangian $\cL^{(2)}$ that is a homogeneous
function of weight-one polar multiplets, $\U^{(1)I}$ and $\breve{\U}^{(1)\bar{J}}$:
 \bea
 \Big(\U^{(1)I} \frac{\pa}{\pa\U^{(1)I}}
+\breve{\U}^{(1)\bar{J}}
\frac{\pa}{\pa\breve{\U}^{(1)\bar{J}} }\Big) \cL^{(2)} =2\cL^{(2)}~.
\eea
We impose
\begin{align}
\delta_\Omega \U^{(1)I} &=
	\D^{(4)}\Big(\Omega^{(-2)} {\bm\pa}^{(-2)}  
	+ \frac{1}{2} ({\bm\pa}^{(-2)} \Omega^{(-2)})\Big)\U^{(1)I}~.
\end{align}
The parameter $\Omega^{(-2)}$ may correspond to either a (4,0) or a (3,1)
extended supersymmetry. Depending on the situation, the parameters
$\Xi_\alpha$ and $\tilde \L^{ij}$ defined in \eqref{useful-X-O}
obey different conditions \eqref{N3uuuseful-2}--\eqref{N3uuuseful-3}.
Regardless of the case in question, it is easy to prove that the
off-shell $\cN=3$ action possesses an extended supersymmetry.
Due to the homogeneity condition, the Lagrangian $\cL^{(2)}$ must obey
\begin{align}
\delta_\Omega \cL^{(2)} &=
	\D^{(4)}\Big(\Omega^{(-2)} {\bm\pa}^{(-2)}  
	+ ({\bm\pa}^{(-2)} \Omega^{(-2)})\Big)\cL^{(2)}
	= \D^{(4)} {\bm\pa}^{(-2)} \Big(\Omega^{(-2)} \cL^{(2)}\Big)
\end{align}
and one can prove invariance of the action along the same lines as
in section \ref{projective-4-3}. We therefore conclude that any
off-shell (3,0) Lagrangian is actually invariant under both off-shell
(4,0) and (3,1) supersymmetries.

\subsubsection{(2,1) AdS supersymmetry implies (2,2) AdS supersymmetry} 
A general off-shell $\sigma$-model with (2,1) supersymmetry is given by
a Lagrangian
\begin{align}
\cL^{(2)} = w^{(2)} \cL(\U^I, \breve \U^{\bar J})
\end{align}
where $\U^I$ and $\breve \U^{\bar J}$ are weight-zero polar multiplets.
No homogeneity condition is imposed on $\cL$, and so the action need not be
superconformal.
To perform a (2,2) lift, we require the extended supersymmetry transformation
\begin{align}
\delta_\Omega \U^{I} &=
	\D^{(4)}\Big(\Omega^{(-2)} {\bm\pa}^{(-2)} \U^I\Big)~,
\end{align}
which leads to
\begin{align}
\delta_\Omega \cL^{(2)} = w^{(2)} \delta_\Omega \cL =
	\D^{(4)}{\bm\pa}^{(-2)} \Big(\Omega^{(-2)} \cL^{(2)}\Big)
	- v_i v_j \tilde\L^{(i}{}_k w^{j) k} \cL~.
\end{align}
The first term involving the analytic projector vanishes under the $\cN=3$ measure
as before. The second term, which involves $w^{ij}$ explicitly, vanishes for the (2,2)
case alone where $\tilde \L^{ij} \propto w^{ij}$.


\section{(4,0) supersymmetric sigma models with $X\neq 0$: Off-shell approach}

The results of section \ref{section9} imply that there is only one class of $\cN=4$ supersymmetric 
$\s$-models in AdS${}_3$ which requires a separate study. It consists of all $\s$-models 
possessing (4,0) AdS supersymmetry with $X\neq 0$.  These theories cannot be realized as 
$\cN=3$ supersymmetric $\s$-models in AdS${}_3$, and therefore 
the analysis of sections 3 -- 6 is not applicable.  
In this and subsequent sections we provide a detailed study
of the $\s$-models possessing (4,0) AdS supersymmetry with $X\neq 0$. 

Let us begin by recalling the algebra of covariant derivatives 
for the (4,0) AdS superspace with $X\neq 0$:
\begin{subequations}
\begin{align}
\{\cD_\a^{i\bai},\cD_\b^{j\baj }\}&=
	2\ri\,\ve^{ij}\ve^{\bai \baj }\cD_{\a\b}
	-\,4\ri \ve^{ij}\ve^{\bai \baj }S \cM_{\a\b}
	\non\\&\quad
	+\,{2\ri}\ve_{\a\b}\ve^{\bai \baj }(2S+X)\bL^{ij}
	+\,2\ri\ve_{\a\b}\ve^{ij}(2S-X)\bR^{\bai\baj}~,\\
{[}\cD_{\a\b},\cD_\g^{k\bak}{]}&=
-\,2 S\ve_{\g(\a}\cD_{\b)}^{k\bak}~,\\
{[}\cD_a,\cD_b{]}&=-\,4\,S^2\cM_{ab}~.
\end{align}
\end{subequations}
For $|X|\neq 2S$, the $R$-symmetry part of the superspace holonomy group 
(or simply the $R$-holonomy group in what follows) is 
 $\rm SU(2)_L \times SU(2)_R$. However, when
$X=\pm 2S$, which we will refer to as the critical cases,
either the $\rm SU(2)_L$ or the $\rm SU(2)_R$ curvature vanishes and
one may adopt a gauge where the corresponding connection is zero.

This algebra possesses a (2,0) AdS subalgebra corresponding to the choice
$\cD_\alpha = \cD_\alpha^{1\bar 1}$ and $\bar\cD_\alpha = -\cD_\alpha^{2\bar 2}$:
\begin{subequations}
\begin{gather}
\{\cD_\a,\cD_\b\} = \{\bar\cD_\a,\bar\cD_\b\} = 0
~,\qquad
\{\cD_\a,\bar \cD_\b\}=
-2\ri\,\cD_{\a\b}
+\,4\ri S \cM_{\a\b}
-\,{4\ri} S\ve_{\a\b}  { \cJ}
~,
\\
{[}\cD_{\a\b},\cD_\g{]}= -\,2 S \ve_{\g(\a}\cD_{\b)}~,\qquad
{[}\cD_a,\cD_b{]}=-\,4\,S^2\cM_{ab}~,
\end{gather}
\end{subequations}
where the U$(1)_R$ generator $ \cJ$ is given by \eqref{mod-cJ},
which we repeat here in a slightly modified form:
\begin{align}
\cJ = \left(1+\frac{X}{2S}\right) \bL^{12} + \left(1-\frac{X}{2S}\right) \bR^{\bar 1 \bar 2}~.
\end{align}
For later convenience, we will introduce operators
\begin{align}
\cJ_\rL := {\bf L}^{12}~, \qquad \cJ_\rR:={\bf R}^{\bar 1 \bar 2}~.
\end{align}
In the critical cases of $X=\pm 2S$, we find the particularly
simple results $ \cJ = 2 \cJ_\rL$ or $ \cJ = 2 \cJ_\rR$.

Because of the potential truncation of the $R$-holonomy group, it is clear that
there are essentially two cases of interest. The first is the general
non-critical case where $|X|\neq 2S$ and the full $R$-holonomy  group is
$\rm SU(2)_L \times SU(2)_R$, and the second is when
$X=2S$ and only $\rm SU(2)_L$ remains. 
(The case $X=-2S$ is found by applying the mirror map.)

\subsection{Off-shell non-critical (4,0) sigma models }

Let us first discuss the non-critical case, where $|X|\neq 2S$.
The general off-shell $(4,0)$ supersymmetric $\s$-model can be written in (2,0) AdS 
superspace as
\begin{align}\label{eq_40action}
S = \int \rd^3x\, \rd^4\q\, E\, (\mathbb L_\rL + \mathbb L_\rR)~.
\end{align}
The term $\mathbb L_\rL$ is a Lagrangian constructed via a contour integral
\begin{align}
\mathbb L_\rL = \oint_C \frac{\rd\zeta}{2\pi \ri \z} \cL^{[2]}_\rL~, 
\end{align}
from the left Lagrangian 
$\cL^{[2]}_\rL (\U_\rL, \breve{\U}_\rL;\z)$
which depends on
left arctic {\it weight-one} superfields $\U_\rL^{I} (\z) $ and their smile-conjugates
$\breve \U_\rL^{\bar I}(\z)$. In some cases, $ \cL^{[2]}_\rL$ may also explicitly depend 
on the complex variable $\z$.
Similar statements pertain to the right Lagrangian $\mathbb L_\rR$.
The left and right sectors are completely decoupled (without higher-derivative couplings); 
we may even choose one of the sectors to vanish. 
For this reason, we can focus our analysis for the moment on the left sector.

We have to add some projective-superspace details regarding the left $\s$-model sector.  
The dynamical variables are weight-one arctic multiplets $\U_\rL^{(1) I} (v_\rL) $
and their smile-conjugate antarctic multiplets $\breve \U_\rL^{(1) \bar I}(v_\rL)$.
The $\s$-model Lagrangian $\cL^{(2)}_\rL$ defining the (4,0) supersymmetric 
action  \eqref{InvarAc} is 
\bea
\cL^{(2)}_\rL = \ri {\mathfrak K}_\rL (\U^{(1)}_\rL , \breve \U^{(1)}_\rL) ~,
\eea
where the function ${\mathfrak K}_\rL$ obeys the homogeneity condition 
\begin{align}\label{eq_40homogeneity}
 \Big(\U_\rL^{(1)I} \frac{\pa }{\pa \U_\rL^{(1)I}}
	+ \breve\U_\rL^{(1) \bar I} \frac{\pa }{\pa \breve\U_\rL^{(1) \bar I}} \Big){\frak K}_\rL  
	=2 {\frak K}_\rL 
\end{align}
which guarantees that  $\cL^{(2)}_\rL$ has weight two. The  Lagrangian $\cL^{(2)}_\rL$
has to be real under the smile-conjugation, which restricts ${\frak K}_\rL$ to obey 
a reality condition of the type \eqref{reality_condition_strange}.
To reformulate the (4,0) supersymmetric action  \eqref{InvarAc}  in (2,0) AdS superspace, 
as in equation (\ref{AdS-N4-2_0_action}),
we recall that all projective supermultiplets should be recast  as functions of $\z$, using a prescription
$Q^{(n)}_\rL (v_\rL) \to Q^{[n]}_\rL (\z) \propto Q^{(n)}_\rL (v_\rL)$. 
For the Lagrangian $\cL^{(2)}_\rL (v_\rL)$, 
we have to use the rule \eqref{8.11}. For the dynamical superfields  $\U_\rL^{(1) } (v_\rL) $ and
 $\breve \U_\rL^{(1) }(v_\rL)$, we have to use the rules \eqref{arctic1} 
 and \eqref{antarctic1} respectively. 
 Since we are going to work only with weight-one hypermultiplets, 
 we will denote  $\U_\rL^{ [1] } (\z) $ and $\breve \U_\rL^{ [1] }(\z)$ 
 simply as $\U_\rL(\z) $ and  $\breve \U_\rL(\z)$ respectively. 
 As a result, for the Lagrangian $\cL^{[2]}_\rL (\z)$
 we end up with the following expression: 
 \bea
 \cL^{[2]}_\rL (\U_\rL, \breve{\U}_\rL;\z) = \frac{1}{\z} 
 {\mathfrak K}_\rL (\U_\rL , \z\breve \U_\rL) ~.
 \eea

The arctic  $\U_\rL^I (\z) $ and antarctic $\breve\U_\rL^{\bar I} (\z)$ multiplets
are given by Taylor and Laurent  series in $\z$ respectively, with the coefficients being 
$\cN=2$ superfields,
\begin{align}
\U_\rL^I(\z) = \sum_{n=0}^\infty \z^n \U_{\rL n}^I = \Phi_\rL^I + \z \Sigma_\rL^I + \cdots~, \qquad
\breve{\U}_\rL^{\bar I}(\z) = \sum_{n=0}^\infty (-\z)^{-n} \bar \U_{\rL n}^{\bar I} ~.
\end{align}
The component superfields $\Phi_\rL^I:=\U_{\rL 0}^I$ and $\S_\rL^I := \U_{\rL 1}^I$
are chiral and complex linear respectively. These correspond to the
physical fields while the remaining superfields are auxiliary.
The extended supersymmetry transformation of $\U_\rL^I$ \eqref{eq_dUpL}
implies the transformations
\begin{subequations}
\begin{align}
\delta \Phi_\rL^I &= - \nrho_\alpha \bar \cD^\alpha \Sigma_\rL^I 
-4S \ve_\rL \Sigma_\rL^I
	= - \frac{1}{2} \bar \cD^2 (\bar \rho_\rL \Sigma_\rL^I)~, \\
\delta \Sigma_\rL^I &= (\bar\nrho^\alpha \cD_\alpha 
+4S\bar \ve_\rL
) \Phi_\rL^I
	- \bar \cD_\alpha (\nrho^\alpha \U_{\rL 2}^I)
	\end{align}
\end{subequations}
on the physical fields. For convenience, we have introduced the
parameter $\bar\rho_\rL$ which is defined by the condition
\begin{align}
\ve_\alpha = \bar \cD_\alpha \bar \rho_\rL~, \qquad
\bar\ve_\alpha = \cD_\alpha \rho_\rL~.
\end{align}
In the non-critical case we are discussing here, 
due to eq. \eqref{extra4020-2}
we can always choose
\begin{align}
\rho_\rL = \frac{2\ri}{2-X/S} \,\veps_\rR~, \qquad
\bar\rho_\rL = -\frac{2\ri}{2-X/S} \,\bar\veps_\rR~.
\end{align}

This off-shell formulation with an infinite number of auxiliary fields is
rather elaborate. We can simplify the theory by expressing  the auxiliary superfields
$ \U_{\rL 2}^I,  \U_{\rL 3}^I, \dots$ in terms of the physical ones, 
using their equations of motion 
\bea
\frac{\pa \mathbb L}{\pa \U^I_{\rL n}} = 0 ~, \qquad n = 2,3, \dots
\eea
The price for such a simplification 
is that $\cN=4$ supersymmetry is no longer off-shell.
This leaves the Lagrangian $\mathbb L_\rL$ depending only on $\Phi_\rL$, $\Sigma_\rL$ and
their complex conjugates. As a consequence of the homogeneity condition
\eqref{eq_40homogeneity},
\begin{align}
\Big( \Phi_\rL^I \frac{\pa}{\pa \Phi_\rL^I}
	+ \bar\Sigma_\rL^{\bar I} \frac{\pa  }{\pa \bar\Sigma_\rL^{\bar I}} \Big)  \mathbb L_\rL
	=\mathbb L_\rL 	~.
	\label{10.12}
\end{align}
Next, we dualize the complex linear superfields $\Sigma_{\rL}^{ I}$ and their conjugates
$\bar \Sigma_{\rL}^{\bar I}$  into chiral superfields $\Psi_{\rL I}$ 
and their conjugates $\bar \J_{\rL \bar I}$  via
a Legendre transformation and arrive at the dual action
\begin{align}
S_{\rm dual} = \int \rd^3x\, \rd^4\q\, E \, \mathbb K_\rL(\Phi_\rL,  \Psi_\rL, \bar\Phi_\rL,\bar\Psi_\rL)
~, \qquad	\mathbb K_\rL = \mathbb L_\rL + \S_\rL^I \Psi_{\rL I} 
+ \bar\S_\rL^{\bar J} \bar\Psi_{\rL \bar J}~.
\label{10.13}
\end{align}
The target space of this $\s$-model is a hyperk\"ahler cone. 
The cone condition follows from the fact that K\"ahler potential $\mathbb K_\rL$ 
obeys the homogeneity condition
\begin{align}\label{eq_Kcone}
 \chi_\rL^\ra \frac{\pa}{\pa \phi_\rL^\ra}   \mathbb K_\rL = \mathbb K_\rL ~, 
\qquad \chi_\rL^\ra = \phi_\rL^\ra = (\Phi_\rL^I, \Psi_{\rL I})~,
\end{align}
as a consequence of \eqref{10.12}.
The target space is hyperk\"ahler since it possesses a 
covariantly constant holomorphic two-form $\omega_{\ra\rb}$  given by
\begin{align}\label{eq_canonicalOmega}
\omega_{\ra\rb} =
\begin{pmatrix}
0 & \delta_I{}^J \\
-\delta^I{}_J & 0
\end{pmatrix}
\end{align}
with the property $\omega^{\ra\rb} \omega_{\rb\rc} = -\delta^\ra_\rc$, 
where $ \omega^{\ra \rb} = g^{\ra \bar \rc } g^{\rb \bar \rd} \bar \o_{\bar \rc \bar \rd}$.
The action \eqref{10.13} proves to be invariant under the extended supersymmetry
transformation
\begin{align}
\delta \Phi_\rL^I
	= - \frac{1}{2} \bar \cD^2 \Big(\bar \rho_\rL \frac{\pa\mathbb K_\rL}{\pa \Psi_{\rL I}}\Big)~, \qquad
\delta \Psi_{\rL I}
	= \frac{1}{2} \bar \cD^2 \Big(\bar \rho_\rL \frac{\pa\mathbb K_\rL}{\pa \Phi_{\rL}^I}\Big)~,
\end{align}
which we can cast in the target-space reparametrization-covariant form
\begin{align}
\delta \phi_\rL^\ra = - \frac{1}{2} \bar \cD^2 \Big(\bar \rho_\rL \omega^{\ra\rb} 
\pa_\rb \mathbb K_{\rL}\Big)~.
\end{align}
A similar calculation may be carried out for the right sector; one finds an
identical expression with the parameters $\rho_\rR$ and $\bar\rho_\rR$ given by
\begin{align}
\rho_\rR = \frac{2\ri}{2+X/S} \,\veps_\rL~, \qquad
\bar\rho_\rR = -\frac{2\ri}{2+X/S} \,\bar\veps_\rL~.
\label{10.18}
\end{align}

\subsection{Off-shell critical (4,0) sigma models}

Now we turn to the critical case of $X=2S$. Because the $\rm SU(2)_R$ factor
in the structure group has vanished, it is possible to introduce a
frozen vector multiplet.

Let us consider a {\it left } $\rm U(1)$ vector multiplet in (4,0) AdS superspace.\footnote{There are two 
types of vector multiplets in $\cN=4$ conformal supergravity, left and right ones \cite{KLT-M11}.
The left vector multiplets are used to gauge left hypermultiplets and do not couple to right 
hypermultiplets. The field strength of a left Abelian vector 
multiplet is a right $\cO(2) $ multiplet.
The  right vector multiplets are obtained from the left ones by applying the mirror map. }
It is described by the covariant derivatives
\begin{align}
{\bm \cD}_A = ({\bm \cD}_a , {\bm \cD}^{i \bai}_\a ) = 
\cD_A  + \ri \,V_A \cZ~, \qquad [\cZ,\cD_A] =[\cZ, {\bm \cD}_A ]=[\cZ, \cJ_{ij}]=0~,
\end{align}
where $V_A$ and $\cZ$ are the $\rm U(1)$ gauge connection and generator respectively.
The anti-commutator of two spinor derivatives is given by
\begin{align}
\{\bm\cD_\alpha^{i \bar i}, \bm\cD_\beta^{j \bar j}\} =
	2 \ri \veps^{ij} \veps^{\bar i \bar j} \bm\cD_{\alpha \beta}
	- 4 \ri S \veps^{ij} \veps^{\bar i \bar j} \cM_{\alpha \beta}
	+ 8 \ri S \veps_{\alpha \beta} \veps^{\bar i \bar j} {\bf L}^{ij}
	+ 2 \veps_{\alpha \beta} \veps^{ij} W^{\bar i \bar j} \cZ_\rL
\end{align}
where the gauge invariant field strength $W^{\bar i \bar j} = W^{\bar i \bar j}$ is real,
$\overline{W^{ \bar i \bar j } } = W_{\bar k \bar l} =
	\ve_{\bar i \bar k} \ve_{\bar j \bar l} W^{\bar k \bar l}$,
and obeys the Bianchi identity
\bea
\bm\cD^{i (\bar j}_\a W^{\bar k \bar l)} = 0~.
\eea 

We may take the field strength $W^{\bar i \bar j}$ to be covariantly constant, 
$\bm\cD^{i\bar j}_\a W^{\bar k \bar l} =0$.
Remarkably, unlike the situation in (2,1) AdS, there is no integrability condition to be
satisfied since the $\rm SU(2)_R$ factor has been removed from the
structure group. Thus $W^{\bar i \bar j}$ can be chosen completely
arbitrarily. In fact, since we can assume that we have gauged away the $\rm SU(2)_R$
connection, we can take $W^{\bar i \bar j}$ to be any constant isotriplet.
The remainder of the (4,0) algebra is then simply
\begin{align}
{[}\bm\cD_{\a\b},\bm\cD_\g^{i \bar i}{]}= -\,2 S \ve_{\g(\a} \bm\cD^{i\bar i}_{\b)}~,\qquad
{[}\bm\cD_a,\bm\cD_b{]}=-\,4\,S^2\cM_{ab}~.
\end{align}
The presence of the gauging requires a minor alteration to the (4,0)  isometry,
\begin{align}
\delta = \xi^A {\bm\cD}_A + \frac{1}{2} \Lambda^{ab} \cM_{ab}
	+ \L^{ij} {\bf L}_{ij} + \ri \lambda \cZ
\end{align}
where $\lambda$ is the U(1)  parameter. All the other Killing parameters
obey the same conditions while $\lambda$ obeys
\begin{align}
{\bm\cD}_\alpha^{i \bar j} \lambda = 2\ri \,\xi_\alpha^{i \bar k} \,W_{\bar k}{}^{\bar j}~.
\end{align}

As before, we can recover a (2,0) subalgebra, which possesses the spinor anticommutator
\begin{align}
\{\bm\cD_\alpha, \bar{\bm\cD}_\beta\} &= -2\ri \bm\cD_{\alpha \beta} + 4 \ri S \cM_{\alpha \beta}
	- 8 \ri S \veps_{\alpha \beta} {\bf L}^{12}
	- 2 \veps_{\alpha \beta} W^{\bar 1 \bar 2} \cZ \eol
	&= -2\ri \bm\cD_{\alpha \beta} + 4 \ri S \cM_{\alpha \beta}
	- 4 \ri S \veps_{\alpha \beta} \bm\cJ~,
\end{align}
where the effective $\rm U(1)$ generator $\bm\cJ$ is given by
\begin{align}
\bm \cJ = 2 \cJ_\rL - \frac{\ri}{2 S} W^{\bar 1 \bar 2} \cZ
\end{align}
The U(1) parameter $\lambda$ decomposes as
\begin{align}
\lambda = -\frac{\ri}{2S} W^{\bar 1 \bar 2} t + \lambda'
\end{align}
where $t$ is the ${\rm U}(1)_R$ parameter of the (2,0) AdS isometry group
and $\lambda'$ is associated with the extended supersymmetry.
The parameter $\lambda'$ obeys
\begin{align}
\cD_\alpha \lambda' = -2 \ri \bar \ve_\alpha W^{\bar 1 \bar 1}~, \qquad
\bar\cD_\alpha \lambda' = 2 \ri \ve_\alpha W^{\bar 2 \bar 2}~.
\end{align}
One may check that $\cD_{\alpha \beta} \lambda' = 0$.

Let us now consider the general off-shell $\s$-models in this superspace.
In contrast to the non-critical case, there are two conceptually new features which
occur. The first is that we may introduce \emph{gauge-covariant} left arctic weight-one
multiplets $\U_\rL^{(1)} (v)$ obeying the covariant analyticity condition
\bea
\bm \cD_\alpha^{(1) \bar i} \U_\rL^{(1)} = 0~, \qquad 
\bm \cD_\alpha^{(1) \bar i}:= v_i \bm \cD_\alpha^{i \bar i}~.
\eea 
However, if we try
to do the same for the right arctic weight-one multiplets,
we encounter a nontrivial integrability condition
\begin{align}
0 = \{\bm \cD_\alpha^{i (1)}, \bm \cD_\beta^{j (1)}\} \U_\rR^{(1)}
	= 2 \veps_{\alpha \beta} \veps^{i j} W^{(2)}_\rR \cZ \U_\rR^{(1)} \quad\implies\quad
\cZ \U_\rR^{(1)} = 0~.
\end{align}
In other words, we can take the left arctic multiplets to be charged
under the frozen vector multiplet $\rm U(1)$, but the right arctic
multiplets \emph{must be neutral}. The second elaboration is that
the nonzero vector multiplet field strength $W^{\bar i \bar j}$
can be used to construct a covariantly constant $\cO(2)$ multiplet $W^{(2)}_\rR$,
upon which the right  Lagrangian $\cL_\rL^{(2)} $ may depend.

Let us see how these modifications alter the analysis.
We begin with the left sector, where the Lagrangian is given by
 \bea
 \cL^{[2]}_\rL (\U_\rL, \breve{\U}_\rL;\z) = \frac{1}{\z} 
 {\mathfrak K}_\rL (\U_\rL , \z\breve \U_\rL) ~,
\label{10.34}
 \eea
 where $ {\mathfrak K}_\rL$
still obeys the homogeneity
condition \eqref{eq_40homogeneity}. However, now $\U_\rL^I$ is charged under
the  $\rm U(1)$ gauge group associated with the frozen vector multiplet. 
We require the Lagrangian to be invariant under  U(1) gauge transformations.
Let us denote
\begin{align}\label{eq_ZUps}
\cZ \U_\rL^I = - \ri\, Z^I(\U_\rL)
\end{align}
where $Z^I(\U_\rL)$ is some function of the left arctic multiplets.
Thus, the condition of gauge invariance is 
\bea
\Big( Z^I (\U_\rL) \frac{\pa }{\pa \U_\rL^I}
+ \bar Z^{\bar I} (\breve\U_\rL ) \frac{\pa }{\pa \breve\U_\rL^{\bar I}} \Big) \cL^{[2]}_\rL   =0~.
\label{10.36}
 \eea
 An important special case corresponds to the situation when $ {\mathfrak K}_\rL $
 obeys a homogeneity conditions of the form \eqref{hom_cond_strange2}. 
In this case the Lagrangian \eqref{10.34} turns into
 \bea
 \cL^{[2]}_\rL (\U_\rL, \breve{\U}_\rL) =
 {K}_\rL (\U_\rL , \breve \U_\rL) ~,
 \eea
where $K_\rL (\F^I, \bar \F^{\bar J}) $ is the preferred K\"ahler potential of a K\"ahler cone $\cX$
(see appendix A). Then the invariance condition \eqref{10.36} 
means that the holomorphic vector field 
\eqref{eq_ZUps} is a Killing vector field on $\cX $. 

Because $\U_\rL^I$ is a covariant arctic multiplet, its lowest two
components $\Phi_\rL$ and $\Sigma_\rL$ turn out to obey
\begin{align}
\bar{\bm\cD}_\alpha \Phi_\rL = 0~, \qquad
\bar {\bm\cD}^2 \Sigma_\rL = - 4\ri W^{\bar 2 \bar 2} Z^I(\Phi_\rL)~.
\end{align}
That is, $\Phi_\rL$ is covariantly chiral while $\Sigma_\rL$ obeys a modified
complex linearity condition. They inherit from \eqref{eq_ZUps} the gauge
transformations
\begin{align}
\cZ \Phi_\rL^I = - \ri\, Z^I(\Phi_\rL)~, \qquad
\cZ \Sigma_\rL^I = - \ri\, \Sigma_\rL^J \pa_J Z^I(\Phi_\rL)~.
\end{align}
Their transformation rules differ slightly from before. 
The extended supersymmetry transformation of $\Phi_\rL^I$ is
\begin{align}
\delta \Phi_\rL^I
	&= - \ve_\alpha \bar {\bm\cD}^\alpha \Sigma_\rL^I
	-4 S \veps_\rL \Sigma_\rL + \lambda' Z^I(\Phi_\rL)~.
\end{align}
For the critical case $X=2S$, recall that we have
\begin{gather*}
\cD_\alpha \veps_\rL = 2 \ri  \ve_\alpha~, \qquad
\bar{\cD}_\alpha \bar\veps_\rL = -2 \ri \bar\ve_\alpha
\end{gather*}
(the parameter $\veps_\rR$ may be consistently set to zero)
while we introduce $\bar\rho_\rL$ and $\bar\rho_\rR$ via the equations
\begin{gather*}
\nrho_\alpha = \bar {\cD}_\alpha \bar \rho_\rL = -{\cD}_\alpha \rho_\rR~, \qquad
\bar\nrho_\alpha = {\cD}_\alpha \rho_\rL = -\bar{\cD}_\alpha \bar\rho_\rR~.
\end{gather*}
We may choose $\bar \rho_\rR = -\dfrac{\ri}{2} \bar \veps_\rL$
and choose $\bar\rho_\rL$ to be antichiral. 
It follows that (up to a constant)
\begin{align}
\lambda' = - 2\ri \rho_\rL W^{\bar1\bar1} + 2\ri \bar\rho_\rL W^{\bar 2 \bar 2}~.
\end{align}
Then the extended supersymmetry transformation becomes
\begin{subequations}
\begin{align}
\delta\Phi_\rL^I
	=& - \frac{1}{2} \bar {\bm\cD}^2 (\bar \rho_\rL \Sigma_\rL)
	- 2 \ri \rho_\rL W^{\bar 1 \bar 1} Z^I(\Phi_\rL)~, \\
\delta \Sigma_\rL^I
	=& -\bar {\bm\cD}_\alpha \Big(\frac{\ri}{2} \bar\veps_\rL {\bm\cD}^\alpha \Phi_\rL^I
+	\ve^\alpha \U_{\rL 2}^I\Big)
		- 2\bar \veps_\rL W^{\bar 1 \bar 2} Z^I(\Phi_\rL) 
	+ \lambda' \Sigma_\rL^J \pa_J Z^I(\Phi_\rL)~.~~~~
\end{align}
\end{subequations}

Because of the modified complex linearity condition $\S_\rL$ obeys, the dual action receives
a superpotential contribution:
\begin{align}
S_{\rm dual} &= \int \rd^3x\, \rd^4\q\, E \, \mathbb K_\rL
	+ \Big(\int \rd^3x\, \rd^2\q\, \cE \,W_\rL + \HC\Big)~, \eol
	\mathbb K_\rL &= \mathbb L_\rL + \S_\rL^I \Psi_{\rL I} 
	+ \bar\S_\rL^{\bar J} \bar\Psi_{\rL \bar J}~, \qquad
	W_\rL = - \ri \,W^{\bar 2 \bar 2} \Psi_{\rL I} Z^I(\Phi_\rL)~.
\end{align}
Here $\Psi_{\rL I}$ is gauge covariantly chiral, ${\bm \cDB}_\a\Psi_{\rL I}=0$.
The action of $\cZ$ on $\Psi_{\rL I}$ is 
\bea
\cZ\Psi_{\rL I}=\ri\pa_I Z^J(\F_\rL)\Psi_{\rL J}~.
\eea
As before, the left target space 
comes equipped with a holomorphic two-form \eqref{eq_canonicalOmega}
which identifies the target space as hyperk\"ahler. Moreover, it 
is a hyperk\"ahler cone,  since the K\"ahler potential $\mathbb K_\rL$
obeys the homogeneity condition  \eqref{eq_Kcone}. 
 However, the $\s$-model is not superconformal due to the presence of 
the  superpotential with the property
\begin{align}
\chi_\rL^\ra W_{\rL\ra} = 2 W_\rL~, \qquad \chi_\rL^\ra = (\Phi_\rL^I, \Psi_{\rL I})~.
\end{align}
The fact that the $\s$-model is not superconformal becomes obvious if we
interpret $W^{\bar 2 \bar 2}$ as a frozen $\cN=2$ chiral superfield
carrying dimension.

The extended supersymmetry is also modified:
\begin{subequations}
\begin{align}
\delta \Phi_\rL^I
	&= -\frac{1}{2} \bar {\bm\cD}^2 \Big(\bar\rho_\rL \frac{\pa\mathbb K_\rL}{\pa\Psi_{\rL I}}\Big)
	- 2 \ri \rho_\rL W^{\bar 1 \bar 1} Z^I~, \\
\delta \Psi_{\rL I}
	&= + \frac{1}{2} \bar {\bm\cD}^2 \Big(\bar\rho_\rL \frac{\pa\mathbb K_\rL}{\pa\Phi_\rL^I}\Big)
	- 2\ri \rho_\rL W^{\bar 1 \bar 1} \pa_I Z^J(\Phi_\rL) \Psi_{\rL J}~,
\end{align}
\end{subequations}
which can be written as
\begin{align}
\delta \phi^\ra
	&= -\frac{1}{2} \bar {\bm\cD}^2 \Big(\bar\rho_L \omega^{\ra\rb} \mathbb K_\rb\Big)
	- 2 \ri \rho_L W^{\bar 1 \bar 1} Z^\ra~,
\end{align}
where $\omega_{\ra \rb}$ is given by \eqref{eq_canonicalOmega} and
$Z^\ra = \ri \cZ \phi^\ra = (Z^I, -\pa_I Z^J \Psi_{LJ})$ is a tri-holomorphic
Killing vector with the properties
\begin{align}
Z^\ra \mathbb K_{\rL\,\ra} + Z^{\bar \ra} \mathbb K_{\rL\,{\bar \ra}} = 0~, \qquad
\cL_Z \omega_{\ra\rb} = 0~.
\end{align}

Now we turn to the right sector. 
The corresponding  Lagrangian $\cL_\rR^{(2)} $ may possess an additional
$\z$-dependence through a frozen hypermultiplet $q^{(1)}$ and its smile-conjugate
$\breve{q}^{(1)}$, 
\bea
 q_\rR^{(1)} := q_{\bai} v^{\bai} ~, \qquad \breve{q}_\rR^{(1)} := \bar q_{\bai} v^{\bai}~, 
 \qquad q_{\bai} = \text{const}~.
 \eea
 Since in the critical case there is no $\rm SU(2)_R$ factor
in the structure group, the conditions ${\bm\cD}_A q_{\bai} = 0$ are integrable. 
The most general $\s$-model Lagrangian is 
\begin{align}
\cL_\rR^{(2)} = \cL_\rR(\U_\rR^{(1)}, \breve \U_\rR^{(1)}, q_\rR^{(1)}, \breve{q}_\rR^{(1)} )~.
\label{10.43}
\end{align}
The Lagrangian must be of weight two, which means
 \begin{align}
 \Big(\U_\rR^I \frac{\pa }{\pa \U_\rR^I}
	+ \breve\U_\rR^{\bar I} \frac{\pa }{\pa \breve\U_\rR^{\bar I}}
	+ q_\rR \frac{\pa}{\pa q_\rR} +  \breve{q}_\rR \frac{\pa}{\pa \breve{q}_\rR} 
	\Big)\cL_\rR
	=2 \cL_\rR  ~.
\label{10.44}
\end{align}
Here we have omitted the weight superscripts of the physical and frozen hypermultiplets, 
to make the equations less cluttered.
The relation \eqref{10.44} is a generalization of  the homogeneity condition 
\eqref{eq_40homogeneity}.
The right multiplets $\U_\rR^{(1)}$ and $\breve \U_\rR^{(1)}$ are neutral under the $\rm U(1)$
 gauge group and so they possess the same transformation laws and obey the same
constraints as before. We find $\Phi_\rR^I$ and $\Sigma_\rR^I$ to be chiral
and complex linear, respectively, and to possess the extended supersymmetry
transformations
\begin{subequations}
\begin{align}
\delta \Phi_\rR^I &= \bar\nrho_\alpha \bar {\bm\cD}^\alpha \Sigma_\rR^I 
-4S \ve_\rR
\Sigma_\rR^I
	= - \frac{1}{2} \bar {\bm\cD}^2 (\bar \rho_\rR \Sigma_\rR^I)~, \\
\delta \Sigma_\rR^I &= -(\nrho^\alpha {\bm\cD}_\alpha 
 -4S \bar \ve_\rR)
 \Phi_\rR^I
	- \bar {\bm\cD}_\alpha (\nrho^\alpha \U_{\rR 2}^I)~.
\end{align}
\end{subequations}
We may perform the duality transformation as before, with the result
\begin{align}
S_{\rm dual} = \int \rd^3x\, \rd^4\q\, E \, 
\mathbb K_\rR(\Phi_\rR, \bar\Phi_\rR, \Psi_\rR, \bar\Psi_\rR)~, \qquad
	\mathbb K_\rR = \mathbb L_\rR + \S_\rR^I \Psi_{\rR I} + \bar\S_\rR^{\bar J} \bar\Psi_{\rR \bar J}~.
\end{align}
The target space is  an {\it arbitrary}  hyperk\"ahler manifold,  with
a covariantly constant holomorphic two-form $\omega_{\ra\rb}$ given by
\begin{align}
\omega_{\ra\rb} =
\begin{pmatrix}
0 & \delta_I{}^J \\
-\delta^I{}_J & 0
\end{pmatrix}~,
\end{align}
obeying $\omega^{\ra\rb} \omega_{\rb\rc} = -\delta^\ra_\rc$.
The K\"ahler potential need no longer satisfy a homogeneity condition. 
The extended supersymmetry transformations of the fields are
as before
\begin{align}
\delta \Phi_\rR^I
	= - \frac{1}{2} \bar {\bm\cD}^2 
	\Big(\bar \rho_\rR \frac{\pa\mathbb K_\rR}{\pa \Psi_{\rR I}}\Big)~, \qquad
\delta \Psi_{\rR I}
	= \frac{1}{2} \bar {\bm\cD}^2 \Big(\bar \rho_\rR \frac{\pa\mathbb K_\rR}{\pa \Phi_{\rR}^I}\Big)~,
\end{align}
with $\bar \rho_\rR = -\dfrac{\ri}{2}  \bar \veps_\rL$.


\section{(4,0) supersymmetric sigma models with $X\neq 0$: On-shell approach}\label{section11}

In the previous section, we discussed the off-shell approach to
general (4,0) supersymmetric $\s$-models in AdS${}_3$. In this section, we turn to developing
the on-shell formulation for these $\s$-models in terms of chiral superfields in $(2,0)$ AdS 
superspace.

Before diving directly into the specifics of the (4,0)  situation,
we begin with a brief discussion of the universal details
we will encounter, which are valid for any $\cN=4$ supersymmetric $\s$-model in
either AdS or Minkowski. After that, we will focus on the (4,0)
case specifically.

\subsection{General features of $\cN=4$ supersymmetry}\label{section11.1}
From the discussion of the off-shell supersymmetric $\s$-models in projective
superspace 
both in the previous section and in prior papers \cite{KPT-MvU, KLT-M11}
any $\cN=4$ supersymmetric $\s$-model naturally involves two sectors: a left sector constructed
of left analytic multiplets and a right sector involving right analytic multiplets.
When the auxiliary fields are eliminated, one recovers two separate sectors
involving left and right hypermultiplets which are transformed into each other
under mirror symmetry.\footnote{The two classes of
hypermultiplets have been distinguished in the literature by referring to one type
as a ``twisted hypermultiplet.'' From an AdS perspective, the left / right
nomenclature is more precise.}
Interactions between the two sectors can be mediated by vector multiplets,
but we will avoid discussing these.

The existence of these decoupled sectors can be deduced from the $\s$-model
by the presence of two copies of the covariantly constant
holomorphic two-form $\omega_{\ra\rb}$, which we may denote
$\omega_{\rL\ra\rb}$ and $\omega_{\rR\ra\rb}$. They obey both an orthogonality condition
and a completeness condition
\begin{align}\label{eq_40omegas}
\omega_L^{\ra\rb} \omega_{R\rb\rc} = 0~, \qquad
\omega_L^{\ra\rb} \omega_{L\rb\rc} + \omega_R^{\ra\rb} \omega_{R\rb\rc} = -\delta^\ra{}_\rc~.
\end{align}
These conditions allow us to construct covariantly constant projection operators
\begin{gather}
(P_\rL)^\ra{}_\rb = - \omega_\rL^{\ra\rc} \omega_{\rL \rc\rb}~, \qquad
(P_\rR)^\ra{}_\rb = - \omega_\rR^{\ra\rc} \omega_{\rR \rc\rb}~, \eol
P_\rL P_\rR = 0~, \qquad P_\rL + P_\rR = {\mathbbm 1}~.\label{eq_40projs}
\end{gather}
Naturally, one may adopt a coordinate system where
$P_\rL = \textrm{diag}(1,\cdots,1,0,\cdots,0)$ and similarly
for $P_\rR$, which allows us to separate the fields into a ``left sector''
and a ``right sector.'' The covariant constancy of both operators
then allows one to prove that the K\"ahler potential should
decouple into two sectors. Naturally, each sector is separately
hyperk\"ahler. One may interpret $\omega_\rL$ and $\omega_\rR$
as arising from the covariantly constant
$\omega_{\ra\rb} := \omega_{\rL\ra\rb} + \omega_{\rR\ra\rb}$ which can be
easily seen to obey $\omega^{\ra\rb} \omega_{\rb\rc} = -\delta^\ra_\rc$.
One finds $\omega_{\rL\ra\rb} = (P_\rL)_\ra{}^\rc \omega_{\rc\rb}$ and similarly
for $\omega_{\rR\ra\rb}$.

We note that these same conditions (rephrased in slightly
different language) were found by de Wit, Tollst\'en and Nicolai in
the context of locally supersymmetric $\cN=4$ $\s$-models \cite{deWTN}.
There, the target space was found to be the product of two quaternionic
K\"ahler manifolds; this naturally reduces to a product of two
hyperk\"ahler manifolds when supergravity is turned off.

\subsection{Formulation of (4,0) supersymmetric sigma models in (2,0) superspace}

When the Killing vectors of (4,0) AdS superspace are recast in
(2,0) language, one recovers the usual (2,0) Killing vectors plus
additional parameters associated with the extended supersymmetry.
These parameters are the complex spinor $\nrho_\alpha$, whose
real and imaginary parts correspond to the extra two supersymmetries,
and the complex chiral parameters $\veps_L$ and $\veps_R$ which correspond
to the off-diagonal $\rm SU(2)_L$ and $\rm SU(2)_R$ transformations.
As in the previous section, it is useful to introduce the antichiral
parameters $\bar\rho_\rL$ and $\bar\rho_\rR$ which obey
\begin{subequations}\label{eq_40KillingParams}
\begin{gather}
0 = \bar\cD_\alpha \rho_\rL = \bar\cD_\alpha \rho_\rR 
= \cD_\alpha \bar\rho_\rL = \cD_\alpha \bar\rho_\rR~, \\
\ve_\alpha = \bar \cD_\alpha \bar\rho_\rL = -\cD_\alpha \rho_\rR~, \qquad
\bar\ve_\alpha = -\bar \cD_\alpha \bar\rho_\rR = \cD_\alpha \rho_\rL~, \\
8S \veps_\rL = \bar \cD^2 \bar\rho_\rL~, \qquad
8S \veps_\rR = \bar \cD^2 \bar\rho_\rR~.
\end{gather}
\end{subequations}
For the noncritical cases where $|X| \neq 2S$, we can choose
\begin{align}
\bar\rho_\rL = - \frac{2\ri}{2-X/S} \,\bar\veps_\rR~, \qquad
\bar\rho_\rR = - \frac{2\ri}{2+X/S} \,\bar\veps_\rL~.
\end{align}

Such a choice is not possible when $|X| = 2S$.
For the case $X=2S$ (the case $X=-2S$ is similar),
the elimination of the $\rm SU(2)_R$ factor in the structure group
means one may take $\veps_\rR = \bar\veps_\rR = 0$.
This means that $\bar \rho_\rL$ cannot be given explicitly in terms of any
other parameters, but only implicitly through the equations
\eqref{eq_40KillingParams}.
However, the choice $\bar\rho_\rR = -\dfrac{\ri}{2} \bar\veps_\rL$ remains possible.

Let us make an ansatz for the extended supersymmetry which is consistent
with the off-shell analysis in the previous section:
\begin{align}\label{eq_40susy}
\delta \phi^\ra = -\frac{1}{2} \bar \cD^2 (\bar \rho_\rL \Omega_\rL^\ra)
	-\frac{1}{2} \bar \cD^2 (\bar \rho_\rR \Omega_\rR^\ra)
\end{align}
where $\Omega_\rL^\ra$ and $\Omega_\rR^\ra$ are functions of $\phi$ and $\bar\phi$.
Requiring that the action be invariant under this transformation
(we leave the details again to appendix \ref{SUSY_Derivs}),
we recover
the following feature common to both critical and non-critical cases:
the objects $\omega_{\rL \ra\rb} = g_{\ra \bar \ra} \pa_\rb \Omega_\rL^{\bar \ra}$
and $\omega_{\rR \ra\rb} = g_{\ra \bar \ra} \pa_\rb \Omega_\rR^{\bar \ra}$ prove to be
covariantly constant holomorphic two-forms obeying the conditions
\eqref{eq_40omegas}. As discussed in the previous subsection, these
allow us to define the projector operators $P_\rL$ and $P_\rR$,
obeying \eqref{eq_40projs}, which separate the hyperk\"ahler target
space into distinct left and right sectors. Now let us address the features
which distinguish the noncritical and critical cases.

\subsubsection*{Noncritical case: $|X| \neq 2S$}

We first observe that the structure group possesses the full $\rm SU(2)_L \times SU(2)_R$.
This ensures that both the left and right sectors of the hyperk\"ahler manifold
will possess the full $\rm SO(3)$ Killing vectors. Moreover, it turns out that
both sectors are hyperk\"ahler cones.
Indeed, the explicit analysis reveals that there must exist
a homothetic conformal Killing vector $\chi^\ra$, which can be decomposed into
left and right sectors via
\begin{align}
\chi_\rL^\ra = (P_\rL)^\ra{}_\rb \chi^\rb~, \qquad
\chi_\rR^\ra = (P_\rR)^\ra{}_\rb \chi^\rb~,
\end{align}
so that the K\"ahler potential is given by
\begin{align}
K = \chi^\ra \chi_\ra = \chi_\rL^\ra \chi_{\rL \ra} + \chi_\rR^\ra \chi_{\rR \ra}~.
\end{align}
The term $\chi_\rL^\ra \chi_{\rL \ra}$ is the hyperk\"ahler potential for the
left sector as is $\chi_\rR^\ra \chi_{\rR \ra}$ for the right sector.
The vector $\bm J^\ra$ on the target space turns out to be given by
\begin{align}\label{eq_Jnoncrit}
\bm J^\ra &= \left(1+\frac{X}{2S}\right) J_\rL^\ra + \left(1-\frac{X}{2S}\right) J_\rR^\ra~, \eol
J_\rL^\ra &= -\frac{\ri}{2} \chi_\rL^\ra~, \qquad
J_\rR^\ra = - \frac{\ri}{2} \chi_\rR^\ra~.
\end{align}

Of course, the K\"ahler cone is also hyperk\"ahler, possessing a covariantly
constant holomorphic two-form $\omega_{\ra\rb}$ obeying the condition
$\omega^{\ra\rb} \omega_{\rb\rc} = -\delta^\ra_\rc$. Constructing the complex
structures as usual \eqref{eq_ComplexStructures}, one finds a full set of $\rm SU(2)$ Killing vectors
\begin{align}\label{eq_su2KillingVects}
V_{A}^\mu = -\frac{1}{2} (J_A)^\mu{}_\nu \chi^\nu~.
\end{align}
These decompose, using the projection operators, into $\rm SU(2)_L$ and $\rm SU(2)_R$
Killing vectors which act separately on the left and right sectors.

As in the $(3,0)$ case, the component Lagrangian is quite constrained,
and is given by
\begin{align}
\cL &= -g_{\ra \bar \ra} \cD_m \varphi^\ra \cD^m \bar\varphi^{\bar \ra}
	- \ri g_{\ra \bar \ra} \bar\psi_\alpha^{\bar \ra} \hat \cD^{\alpha \beta} \psi_\beta^{\ra}
	+ \frac{1}{4} R_{\ra \bar \ra \rb \bar \rb} (\psi^{\ra} \psi^\rb) (\bar\psi^{\bar \ra} \bar\psi^{\bar \rb})
	\eol & \quad
	- \frac{\ri}{2} X (\psi^\ra \bar\psi^{\bar \rb}) (P_\rL - P_\rR)_{\ra \bar \rb}
	+ (3 S^2 - \frac{1}{4} X^2) (K_\rL + K_\rR)
	+ X S (K_\rL - K_\rR)~.
\end{align}
It is not possible to deform the mass terms.

\subsubsection*{Critical case: $|X| = 2S$}

The critical case proves to possess a richer structure.
Without loss of generality, we take $X = 2S$. For this choice, we can consistently
remove $\rm SU(2)_R$ from the structure group, and so we consider only the Killing
parameters $\nrho_\alpha$, $\L^{22}$ and their complex conjugates.

Since the structure group is simply $\rm SU(2)_L$ it follows that we should expect stringent
conditions on the target space geometry only for the left case. Indeed, this is what
we find. The left sector, \emph{but only the left sector}, must be a hyperk\"ahler cone.
That is, there exists a holomorphic vector $\chi_\rL^a$ obeying
\begin{align}
(P_\rL \chi_\rL)^\ra = \chi_\rL^\ra~, \qquad \tsD_\rb \chi_\rL^\ra = (P_\rL)^\ra{}_\rb~, 
\qquad \tsD_{\bar \rb} \chi_\rL^\ra = 0~.
\end{align}
These conditions imply that the K\"ahler potential in the left sector is given by
$K_\rL = \chi_\rL^\ra \chi_{\rL \ra}$ as usual for a cone.

Similarly, the $\rm U(1)$ Killing vector $\bm J^\ra$ required by the (2,0)
algebra turns out to be given by $\bm J^\ra = 2 J_\rL^\ra$, where
$J_\rL^\ra$ obeys
\begin{align}
(P_\rL J_\rL)^\ra = J_\rL^\ra~, \qquad 
\cL_{J_\rL} \omega_{\rL \ra\rb} = -\ri \omega_{\rL \ra\rb}~.
\end{align}
This implies that $J_\rL^\ra$ can be decomposed as
\begin{align}
J_\rL^\ra = -\dfrac{\ri}{2} \chi_\rL^\ra + r Z_\rL^\ra~,
\end{align}
where $Z_\rL^\ra$ is a tri-holomorphic Killing vector. We have inserted a real parameter
$r$ for later convenience. No further restrictions are imposed on $Z_\rL^\ra$, and
its presence represents a consistent deformation of the mass parameters of the theory.
The three $\rm SU(2)_\rL$ Killing vectors remain defined as in
\eqref{eq_su2KillingVects} but with $\chi$ replaced by $\chi_\rL$.

Remarkably, the (4,0) critical case also allows a superpotential to be introduced.
Recall that a tri-holomorphic isometry $Z_\rL^\ra$ can be written (at least locally) as
$Z_\rL^\ra = \omega^{\ra\rb} \pa_\rb \L_\rL$ where $\L_\rL$ is a holomorphic
function depending only on the left sector. We may introduce a superpotential
$W = w \L_\rL$ where $w$ is some complex number.
One can check that the action is invariant if we modify the
transformation law of the fields as
\begin{align}
\delta \phi^\ra = -\frac{1}{2} \bar \cD^2 (\bar\rho_\rL \Omega_\rL^\ra)
	-\frac{1}{2} \bar \cD^2 (\bar\rho_\rR \Omega_\rR^\ra)
	- 2 \rho_\rL \bar w Z_\rL^\ra~.
\end{align}
On-shell, this transformation reduces to
\begin{subequations}
\begin{align}
(P_\rL)^\ra{}_\rb \delta \phi^\rb &=
	- \rho_\alpha (\omega_\rL)^\ra{}_{\bar \rb} \bar \cD^\alpha \bar\phi^{\bar \rb}
	-4 S \veps_\rL \omega^{\ra\rb} \chi_{\rL \rb}
	- 2 (\bar w \rho_\rL + w \bar\rho_\rL) Z_\rL^\ra~, \\
(P_\rR)^\ra{}_\rb \delta \phi^\rb &=
	\bar\rho_\alpha (\omega_\rR)^\ra{}_{\bar \rb} \bar \cD^\alpha \bar\phi^{\bar \rb}
\end{align}
\end{subequations}
for the left and the right sectors.

From our experience with off-shell $(4,0)$ models, we may identify
\begin{align}
r = -\dfrac{\ri}{4S} W^{\bar 1 \bar 2}~, \qquad
w = -\ri W^{\bar 2 \bar 2}~, \qquad
\bar w = \ri W^{\bar 1 \bar 1}~,
\end{align}
where $W^{\bar i \bar j}$ can be interpreted as a frozen right vector multiplet.

The component Lagrangian is
\begin{align}
L &= -g_{\ra \bar \ra} \cD_m \varphi^\ra \cD^m \bar\varphi^{\bar \ra}
	- \ri g_{\ra \bar \ra} \bar\psi_\alpha^{\bar \ra} \hat \cD^{\alpha \beta} \psi_\beta^{\ra}
	+ \frac{1}{4} R_{\ra \bar \ra \rb \bar \rb} (\psi^{\ra} \psi^\rb) (\bar\psi^{\bar \ra} \bar\psi^{\bar \rb})
	\eol & \quad
	- \ri \,S (\psi^\ra \bar\psi^{\bar \rb}) (P_\rL - P_\rR)_{\ra \bar \rb}
	+ 4 S^2 K_\rL
	\eol & \quad
	- \frac{1}{2} W^{\bar i \bar j} W_{\bar i \bar j} Z_\rL^\ra Z_\rL^{\bar \rb} g_{\ra \bar \rb}
	- \frac{\ri}{2} W^{\bar 1 \bar 2} (\psi^{\ra } \bar\psi^{\bar \rb}) 
	(\tsD_\ra Z_{\rL\bar \rb} - \tsD_{\bar \rb} Z_{\rL\ra}) 
	\eol & \quad
	- \frac{\ri}{2} W^{\bar 2 \bar 2} (\psi^{\ra } \psi^\rb) \omega_\rb{}^{\bar \rc}
	 \tsD_\ra Z_{\rL\bar \rc}
	+ \frac{\ri}{2} W^{\bar 1 \bar 1} (\psi^{\bar \ra} \psi^{\bar b}) 
	\omega_{\bar \rb}{}^{\rc} \tsD_{\bar \ra} Z_{\rL\rc}~.
\end{align}


\section{Sigma models with non-centrally extended  $\cN=4$ Poincar\'e supersymmetry}

As reviewed in the introduction, following \cite{KLT-M12},
the $(p,q)$ AdS supergeometry \eqref{alg-AdS}
is completely determined in terms of the scalar parameter $S$ defined 
as $S=\sqrt{(S^{IJ}S_{IJ})/\cN} $  (we recall that $\cN=p+q$ and $p\geq q \geq0$) 
provided  (i) $p+q<4$;  or (ii)  $p+q\geq 4$ and $q>0$. 
In the limit $S\to 0$,
this supergeometry reduces to that of ordinary $\cN$-extended Minkowski superspace. 
However, the situation is different in the case of $(\cN , 0)$ AdS supergeometry with $\cN\geq 4$, 
which allows for a second parameter -- the v.e.v. of the supersymmetric Cotton  
tensor $X^{IJKL}= X^{[IJKL]}$.  In the limit $S\to 0$,  this
supergeometry reduces to 
that of  the deformed $\cN$-extended Minkowski superspace \cite{KLT-M12}
\bsubeq \label{12.1}
\bea
\{\cD_\a^I,\cD_\b^J\}&=&
2\ri\d^{IJ}\cD_{\a\b}+\ri\ve_{\a\b}
X^{IJKL} \cN_{KL} ~,\\
{[}\cD_{a},\cD_\b^J{]}
&=& 0~, 
\qquad
{[}\cD_{a},\cD_b{]} =0~,
\eea
\esubeq
where the constant antisymmetric tensor $X^{IJKL}$ is constrained by 
\bea
X_N{}^{IJ[K}X^{LPQ]N} =0 ~.
\label{12.2}
\eea
If $\cN=4$ we simply have  $X^{IJKL} = X \ve^{IJKL} $,  
and eq. \eqref{12.2} is  identically satisfied. This solution can trivially be generalized
to the case of $\cN= 4 n$, with $n$ an integer, 
by considering $n$ copies of the $\cN=4 $ superalgebra.

For $X^{IJKL}\neq 0$, the isometry group of the superspace \eqref{12.1} 
is a deformation of the super Poincar\'e group with $\cN\geq 4$.
The corresponding superalgebra is 
a non-central extension of the standard Poincar\'e superalgebra in three dimensions.
Such non-centrally extended superalgebras   do not exist in four and higher  dimensions. 
Although the existence of these superalgebras was pointed out  
by Nahm \cite{Nahm} many years ago, only in the last decade
have they appeared explicitly in various string- and field-theoretic applications  
\cite{Blau:2001ne,Itzhaki:2005tu,LM,Gomis:2008cv,Hosomichi:2008qk,Bergshoeff:2008ta}.
${}$From the point of view of extended conformal supergravity  in three dimensions
\cite{HIPT,KLT-M11}, the deformation parameter $X^{IJKL}$ in \eqref{12.1} is an expectation value 
of the supersymmetric Cotton tensor.

Unitary representations of the non-centrally extended Poincar\'e superalgebras 
for certain choices of $X^{IJKL} $
have been studied, e.g., in \cite{Bergshoeff:2008ta}. In general, the presence of $X^{IJKL} \neq 0$
makes supermultiplets massive. In particular, in the case $\cN=4$ there are no massless 
representations if $X\neq 0$.
Here  we study the most general nonlinear $\s$-models possessing
the non-centrally extended  $\cN=4$ Poincar\'e supersymmetry. 
We will show that the non-central extension has the physical effect of introducing
a massive deformation proportional to $X$ when one begins with an otherwise massless
superspace action; equivalently, any component action invariant under deformed
Minkowski supersymmetry must have an $X$-dependent mass term.

We first rewrite the algebra of covariant derivatives \eqref{12.1} for $\cN=4$ 
in a  form compatible with the notation used in the previous sections. 
This is found by taking the formal $S=0$ limit
of the (4,0) AdS algebra given in section 
\ref{subsection7.1}:\footnote{We recall that the (4,0) AdS algebra is given by
the relations \eqref{N=4alg} with $\cS=S$ and $\cS^{ij\bai\baj}=0$.}
\bsubeq \label{12.3}
\bea
\{\cD_\a^{i\bai},\cD_\b^{j\baj }\}&=&
2\ri\,\ve^{ij}\ve^{\bai \baj }\cD_{\a\b}
+\,{2\ri}\ve_{\a\b}X (\ve^{\bai \baj } \bL^{ij} - \ve^{ij} \bR^{\bai\baj})
~,
\\
{[}\cD_{\a\b},\cD_\g^{k\bak}{]}&=&
0~,\qquad \qquad
{[}\cD_a,\cD_b{]}=0~.
\eea
\esubeq
Embedded in this $\cN=4$ superspace is a centrally extended  $\cN=2$ Minkowski superspace.
The latter is characterized by spinor covariant derivatives which are obtained by $\cN=2$ projection
from the operators
$\cD_\alpha := \cD_\alpha^{1 \bar 1}$ and $\bar\cD_\alpha := -\cD_\alpha^{2 \bar 2}$
obeying the algebra
\begin{subequations}
\begin{align}
\{\cD_\alpha, \bar\cD_\beta\} &= -2\ri\, \cD_{\alpha \beta} - 2 \veps_{\alpha \beta} X \Delta~, \qquad
\{\cD_\alpha, \cD_\beta\} = 0~, \\
[\cD_\alpha, \cD_b] &= 0~, \qquad [\cD_a, \cD_b] = 0~.
\end{align}
\end{subequations}
Here we have introduced the central charge operator $\Delta := \ri (\bL^{1 2} - \bR^{\bar 1 \bar 2})$.
The existence of the superspace reduction $\cN =4 \to \cN=2$ implies that any nonlinear 
$\s$-model in the deformed $\cN=4$ Minkowski superspace \eqref{12.3}
can be reformulated as a certain $\s$-model in $\cN=2$ Minkowski superspace with a central 
charge. 

It is pertinent to recall that the most general  $\cN=2$ supersymmetric
$\s$-model with a central charge involves a K\"ahler
potential $K (\f , \bar \f) $ and a superpotential $W (\f)$, with a superspace action
formally identical to eq. \eqref{2.8+W}. Both are required
to be invariant under the central charge 
\begin{align}
\Delta^\ra K_\ra + \bar \Delta^{\bar \ra} K_{\bar \ra} = 0~, \qquad \Delta^\ra W_\ra = 0~, \qquad
\Delta^\ra(\f) := \Delta \phi^\ra~.
\end{align}
The component Lagrangian is
\begin{align}
L &= - g_{\ra \bar \ra} \cD_m \varphi^\ra \cD^m \bar\varphi^{\bar \ra}
	- \ri g_{\ra \bar \ra} \bar\psi_\alpha^{\bar \ra} \hat \cD^{\alpha \beta} \psi_\beta^{\ra}
	+ \frac{1}{4} R_{\ra \bar \ra \rb \bar \rb} (\psi^{\ra} \psi^\rb) (\bar\psi^{\bar \ra} \bar\psi^{\bar \rb})
	\eol & \quad
	- g_{\ra \bar \rb} X^2 \Delta^\ra \Delta^{\bar \rb}
	+ \frac{1}{2} X (\psi^{ \rb} \bar\psi^{\bar \rb}) (\tsD_\rb \Delta_{\bar \rb} 
	- \tsD_{\bar \rb} \Delta_{ \rb})
	\eol & \quad
	- g^{\ra \bar \rb} W_\ra \bar W_{\bar \rb}
	- \frac{1}{2} \tsD_\ra W_\rb (\psi^\ra \psi^\rb)
	- \frac{1}{2} \tsD_{\bar \ra} \bar W_{\bar \rb} (\bar\psi^{\bar \ra} \bar\psi^{\bar \rb})
\end{align}
after eliminating the auxiliary fields. Note that masses for the fermions and a scalar
potential are generated from a nonzero constant $X$ as well as from the superpotential.

Since the deformed Minkowski algebra is simply the $S=0$ limit of the $(4,0)$
AdS algebra with $X\neq 0$ discussed in the previous section, we already know the
form the extended supersymmetry should take and its consequences on the target space.
We must have $\delta \phi^\ra$ given by \eqref{eq_40susy} where the parameters
$\bar\rho_\rL$ and $\bar\rho_\rR$ obey \eqref{eq_40KillingParams}. Since $S=0$ but $X \neq 0$,
the target space must obey the same constraints as the non-critical AdS case.
Hence, it must be a hyperk\"ahler cone with separate left and right sectors.
By comparing the deformed Minkowski algebra to the $(4,0)$ AdS algebra, we see that
the Killing vector $\bm J^\ra$ must be related to $\Delta^\ra$ by
$2S \bm J^\ra = X \Delta^\ra$ in the $S=0$ limit. Using
\eqref{eq_Jnoncrit}, the Killing vector $\Delta^\ra$ is given by
\begin{align}
\Delta^\ra = -\frac{\ri}{2} \chi_\rL^\ra + \frac{\ri}{2} \chi_\rR^\ra~.
\end{align}
The hyperk\"ahler cone structure dictates that one actually has a full set of
${\rm SU}(2)$ Killing vectors $V_A$ given by \eqref{eq_su2KillingVects};
the central charge Killing vector $\Delta^\ra$ is related to one of these by
$\Delta^\ra = (P_\rL - P_\rR)^\ra{}_\rb V_3^\rb$.
One also finds that the superpotential $W$ must vanish. The component action is
\begin{align}
L &= -g_{\ra \bar \ra} \cD_m \varphi^\ra \cD^m \bar\varphi^{\bar \ra}
	- \ri g_{\ra \bar \ra} \bar\psi_\alpha^{\bar \ra} \hat \cD^{\alpha \beta} \psi_\beta^{\ra}
	+ \frac{1}{4} R_{\ra \bar \ra \rb \bar \rb} (\psi^{\ra} \psi^\rb) (\bar\psi^{\bar \ra} \bar\psi^{\bar \rb})
	\eol & \quad
	- \frac{\ri}{2} X (\psi^\ra \bar\psi^{\bar \rb}) (P_\rL - P_\rR)_{\ra \bar \rb}
	- \frac{1}{4} X^2 (K_\rL + K_\rR)~.
\label{12.8}
\end{align}
We emphasize that as in the (4,0) case with $X \neq 0$, the action
is \emph{not} superconformal even though the target space is a cone.

The Lagrangian \eqref{12.8} describes the most general nonlinear $\s$-models with
the non-centrally extended  $\cN=4$ Poincar\'e supersymmetry. 
Setting $X=0$ in  \eqref{12.8}
gives the most general $\cN=4$ superconformal $\s$-model, 
with its target space being a hyperk\"ahler cone $\cM_{\rL} \times \cM_{\rR}$.
The deformation parameter $X$ appears in both terms in the second line of \eqref{12.8}. 
The first structure constitutes the fermionic mass term, while the second gives  
the scalar potential 
\bea
V = \frac{1}{4} X^2 (K_\rL + K_\rR)  ~.
\label{12.9}
\eea
For each of the left and right $\s$-model sectors, 
the scalar potential is constructed in terms of the 
homothetic  conformal Killing vector associated with the corresponding target space. 
This follows from the general result that for any hyperk\"ahler cone 
the preferred K\"ahler potential is given in terms of the homothetic 
conformal Killing vector $\c$ as follows:
\bea
K (\f, \bar \f) :={ g}_{\ra \bar \rb}(\f, \bar \f)  \, \c^\ra (\f) {\bar \c}^{\bar \rb} (\bar \f)~,
\eea
see Appendix A.
Therefore the scalar potential \eqref{12.9} is positive except at the tip of the left and right cones. 
Eq. \eqref{12.9} provides a new mechanism to generate massive $\s$-models. 
In the case of 4D $\cN=2$ or 5D $\cN=1$ Poincar\'e supersymmetries,
the standard mechanism to construct massive $\s$-models 
 \cite{A-GF83,HKLR,GTT,AIN,BX2006,K-potential} 
makes use of a tri-holomorphic Killing vector $Z^\ra (\f)$ on the hyperk\"ahler target space
(provided such a  tri-holomorphic Killing vector exists). The superpotential generated, $W(\f)$, 
is related to the tri-holomorphic Killing vector as $\pa_\ra W \propto \o_{\ra \rb} Z^\rb$,  
with $\o_{\ra\rb} (\f)$ the holomorphic two-form on the hyperk\"ahler target space
(we assume that the formulation in terms 4D $\cN=1$ chiral superfields is used);
 the corresponding scalar potential is 
$V \propto   g_{\ra \bar \rb}  \, Z^\ra  {\bar Z}^{\bar \rb} $.
Technically, any massive $\s$-model can be obtained 
by gauging an off-shell massless $\s$-model in projective
superspace and then freezing the background vector multiplet to have a constant field strength 
\cite{BX2006,K-potential}.
In the case of non-centrally extended 3D $\cN=4$ supersymmetry, the scalar potential 
\eqref{12.9} is constructed solely in terms of the homothetic conformal Killing vector 
each hyperk\"ahler cone possesses; no  tri-holomorphic Killing vector is involved. 
Technically, any  $\s$-model with the deformed $\cN=4$ supersymmetry
can be  obtained by coupling an $\cN=4$ 
superconformal $\s$-model to $\cN=4$ conformal supergravity and then freezing the Weyl 
multiplet to have a constant supersymmetric Cotton tensor and zero values for the other components 
of the supergravity torsion and curvature.

\section{Conclusion}

In this paper we have thoroughly studied the nonlinear $\s$-models in AdS${}_3$ with six and eight 
supercharges. With the exception of two {\it critical} (4,0) AdS supersymmetries, for which 
$X=\pm 2S$, all $\s$-model target spaces belong to the following large families of 
{\it non-compact} hyperk\"ahler manifolds:
(i) hyperk\"ahler cones; and  (ii) hyperk\"ahler spaces with a U(1) isometry group which acts 
non-trivially 
on the two-sphere of complex structures (and necessarily leaves one complex structure invariant). 
It is obvious that all hyperk\"ahler cones belong to the family (ii). The target spaces of arbitrary 
nonlinear 
$\s$-models with (3,0), (3,1) and non-critical (4,0) supersymmetries in AdS${}_3$ are hyperk\"ahler 
cones. 
The main reason for this is the property that the $R$-symmetry group of such supersymmetric 
$\s$-models includes SU(2) as a subgroup, see the discussion in \cite{KLT-M12}.
The target spaces of arbitrary nonlinear 
$\s$-models with (2,1) and  (2,2) supersymmetries in AdS${}_3$ are hyperk\"ahler manifolds 
belonging to the family (ii). 

We have demonstrated that 
{\it compact} target spaces are allowed only for those nonlinear $\s$-models in AdS${}_3$ 
which possess critical (4,0) supersymmetry such that $X=\pm 2S$. 
For concreteness, let us choose $X=2S$.
Then the target space of any supersymmetric $\s$-model has the form
\bea
\cM_{\rL} \times \cM_{\rR}~,
\eea
where $\cM_{\rL} $ is a hyperk\"ahler cone, while $\cM_{\rR}$ is an arbitrary hyperk\"ahler 
manifold. 

It is well known that the target spaces of superconformal $\s$-models with six and eight  
supercharges
are hyperk\"ahler cones, see e.g. \cite{deWKV,deWRV}.
All nonlinear $\s$-models in AdS${}_3$ with (3,0) and (3,1) supersymmetries are 
actually $\cN=4$ superconformal. As concerns the nonlinear $\s$-models in AdS${}_3$ 
with (4,0) supersymmetry, they are superconformal only in the case $X=0$. 

We have constructed the most general nonlinear $\s$-model in Minkowski space with  
a non-centrally extended $\cN=4$ Poincar\'e supersymmetry.
Its target space is a hyperk\"ahler cone, but the $\s$-model is massive. 
The Lagrangian includes a positive potential 
proportional to the norm squared  
of the homothetic conformal Killing vector the target space is endowed with. 
This mechanism of mass generation differs from the standard one which corresponds to 
 $\s$-model with the ordinary  $\cN=4$ Poincar\'e supersymmetry 
 and which makes use of a tri-holomorphic Killing vector. 
\\

\noindent
{\bf Acknowledgements:}\\
GT-M thanks the Department of Physics of 
Milano University for the kind hospitality during part of this work.
The work of SMK  and DB was supported in part by the Australian Research Council
under Grant No. DP1096372.  The work of DB was also supported by
ERC Advanced Grant No. 246974, ``{\it Supersymmetry: a window to
non-perturbative physics}.''
The work of GT-M was supported by the Australian Research Council's Discovery Early Career 
Award (DECRA), project No. DE120101498.


\appendix

\section{(Hyper) K\"ahler cones} \label{AppA}
\setcounter{equation}{0}

Consider a K\"ahler manifold $(\cM, g_{\m\n}, J^\m{}_\n )$, where $\m,\n=1,\dots, 2n$,  
and introduce local complex coordinates
$\f^\ra$ and their conjugates $\bar \f^{\bar \ra}$, in which the complex structure 
$J^\m{}_\n$ is diagonal.
It is called a K\"ahler cone \cite{GR} if it possesses
a homothetic conformal Killing vector $\c$
 \bea 
\c = \c^\ra \frac{\pa}{\pa \f^\ra} + {\bar \c}^{\bar \ra}  \frac{\pa}{\pa {\bar \f}^{\bar \ra}}
\equiv \c^\m \frac{\pa}{\pa \vf^\m} 
\eea
which is the gradient of a function. These conditions mean that 
\bea
\tsD_\n \c^\m = \d_\n{}^\m \quad \Longleftrightarrow \quad 
\tsD_\rb \c^\ra= \d_\rb{}^\ra~, \qquad 
\tsD_{\bar \rb} \c^\ra = \pa_{\bar \rb} \c^\ra = 0~.
\label{hcKv}
\eea
In particular,  $\c $ is holomorphic. The properties of $\c$ include the following:
\bea
\c_\ra := {g}_{\ra \bar \rb} \,{\bar \c}^{\bar \rb} = \pa_\ra { K}\quad \Longrightarrow \quad 
\c^\ra K_\ra = K~, 
\label{hcKv-pot}
\eea
where 
\bea
K:={ g}_{\ra \bar \rb} \, \c^\ra {\bar \c}^{\bar \rb}
\label{hcKv-pot2}
\eea
can be used as a  K\"ahler potential, $ {g}_{\ra \bar \rb} = \pa_\ra \pa_{\bar \rb} K$.
Associated with $\c$ is the U(1) Killing vector field 
\bea 
V^\m = -\hf J^\m{}_\n \c^\n~, \qquad \nabla_\m V_\n + \nabla_\n V_\m =0~.
\eea

Local complex coordinates for $\cM$ can always be
chosen such that 
 \bea
\c = \f^\ra \frac{\pa}{\pa \f^\ra} + {\bar \f}^{\bar \ra}  \frac{\pa}{\pa {\bar \f}^{\bar \ra}}~, 
\eea
and then the second relation in \eqref{hcKv-pot} turns into the homogeneity condition 
\bea
\f^\ra K_\ra (\f, \bar \f)= K(\f, \bar \f)~.
\eea

A hyperk\"ahler manifold $(\cM, g_{\m\n}, (J_A)^\m{}_\n )$, where $\m,\n=1,\dots, 4n$
and $A=1,2,3$, 
is called a hyperk\"ahler cone \cite{deWRV}
if it is a K\"ahler cone with respect to each complex structure. 
Using $J_A$ and $\chi$, we may construct three $\rm SU(2)$ Killing vectors
\begin{align}
V_A^\mu := -\frac{1}{2} (J_A)^\mu{}_\nu \chi^\nu~.
\end{align}
These vectors commute with $\c$  and obey an $\rm SU(2)$ algebra amongst themselves,
\begin{align}
[V_A, \chi] = 0~, \qquad [V_A, V_B] = \epsilon_{ABC} V_C~.
\end{align}
They generate a transitive  action of SO(3)  on the two-sphere of complex structures, 
\begin{align}
\cL_{V_A} J_B = \epsilon_{ABC} J_C~.
\end{align}
More information about the hyperk\"ahler cones can be found in \cite{deWRV}.


\section{Derivation of  the action (\ref{components-Ac})} 
\label{AppB}
\setcounter{equation}{0}

Here we sketch the derivation of the action (\ref{components-Ac}) 
by requiring its invariance under the projective transformations (\ref{proj-inv}).
The derivation is analogous to
the analysis
given in  \cite{KT-M,KT-M-normal,KT-M_5D}
and more recently in \cite{KLT-M12} for the 3D $\cN=3$ AdS case.
The interested reader is referred to those papers for more technical details
regarding the general procedure.  

To derive the action we start  from the term
$S_0$ in (\ref{comp-Ac-000}).
As a first step, we vary $S_0$ with respect to the infinitesimal transformation (\ref{proj-inv}).
Then, we  iteratively add to the action extra terms which cancel the variation order by order
such that the final action is invariant.
The $a_\rL$ and $c_\rL$ variations do not give important informations.
The non-trivial terms are generated by the $b_\rL$ variation in (\ref{proj-inv}),
\be
\d u_i=\,b_\rL\,v_i~.
\label{dubv}
\ee
The transformation (\ref{dubv}) induces the variations: 
\bea
\d\cD_\a^{(-1)\bai}=\frac{b_\rL}{(v_\rL,u_\rL)}\cD_\a^{(1)\bai}~,~~~
\d\cS^{(-2)\bai\baj}=\frac{2b_\rL}{(v_\rL,u_\rL)}\cS^{(0)\bai\baj}~,~~~
\d\cS^{(0)\bai\baj}=\frac{b_\rL}{(v_\rL,u_\rL)}\cS^{(2)\bai\baj}~,~~~
\label{d2}
\eea
with $\cS^{(0)\bai\baj}:=(v_iu_j\cS^{ij\bai\baj})/{(v_\rL,u_\rL)}$.

Let us compute the variation of $S_0$ defined by  (\ref{comp-Ac-000}).
By making use of (\ref{d2}) and the analyticity condition $\cD_\a^{(2)}\cL^{(2)}=0$
we obtain
\bea
\d S_0&=&
\frac{1}{96\p} \oint_{\g}  \frac{ v_i {\rm d} v^i }{(v_\rL,u_\rL)}
\int {\rm d}^3x\, e\, b_\rL\Big{[}
\{\cD^{(1)\g\bak},\cD^{(-1)}_{\g}{}^{\bal}\}\cD^{(-2)}_{\bak\bal}
+\cD^{(-2)}_{\bak\bal}\{\cD^{(1)\g\bak},\cD^{(-1)}_{\g}{}^{\bal}\}
\non\\
&&~~~~~~~~~~~~~~~~~~~~~
+\cD^{(-1)\g}_{\bal}\cD^{(-1)\d}_\bak\{\cD^{(1)}_{\g}{}^{\bak},\cD^{(-1)}_{\d}{}^{\bal}\}
+\cD^{(-1)\g}_{\bal}\{\cD^{(1)}_{\g}{}^{\bak},\cD^{(-1)}_{\d}{}^{\bal}\}\cD^{(-1)\d}_{\bak}
\non\\
&&~~~~~~~~~~~~~~~~~~~~~
+\cD^{(-1)\g\bak}{[}\cD^{(-1)\d}_\bak,\{\cD^{(1)}_{\g}{}^{\bal},\cD^{(-1)}_{\d\bal}\}{]}
\Big{]}
\cL_\rL^{(2)} (z,v_\rL)||
~.
\label{var-1}
\eea
Next we use the anti-commutation relations 
\eqref{N=4alg-1}--\eqref{N=4alg-3/2} to transform 
the expression in  the right-hand side of  \eqref{var-1}. 
As a result we obtain a number of  terms dependent on the Lorentz and SU(2) generators 
as well as terms containing the vector covariant derivative $\cD_{a}$.
The next step is to
push all the
Lorentz and SU(2) generators to the right.
Once they hit $\cL_\rL^{(2)}$ we use the identities
$\cM_{\a\b}\cL_\rL^{(2)}=\bR_{\bai\baj}\cL_\rL^{(2)}=v_iv_j\bL^{ij}\cL_\rL^{(2)}=0$ 
and $v_iu_j\bL^{ij}\cL_\rL^{(2)}=-(v_\rL,u_\rL)\cL_\rL^{(2)}$.
To compute the contributions coming from $u_iu_j\bL^{ij}\cL^{(2)}$
one has to use the following formula 
\bea
\oint {v_i\rd{v}^i\over (v_\rL,u_\rL)^6}\,b_\rL\, \cT^{(3)}u_iu_j\bL^{ij} \cL_\rL^{(2)}=
\oint  {v_i\rd{v}^i\over (v_\rL,u_\rL)^5}
\Big\{
b_\rL\Big(u^k\frac{\pa}{\pa v^k}\cT^{(3)}\Big) \cL_\rL^{(2)}
\Big\}
\label{useful111111111111}
\eea
which follows from the analysis of \cite{KT-M_5D,KT-M-normal}.
The equation (\ref{useful111111111111})
holds for any operator $\cT^{(3)}$ which is a function of $v_\rL$ and $u_\rL$ and 
homogeneous in $v_\rL$ of degree three: $\cT^{(3)}(cv_\rL)=c^3\cT^{(3)}(v_\rL)$.
There are also terms containing a vector derivative $\cD_{a}$.
To simplify those terms, we push to the left all the vector derivatives obtaining 
a total derivative, that can be ignored, plus terms involving commutator of spinor and vector 
derivatives.
The final result of the procedure sketched is
\bea
\d S_0&=&
\frac{1}{2\p} \oint_{\g}  \frac{ v_i {\rm d} v^i }{(v_\rL,u_\rL)}
\int {\rm d}^3x \,e\, b_\rL\Big{[}
\,\frac{2\ri}{3}\cS^{(0)}{}^{\bak \bal }\cD^{(-2)}_{\bak\bal}
+\frac{8}{3}\cS^{(-2)}{}^{\bak\bal}\cS^{(0)}{}_{\bak \bal}
\Big{]}
\cL_\rL^{(2)} (z,v_\rL)||
~.~~~~~~
\eea
The only possible functional that can be added to $S_0$ 
to cancel the above variation has the form
\bea
S_{\rm extra}&=&
\frac{1}{2\p} \oint_{\g}  { v_i {\rm d} v^i }
\int {\rm d}^3x \,e\, \Big{[}
a_1\ri\cS^{(-2)}{}^{\bak \bal }\cD^{(-2)}_{\bak\bal}
+a_2\cS^{(-2)}{}^{\bak \bal }\cS^{(-2)}_{\bak \bal }
\,\Big{]} \cL_\rL^{(2)}||~,
\eea
with $a_1$ and $a_2$ some numerical coefficients. 
By using the same procedure described for the computation of $\d S_0$, we derive
\bea
\d S_{\rm extra}&=&
\int {\rm d}^3x\, e 
\oint_{\g}  \frac{ v_i {\rm d} v^i }{2\pi}
\frac{b_\rL}{(v_\rL,u_\rL)}
\Big{[}\,
2\ri a_1\cS^{(0)}{}^{\bak \bal }\cD^{(-2)}_{\bak\bal}
+(4a_2-16a_1)\cS^{(-2)}{}^{\bak \bal }\cS^{(0)}_{\bak \bal }
\Big{]} \cL_\rL^{(2)}||
~.
~~~~~~~
\eea
If we impose the condition  $\d S_0+\d S_{\rm extra}=0$, we fix the coefficients  to be
\bea
a_1=-\frac{1}{3}~,~~~
a_2=-2~.
\eea
The functional given by $S=S_0+S_{\rm extra}$
is the invariant action (\ref{components-Ac}).

\section{Deriving conditions for extended supersymmetry}\label{SUSY_Derivs}

In this appendix we briefly sketch how to establish the conditions imposed
by extended supersymmetry on the target spaces of $\cN=2$ $\sigma$-models.

\subsection{(3,0) AdS supersymmetry in (2,0) AdS superspace}

In section \ref{section4}, we addressed what conditions $\sigma$-models in (2,0)
AdS superspace must obey in order to possess (3,0) AdS supersymmetry. We summarize
below how one goes about establishing the properties of the target space.

\subsubsection*{Deriving the conditions}

Let us first derive a set of necessary conditions for the variation of the action,
eq. \eqref{eq_varyS30}, to vanish.
Instead of analyzing it directly, we can consider its functional
variation with respect to the chiral superfield $\phi^\ra$:
\begin{align*}
0 = \delta_\phi \delta S = \frac{\ri}{2} \int \rd^3x\, \rd^4\q\,E\,\delta \phi^\ra \Big(&
	K_{\ra\rb} \bar \cD^2 (\bar \rho \Omega^\rb)
	+ K_{\rb} \bar \cD^2 (\bar \rho \Omega^\rb{}_{,\ra})
	\\
	&
	- K_{\ra \bar \rb} \cD^2 (\rho \Omega^{\bar \rb})
	- \cD^2 K_{\bar \rb} (\rho \Omega^{\bar \rb}{}_{,\ra})
	\Big)~.
\end{align*}
It turns out that this condition is far simpler to analyze. We must impose
\begin{align}\label{eq_30InvCond}
0 = -\frac{\ri}{8} \bar \cD^2 \Big(
	K_{\ra\rb} \bar \cD^2 (\bar \rho \Omega^\rb)
	+ K_{\rb} \bar \cD^2 (\bar \rho \Omega^\rb{}_{,\ra})
	- K_{\ra \bar \rb} \cD^2 (\rho \Omega^{\bar \rb})
	- \cD^2 K_{\bar \rb} (\rho \Omega^{\bar \rb}{}_{,\ra})
	\Big)~.
\end{align}
Let us consider several classes of terms.
Those involving $\cD^{\alpha \beta} \cD_{\alpha \beta} \phi^\rb$ will be proportional to
$(K_{\ra \bar \rb} \Omega^{\bar \rb}{}_{,\rb} + K_{\rb \bar \rb} \Omega^{\bar \rb}{}_{,\ra})$
and so $\omega_{\ra\rb} := g_{\ra \bar \rc} \Omega^{\bar \rc}{}_{, \rb}$
must be antisymmetric. Making use of this condition,
\begin{align*}
\frac{\ri}{8} \bar \cD^2 \Big(
	K_{\ra \bar \rb} \cD^2 (\rho \Omega^{\bar \rb})
	+ \cD^2 K_{\bar \rb} (\rho \Omega^{\bar \rb}{}_{,\ra})
	\Big) 
	=& \frac{\ri}{8} \bar \cD^2 \Big(
	\cD^2 \rho K_{\ra \bar \rb} \Omega^{\bar \rb}
	+ 2 K_{\ra \bar \rb} \cD^\alpha \rho \omega^{\bar \rb}{}_\rb \cD_\alpha \phi^\rb
	\eol
	&
	+ K_{\ra \bar \rb} \rho \pa_\rc \omega^{\bar \rb}{}_\rb \cD^\alpha \phi^\rc \cD_\alpha \phi^\rb
	+ \pa_\rc K_{\rb \bar \rb} \cD^\alpha \phi^\rc \cD_\alpha \phi^\rb \omega^{\bar \rb}{}_\ra
	\Big)~.
\end{align*}
The third and fourth terms are unique: these will give terms involving
two vector derivatives of $\phi$. They are cancelled only if
$\tsD_\rc \omega_{\rb\ra} = \pa_\rc \omega^{\bar \rb \bar \ra} = 0$.
Imposing both of these conditions, we are left with
\begin{align*}
0 = -\frac{\ri}{8} \bar \cD^2 \Big(
	K_{\ra\rb} \bar \cD^2 (\bar \rho \Omega^\rb)
	+ K_{\rb} \bar \cD^2 (\bar \rho \Omega^\rb{}_{,\ra})
	+ 8 \ri S \bar\rho \,\Omega_\ra
	- 2 \cD^\alpha \rho \,\cD_\alpha \phi^\rb \,\omega_{\ra\rb} 
	\Big)~.
\end{align*}
Taking now all terms proportional to $\bar \cD^2 \bar\phi^{\bar \rb} \bar \cD^2 \bar\phi^{\bar \rc}$,
one finds that $\pa_\ra \omega_{\bar \rb \bar \rc} = 0$.
This establishes the existence of a covariantly constant two-form $\omega_{\ra\rb}$.

Exploiting this result, we can rewrite the condition \eqref{eq_30InvCond} as
\begin{align}\label{eq_30InvCondv2} 
0 = -\frac{\ri}{8} \bar \cD^2 \Big(
	8 \ri S \bar\rho \,\Omega_\ra
	- \bar \cD_\beta \bar \rho \bar \cD^\beta \bar\phi^{\bar \rc} \,\pa_\ra \Omega_{\bar \rc}
	- 2 \cD^\alpha \rho \,\cD_\alpha \phi^\rb \,\omega_{\ra\rb} 
	\Big)~.
\end{align}
The only term involving $\rho \bar \cD^2 \bar\phi^{\bar \rb}$ is proportional to
$\pa_{\ra} \Omega_{\bar \rb}$, so we must require
$\pa_{\ra} \Omega_{\bar \rb} = 0$. Now we observe that $\Omega_{\ra}$ obeys the following
properties:
\begin{align}\label{eq_condOmega30}
\tsD_{\bar \rb} \Omega_{\ra} = 0~, \qquad \tsD_{\rb} \Omega_\ra = \omega_{\ra\rb}~.
\end{align}
We will shortly find that $\omega_{\ra\rb}$ is invertible.
Assuming this now, we immediately observe that the conditions \eqref{eq_condOmega30}
are satisfied if and only if
\begin{align}
\Omega_\ra = \omega_{\ra\rb} \chi^\rb
\end{align}
where $\chi^\rb$ is a homothetic conformal Killing vector.
The remainder of the condition \eqref{eq_30InvCondv2} amounts to
\begin{align*}
0 = 8 \ri S^2 \rho\,\Omega_\ra + 16 \rho S^2 J^\rb \,\omega_{\ra\rb}
\end{align*}
and so we conclude that
$J^\ra = -\dfrac{\ri}{2} \chi^\ra$.

\subsubsection*{Closure of the algebra}
To complete our analysis, we require an additional condition: we must
enforce that the algebra of two extended supersymmetries closes on-shell.
Examining
\begin{align}
[\delta_2, \delta_1] \phi^\ra = \ri \bar\cD^2 (\bar \rho_{[1} \delta_{2]}\Omega^\ra)~,
\end{align}
and using the constraint
\begin{align}
0 = [\tsD_\ra, \tsD_{\bar \ra}] (\omega^\rb{}_{\bar \rb} \chi^{\bar \rb})
	= R_{\ra \bar \ra}{}^\rb{}_{\rc} \omega^\rc{}_{\bar \rb} \chi^{\bar \rb}
	= R_{\ra \bar \ra}{}^\rb{}_{\rc} \Omega^\rc~,
\end{align}
one can straightforwardly check that
\begin{align}
[\delta_2, \delta_1] \phi^\ra
	&= - \rho_{[2\alpha} \bar \cD^\alpha \bar\phi^{\bar \rb} \omega^\rb{}_{\bar \rb} \Big(
	\bar \rho_{1]} R_{\rc \bar \rd \rb}{}^\ra \omega^\rc{}_{\bar \rc}
	\bar \cD_\beta \bar\phi^{\bar \rd} \bar \cD^\beta \bar\phi^{\bar \rc}
	\Big)
	-\ri \bar\cD^2 (\bar\rho_{[2} \delta_{1]} \bar\phi^{\bar \rb} \omega^\ra{}_{\bar \rb})~.
\end{align}
The first term vanishes since $\omega_{\bar \rc}{}^\rc \omega_{\bar \rb}{}^\rb R_{\rc \bar \rd \rb}{}^\ra$
is totally symmetric in the indices $\bar \rc$, $\bar \rb$ and $\bar \rd$ and the symmetrized
product of the three fermionic fields vanishes. For the remaining term, we get
\begin{align}\label{eq_closure30}
[\delta_2, \delta_1] \phi^\ra
	&= -\frac{1}{2} \bar\cD^2 \Big(\bar\rho_{[2} \omega^\ra{}_{\bar \rb} \omega^{\bar \rb \bar \rc}
	 \cD^2 (\rho_{1]} \bar\chi_{\bar \rc}) \Big)
	= -\frac{1}{2} \omega^{\ra\rb} \omega_{\rb \rc} \bar\cD^2 \Big(\bar\rho_{[2}  g^{\rc \bar \rc} 
	\cD^2 (\rho_{1]} \bar\chi_{\bar \rc}) \Big) \eol
	&= -\omega^{\ra\rb} \omega_{\rb \rc}  \Big(
	- 2 \ri \veps_{[2} \rho_{1]}^\alpha \cD_\alpha \phi^\rc
	- 4 \ri \rho_{[2}^\beta \rho_{1]}^\alpha \cD_{\alpha \beta} \phi^\rc
	- \ri \bar \veps_{[2} \veps_{1]} J^\rc
	+ \frac{1}{2} \bar \cD^2 (\bar \rho_{[2} \rho_{1]} g^{\rc \bar \rc} \cD^2 \bar \chi_{\bar \rc})
	\Big)
\end{align}
where we have used $\veps = -8 S \rho$.

We may check this against the extended supersymmetry algebra.
Let $\Psi$ be some $\cN=3$ superfield. Under an extended supersymmetry transformation,
\begin{align}
\delta \Psi = 2 \ri \rho^\alpha \cD_\alpha^{\1\2} \Psi
	+ \frac{1}{2} \bar\veps J_{\1\1} \Psi
	+ \frac{1}{2} \veps J_{\2\2} \Psi~,
\end{align}
and a straightforward calculation leads to
\begin{align}
[\delta_2, \delta_1] \Psi
	&= - 4 \ri \rho_{[2}^\alpha \rho_{1]}^\beta \cD_{\beta \alpha} \Psi
	- 2\ri \veps_{[2} \rho_{1]}^\alpha \cD_\alpha \Psi
	+ \bar\veps_{[2} \veps_{1]} \cJ \Psi~.
\end{align}
In order for the $\cN=2$ projection of this expression to match
\eqref{eq_closure30} for $\phi^\ra = \Psi\vert$, we find that
$\phi^\ra$ must be on-shell, $\cD^2 \chi_{\bar \ra} = \cD^2 K_{\bar \ra} = 0$,
and that the holomorphic two-form $\omega_{\ra\rb}$ must obey
$\omega^{\ra\rb} \omega_{\rb\rc} = -\delta^\ra_\rc$.

\subsubsection*{Invariance of the full action}

We must still check that the conditions we have derived are sufficient to
ensure the invariance of the action. First note that $\delta \phi^\ra$ can be rewritten
\begin{align}
\delta \phi^\ra = \frac{\ri}{2} \bar\cD^2 (\bar \rho \Omega^\ra)
	= \frac{\ri}{2} \bar\cD^2 (\bar \rho \omega^{\ra\rb} \chi_\rb)~, \qquad \chi_\rb = K_\rb~.
\end{align}
This allows the variation of the action to be written
\begin{align}
\delta S &= \int \rd^3x \, \rd^4\q\,E\, \Big(
	- \frac{\ri}{2} \rho_\alpha \bar A^\alpha + \frac{\ri}{2} \rho^\alpha A_\alpha
	\Big)~, \qquad
A_\alpha = \cD_\alpha \phi^\rb \omega_{\rb\rc} \chi^\rc~.
\end{align}
Now note that
\begin{align}
\bar \cD_\beta A_\alpha &=
	-2\ri \cD_{\alpha \beta} \phi^\rb \omega_{\rb\rc} \chi^\rc 
	- 4 \veps_{\alpha \beta} S J^\rb \omega_{\rb\rc} \chi^\rc
	= -2\ri \cD_{\alpha \beta} \phi^\rb \omega_{\rb\rc} \chi^\rc~.
\end{align}
In particular, $\bar \cD^\alpha A_\alpha = 0$. It follows that $\delta S = 0$
by writing $\rho^\alpha A_\alpha = \bar \cD^\alpha \bar \rho \,A_\alpha$ and integrating
by parts.

\subsection{(2,1) AdS supersymmetry in (2,0) AdS superspace}

Our next case involves a (2,1)-supersymmetric $\sigma$-model in (2,0) AdS superspace.

\subsubsection*{Derivation of conditions and invariance of the action}
We proceed as in our analysis of the (3,0) case. Instead of
requiring $\delta S = 0$ directly, we analyze $\delta_\phi \delta S = 0$.
This condition amounts to
\begin{align}\label{eq_21aInvCond}
0 = - \frac{1}{4} \bar \cD^2 \Big(
	\frac{\ri}{2} K_{\ra\rb} \bar \cD^2 (\bar \rho \Omega^\rb)
	+ \frac{\ri}{2} K_{\rb} \bar \cD^2 (\bar \rho \pa_\ra \Omega^\rb)
	- \frac{\ri}{2} g_{\ra \bar \rb} \cD^2 (\rho \Omega^{\bar \rb})
	- \frac{\ri}{2} \cD^2 K_{\bar \rb} \rho \pa_\ra \Omega^{\bar \rb}
	\Big)~.
\end{align}
Analyzing \eqref{eq_21aInvCond} and focusing our attention on terms involving
the highest number of derivatives, we immediately deduce as before that
$\omega_{\ra\rb} := g_{\ra \bar \ra} \pa_\rb \Omega^{\bar \ra}$
is antisymmetric and covariantly constant. Then \eqref{eq_21aInvCond} simplifies to
\begin{align}
0 = -\frac{1}{4} \bar \cD^2 \Big(
	-\ri \omega_{\ra\rb} \rho^\alpha \cD_\alpha \phi^\rb
	- \frac{\ri}{2} \rho_\alpha \bar\cD^\alpha \bar\phi^{\bar \rb} \pa_\ra \Omega_{\bar \rb}
	\Big)
	~.
\end{align}
The first term gives zero when we apply $\bar\cD^2$ since $\rho^\alpha$
is constant, $\omega_{\ra\rb}$ is chiral, and $\bar \cD^2 \cD_\alpha \phi^\rb = 0$.
The remaining term is
\begin{align}
0 &= \frac{\ri}{8} \rho_\alpha (
	\bar\cD^\alpha \bar\phi^{\bar \rb} \bar\cD^2 \phi^{\bar \rc} (\pa_{\bar \rc} \pa_\ra 
	\Omega_{\bar \rb})
	-\bar \cD^2 \bar\phi^{\bar \rb} \bar \cD_\alpha \bar\phi^{\bar \rc}
		\pa_\ra \pa_{\bar \rc} \Omega_{\bar \rb}
	+ \bar\cD^\alpha \bar\phi^{\bar \rb} \bar\cD_\beta \bar\phi^{\bar \rc} 
	\bar\cD^\beta \bar\phi^{\bar \rd}
		(\pa_{\bar \rd} \pa_{\bar \rc} \pa_\ra \Omega_{\bar \rb})
	\Big)~.
\end{align}
The first two terms cancel since they amount to
\begin{align}
\pa_\ra \pa_{\bar \rc} \Omega_{\bar \rb} - \pa_\ra \pa_{\bar \rb} \Omega_{\bar \rc}
	= \pa_\ra (2 \omega_{\bar \rb \bar \rc}) = 0~.
\end{align}
The third term cancels since the product of the three fermionic fields vanishes if totally
symmetrized in $\bar \rb$, $\bar \rc$ and $\bar \rd$, and the 
quantity $(\pa_{\bar \rd} \pa_{\bar \rc} \pa_\ra \Omega_{\bar \rb})$ is indeed totally
symmetric in these indices. To prove this, we first note that it is
already symmetric in $\bar \rc$ and $\bar \rd$ by construction. Then writing
\begin{align}
\pa_{\bar \rd} \pa_\ra \pa_{\bar \rc} \Omega_{\bar \rb}
	= \pa_{\bar \rd} \pa_\ra (\omega_{\bar \rb \bar \rc} 
	+ \Gamma_{\bar \rc \bar \rb}{}^{\bar \rd} \Omega_{\bar \rd})
	= \pa_{\bar \rd} \pa_\ra (\Gamma_{\bar \rc \bar \rb}{}^{\bar \rd} \Omega_{\bar \rd})
\end{align}
we discover it is symmetric in $\bar \rc$ and $\bar \rb$ as well.

Now let us prove these conditions are sufficient. The proof is very similar to
that in $\textrm{AdS}_4$ \cite{BKsigma2}. We begin by writing
\begin{align}
\delta S =\int \rd^3 x\, \rd^4\q\, E\, \Big(
	\frac{\ri}{2} \rho^\alpha A_\alpha - \frac{\ri}{2} \rho_\alpha \bar A^\alpha
	\Big)~, \qquad A_\alpha = \cD_\alpha \phi^\rb \omega_\rb{}^{\bar \rb} K_{\bar \rb}~.
\end{align}
One can check that
\begin{align}
\cD_\alpha A_\beta + \cD_\beta A_\alpha = -2 \cD_\alpha \phi^\rb \cD_\beta \phi^\rc \omega_{\rb\rc}~.
\end{align}
Since $\omega_{\rb\rc}$ is closed, we may take (at least locally)
$\omega_{\rb\rc} = \pa_\rb \Gamma_\rc - \pa_\rc \Gamma_\rb$ for some holomorphic one-form
$\Gamma_\ra$, and rewrite the integrand of $\delta S$ as
\begin{align}
\frac{\ri}{2} \rho^\alpha (A_\alpha + 2 \cD_\alpha \phi^\rb \Gamma_\rb) + \textrm{c.c.}
\end{align}
The additional term we have added is annihilated by $\bar\cD^2$. The term in parentheses
can be denoted $B_\alpha$ and obeys $\cD_{(\alpha} B_{\beta)} = 0$,
implying that $B_\alpha = \cD_\alpha B$ for some complex quantity $B$. Making this
substitution and integrating by parts, we find $\delta S = 0$.

So far we have established only the existence of the covariantly constant two-form.
We must still establish that $\omega^{\ra\rb} \omega_{\rb\rc} = -\delta^\ra_\rc$, and that
the U(1)$_R$ isometry rotates $\omega_{\ra\rb}$. The first condition is completely
straightforward to prove and proceeds analogously as in the (3,0) case from an
analysis of closure. The second condition is slightly more involved, and arises by considering
the algebra of an extended supersymmetry transformation with a (2,0) supersymmetry
transformation -- in particular with a U(1)$_R$ transformation. 
We note that the U(1)$_R$ generator acts as
$\cJ \phi^\ra = -\ri J^\ra$ and $\cJ \phi^{\bar \rb} = -\ri J^{\bar \rb}$.
It follows that
\begin{align}
\cJ \delta \phi^\ra &= \frac{\ri}{2} \cJ \bar \cD^2 (\bar \rho \Omega^\ra)
	= \frac{1}{2} \bar \cD^2 (\bar \rho J^\rb \pa_\rb \Omega^\ra
	 + \bar\rho J^{\bar \rb} \Omega^\ra{}_{\bar \rb}
	- \ri\, \bar \rho \Omega^\ra)
\end{align}
after accounting for the nontrivial U(1)$_R$ transformation properties of
$\bar\cD_\alpha$ and $\bar\rho$. We next note that
\begin{align}
\delta \cJ \phi^\ra = -\ri \,\delta J^\ra
	= \frac{1}{2} \bar\cD^2 (\bar \rho \Omega^\rb \pa_\rb J^\ra)~.
\end{align}
In order for the extended supersymmetry transformation to be an isometry,
it must commute with the generator $\cJ$. In other words,
\begin{align}
0 = [\cJ, \delta] \phi^\ra = \frac{1}{2} \bar \cD^2 \Big(\bar \rho J^\rb \pa_\rb \Omega^\ra
	+ \bar\rho J^{\bar \rb} \Omega^\ra{}_{\bar\rb} - \ri \, \bar \rho \Omega^\ra
	- \bar \rho \Omega^\rb \pa_\rb J^\ra \Big)
	~.
\end{align}
This holds if and only if
\begin{align}
\cL_J \omega_{\ra \rb} = -\ri \,\omega_{\ra \rb}~, \qquad
\cL_J \omega_{\bar \ra \bar \rb} = \ri \,\omega_{\bar \ra \bar \rb}~.
\end{align}

In contrast to the (3,0) $\s$-model, we see here only a single U(1)
Killing vector exists in the hyperk\"ahler target space --
coinciding with that generated by the U(1)$_R$ of the (2,0) algebra -- 
rather than the full triplet of SU(2) Killing vectors which the (3,0)
$\s$-model possesses. This is completely consistent with the off-shell
description and reflects simply the fact that the (2,1) algebra possesses
only a single U(1) isometry.

\subsection{(2,1) AdS supersymmetry in (1,1) AdS superspace}
Now we derive the consequences of imposing (2,1) AdS supersymmetry on
a $\s$-model in (1,1) AdS superspace.

\subsubsection*{Analysis of constraints and invariance of the action}
Recall the postulated transformation law
\begin{align}
\delta \phi^\ra = \frac{1}{2} (\bar {\frak D}^2 - 4 \mu) (\veps \Omega^\ra)~.
\end{align}
As in $\textrm{AdS}_4$ \cite{BKsigma1, BKsigma2}, this is the most general ansatz
available. Also as in $\textrm{AdS}_4$, the most general Lagrangian is just the full superspace
integral of a real function $\cK$. We derive the conditions necessary for $\delta S = 0$ by
analyzing the weaker condition $\delta_\phi \delta S = 0$. Observe that
\begin{align}
\delta_\phi \delta S &= \int \rd^3x\, \rd^4\q\, E\, \delta \phi^\ra \Big(
	\frac{1}{2} K_{\ra\rb} (\bar {\frak D}^2 - 4 \mu) (\veps \Omega^\rb)
	+ \frac{1}{2} K_{\rb} (\bar {\frak D}^2 - 4 \mu) (\veps \pa_\ra \Omega^\rb)
	\eol & \quad
	+ \frac{1}{2} g_{\ra \bar \rb} ({\frak D}^2 - 4 \bar\mu) (\veps \Omega^{\bar \rb})
	+ \frac{1}{2} ({\frak D}^2 - 4 \bar \mu) K_{\bar \rb} \veps \pa_\ra \Omega^{\bar \rb}
	\Big)
	~.
\end{align}
The vanishing of the terms with the highest number of derivatives indicates
that $\omega_{\ra\rb} := g_{\ra \bar \ra} \pa_\rb \Omega^{\bar \ra}$ is a covariantly constant two-form.
Without loss of generality, we can then choose $\Omega^\ra = \omega^{\ra\rb} \cK_\rb$
and one finds that
\begin{align}
\delta_\phi \delta S = \int \rd^3x\, \rd^4\q\, E\, \delta \phi^\ra \Big(
	\frac{1}{2} \tsD_\ra \tsD_\rb \cK \omega^\rb{}_{\bar \rb} \bar{\frak D}_\alpha \veps
	 \bar{\frak D}^\alpha \bar\phi^{\bar \rb}
	+ 2 \bar\mu \veps \omega_\ra{}^{\bar \rb} \cK_{\bar \rb}
	+ {\frak D}^\alpha \veps {\frak D}_\alpha \phi^\rb \omega_{\ra\rb}
	\Big)
	~.
\end{align}
Let us define $V^\ra = \dfrac{\mu}{2S} \omega^{\ra\rb} \cK_\rb$ for shorthand. Then we have
\begin{align}
\delta_\phi \delta S = \int \rd^3x\, \rd^4\q\, E\, \delta \phi^\ra \Big(
	-\frac{S}{\mu } \bar{\frak D}_\alpha \veps \bar{\frak D}^\alpha \bar\phi^{\bar \rb} \, \tsD_\ra 
	V_{\bar \rb} 
	+ 4 S\veps V_\ra
	+ {\frak D}^\alpha \veps {\frak D}_\alpha \phi^\rb \omega_{\ra\rb}
	\Big)
	~.
\end{align}
The third term is zero since we can rewrite
${\frak D}_\alpha \veps = \ri \sqrt{\bar\mu/\mu} \bar {\frak D}_\alpha \veps$ and then
integrate $\bar{\frak D}_\alpha$ by parts to give zero.

The vanishing of the remaining terms requires
\begin{align}
0 = -\frac{1}{4} (\bar{\frak D}^2 - 4 \mu) \Big(
	4 S\veps V_\ra
	-\frac{S}{\mu } \bar{\frak D}_\alpha \veps \bar{\frak D}^\alpha \bar\phi^{\bar \rb} \, \tsD_\ra 
	V_{\bar \rb}
	\Big)~.
\end{align}
Examining all the terms involving $\bar{\frak D}^2 \bar\phi^{\bar \rb}$, we find that
these arise from
\begin{align}
\frac{S}{4\mu } \bar {\frak D}^2 (\bar{\frak D}_\alpha \veps \bar{\frak D}^\alpha \bar\phi^{\bar \rb}) \,
 \tsD_\ra V_{\bar \rb} 
	- S \veps \bar{\frak D}^2 \bar\phi^{\bar \rb} \tsD_{\bar \rb} V_\ra
	&= - S  \veps \bar{\frak D}^2 \bar\phi^{\bar \rb}
		\, \tsD_\ra V_{\bar \rb} 
	- S \veps \bar{\frak D}^2 \bar\phi^{\bar \rb} \tsD_{\bar \rb} V_\ra
\end{align}
so we conclude that $V_\ra$ obeys $\tsD_\ra V_{\bar \rb} + \tsD_{\bar \rb} V_\ra = 0$.
Since it obeys $\tsD_\ra V_{\rb} + \tsD_{\rb} V_\ra = 0$ by construction,
$V^\ra$ must be a Killing vector.

This turns out to be the final condition we require to prove $\delta S = 0$.
The variation of the action can be written
\begin{align}
\delta S
	&= \int \rd^3x\, \rd^4\q\, E\, \Big(
	-\frac{1}{2} g_{\ra \bar \rb} \bar {\frak D}_\alpha \bar\phi^{\bar \rb} 
	\bar{\frak D}^\alpha \veps \omega^{\ra\rb} \cK_\rb
	+ \textrm{c.c.}
	\Big) \eol
	&= \int \rd^3x\, \rd^4\q\, E\, \Big(
	\ri {\frak D}_\alpha \veps \bar {\frak D}^\alpha \bar\phi^{\bar \rb} V_{\bar \rb}
	+ \textrm{c.c.}
	\Big) \eol
	&= \int \rd^3x\, \rd^4\q\, E\, \Big(
	-\ri \veps {\frak D}_\alpha \phi^\rb \bar {\frak D}^\alpha \bar\phi^{\bar \rb} \tsD_\rb V_{\bar \rb}
	+ \textrm{c.c.}
	\Big) = 0
\end{align}
using the Killing condition.

A straightforward analysis of closure of the algebra reveals that
$\omega^{\ra \rb} \omega_{\rb \rc} = -\delta^\ra_\rc$.

\subsection{(4,0) AdS supersymmetry in (2,0) AdS superspace}

Our ansatz for the second supersymmetry is
\begin{align}
\delta \phi^\ra = -\frac{1}{2} \bar \cD^2 (\bar\rho_\rL \Omega_\rL^\ra)
	- \frac{1}{2} \bar \cD^2 (\bar\rho_\rR \Omega_\rR^\ra)
\end{align}
where $\bar\rho_\rL$ and $\bar\rho_\rR$ are antichiral and obey
certain conditions discussed in section \ref{section11}.
Let's require as usual that $\delta_\phi \delta S = 0$. This is quite
similar to the (3,0) case we have already addressed. We find immediately that
$\omega_{\rL\ra\rb} := g_{\ra \bar \rc} \Omega_\rL^{\bar \rc}{}_{,\rb}$ and
$\omega_{\rL\ra\rb} := g_{\ra \bar \rc} \Omega_\rR^{\bar \rc}{}_{,\rb}$
must be antisymmetric and covariantly constant. This implies that
\begin{align}
\delta_\phi \delta S = \int \rd^3x\, \rd^4\q\, \delta \phi^\ra \Big(
	\frac{1}{2} \ve_\alpha \bar\cD^\alpha \bar\phi^{\bar \rb} \pa_\ra \Omega_{\rL \bar \rb}
	- 4S \bar \veps_\rL \Omega_{\rL\ra} - \bar \ve^\alpha \cD_\alpha \phi^\rb \omega_{\rL \ra\rb}
	+ \textrm{m.c.}
	\Big)
\end{align}
where m.c. denotes mirror conjugate. Now let us perform the $\bar\q$ integrals to give
\begin{align}\label{eq_general40}
\delta_\phi \delta S &= \int \rd^3x\, \rd^2\q\, \cE\, \delta \phi^\ra \Big(
	S \veps_\rL \bar \cD_\alpha (\bar\cD^\alpha \bar\phi^{\bar \rb} \pa_\ra \Omega_{\rL \bar \rb})
	+ 2 S \veps_\rR \bar \cD^\alpha (\cD_\alpha \phi^\rb \omega_{\rL \ra\rb})
	\eol & \quad
	+ 4\ri S (2S+X) \veps_\rR \Omega_{\rL\ra}
	- \ri (2S+X) \bar \ve_\beta \bar\cD^\beta \bar\phi^{\bar \rb} \pa_{\bar \rb} \Omega_{\rL \ra}
	+ S \bar \veps_\rL \bar \cD^2 \Omega_{\rL\ra}
	+ \textrm{m.c.}
	\Big)
	~.
\end{align}

Before analyzing this further, we need a result from analyzing closure of the
algebra. As we discussed in section \ref{section11.1}, the holomorphic two-forms
$\omega_{\rL \ra \rb}$ and $\omega_{\rR \ra \rb}$ must obey \eqref{eq_40omegas},
which implies the existence of covariantly constant projection operators
$(P_\rL)^\ra{}_\rb$ and $(P_\rR)^\ra{}_\rb$ \eqref{eq_40projs}.
The K\"ahler metric obeys
\begin{align}
g_{\ra \bar \rb} = (P_\rL)_{\ra \bar \rb} + (P_\rR)_{\ra \bar \rb} \equiv (g_\rL)_{\ra \bar \rb} 
+ (g_\rR)_{\ra \bar \rb}
\end{align}
with $g_{\rL \ra \bar \rb} = \pa_\ra \pa_{\bar \rb} K_\rL$ and 
$g_{\rR \ra \bar \rb} = \pa_\ra \pa_{\bar \rb} K_\rR$, with a K\"ahler potential given by
the sum of two decoupled sectors, $K = K_\rL + K_\rR$.

\subsubsection*{Deriving conditions for the non-critical case}

In the non-critical case, both $\veps_\rL$ and $\veps_\rR$ are non-zero,
so we must arrange for their coefficients in \eqref{eq_general40} to vanish.
This requires two conditions to be satisfied. The first condition,
\begin{align}
\pa_\ra \Omega_{\rL \bar \rb} = \pa_\ra \Omega_{\rR \bar \rb} = 0~,
\end{align}
along with $\nabla_{\ra} \Omega_{\rL \rb} = \omega_{\rL \rb\ra}$, implies that
\begin{align}
\Omega_\rL^\ra = \omega_\rL^{\ra \rb} \chi_\rb~, \qquad
\Omega_\rR^\ra = \omega_\rR^{\ra \rb} \chi_\rb~,
\end{align}
where $\chi^\ra$ is a homothetic conformal Killing vector. Hence, the target space
must be a hyperk\"ahler cone, which decomposes, due to the projection operators,
into a left cone and a right cone. The second condition
\begin{align}
(2S+X) \Omega_{\rL\ra} = 4 \ri S J^\rb \omega_{\rL \ra\rb}~, \qquad
(2S-X) \Omega_{\rR\ra} = 4 \ri S J^\rb \omega_{\rR \ra\rb}~,
\end{align}
implies that
\begin{align}\label{eq_Jdecomp}
J^\ra = \left(1+\frac{X}{2S}\right) J_\rL^\ra + \left(1-\frac{X}{2S}\right) J_\rR^\ra~,
\end{align}
where
\begin{align}
J_\rL^\ra = -\frac{\ri}{2} (P_\rL \chi)^\ra~, \qquad
J_\rR^\ra = - \frac{\ri}{2} (P_\rR \chi)^\ra~.
\end{align}

\subsubsection*{Deriving conditions for the critical case}
In the critical case $X = 2S$, we can consistently take $\veps_\rR = \bar \veps_\rR = 0$.
Then \eqref{eq_general40} leads to
\begin{align}
\delta_\phi \delta S &= \int \rd^3x\, \rd^2\q\, \cE\, \delta \phi^\ra \Big(
	S \veps_\rL \bar \cD_\alpha  
	(\bar\cD^\alpha \bar\phi^{\bar \rb} \pa_\ra \Omega_{\rL \bar \rb})
	+2 S\veps_\rL \bar \cD^\alpha (\cD_\alpha \phi^\rb \omega_{\rR \ra\rb})
	\eol & \quad
	- 4S \ri \bar \ve_\beta \bar\cD^\beta \bar\phi^{\bar \rb} \pa_{\bar \rb} \Omega_{\rL \ra}
	+ S \bar \veps_\rL \bar \cD^2 \Omega_{\rL\ra}
	\Big)~.
\end{align}
Invariance requires that $\pa_\ra \Omega_{\rL \bar \rb} = 0$. Introducing
\begin{align}
\chi_\rL^\ra = \omega_\rL^{\ra \rb} \Omega_{\rL \rb} = \omega_\rL{}^\ra{}_{\bar \rb} 
\Omega_\rL^{\bar \rb}~,
\end{align}
we find that
\begin{align}
\nabla_{\bar \rb} \chi_\rL^\ra = 0~, \qquad \nabla_{\rb} \chi_\rL^\ra = (P_\rL)^\ra{}_\rb~.
\end{align}
This means that the left sector is a hyperk\"ahler cone. We can define
its K\"ahler potential by
\begin{align}
K_\rL := \chi_\rL^\ra \chi_{\rL \ra}~, \qquad \chi_{\rL \ra} = \nabla_\ra K_\rL~.
\end{align}
The remaining variation of the action is
\begin{align}
\delta_\phi \delta S &= \int \rd^3x\, \rd^2\q\, \cE\, \delta \phi^\ra \Big(
	2 S \veps_\rL \bar \cD^\alpha (\cD_\alpha \phi^\rb \omega_{\rR \ra\rb})
	\Big)~.
\end{align}
This vanishes only if $\omega_{\rR \ra\rb} J^\rb = 0$, so we conclude that
\begin{align}
J^\ra = 2 J_\rL^\ra
\end{align}
where $J_\rL^\ra$ is some Killing vector in the left sector. The normalization is
chosen to match \eqref{eq_Jdecomp} and for later convenience.

Further information about the Killing vector $J_\rL^\ra$ can be gleaned by requiring
consistency of the extended supersymmetry transformation with the U(1) isometry.
In other words, we require
\begin{align}
\delta (\ri \cJ_\rL \phi^\ra) = \delta J_\rL^\ra
	= -\frac{1}{2} \bar\cD^2\Big(\bar\rho_\rL \Omega_\rL^\rb \pa_\rb J_\rL^\ra + \bar\rho_\rR 
	\Omega_\rR^\rb \pa_\rb J_\rL^\ra \Big)
\end{align}
to match
\begin{align}
\ri \cJ_\rL \delta \phi^\ra &= -\frac{\ri}{2} \cJ_\rL \bar \cD^2 (\bar \rho_\rL \Omega_\rL^\ra)
	-\frac{\ri}{2} \cJ_\rL \bar \cD^2 (\bar \rho_\rR \Omega_\rR^\ra) \eol
	&= \frac{\ri}{2} \bar \cD^2 (\bar \rho_\rL \Omega_\rL^\ra)
	-\frac{\ri}{2} \bar \cD^2 (\bar \rho_\rL \cJ_\rL \Omega_\rL^\ra)
	-\frac{\ri}{2} \bar \cD^2 (\bar \rho_\rR \cJ_\rL \Omega_\rR^\ra)~.
\end{align}
This forces
\begin{align}\label{eq_generalJL}
0 &= \frac{1}{2} \bar \cD^2 \Big(
	\bar \rho_\rL (J_\rL^\rb \nabla_\rb \Omega_\rL^\ra - \Omega_\rL^\rb \nabla_\rb J_\rL^\ra)
	-\ri \bar \rho_\rL \Omega_\rL^\ra
	+ \bar \rho_\rL J_\rL^{\bar \rb} (\omega_\rL)^\ra{}_{\bar \rb}
	\eol & \quad
	+ \bar \rho_\rR J_\rL^\rb \nabla_\rb \Omega_\rR^\ra
	+ \bar \rho_\rR J_\rL^{\bar \rb} (\omega_\rR)^\ra{}_{\bar \rb}
	- \bar\rho_\rR \Omega_\rR^\rb \nabla_\rb J_\rL^\ra \Big)~.
\end{align}
All the terms involving $\bar\rho_\rR$ cancel using the left
and right projection operators. (Here we take
$\Omega_\rR^\ra = \omega_\rR^{\ra \rb} K_{\rb}$ so that
$\Omega_\rR^\ra$ is a vector in the right sector.)
For the remaining terms, we use $\Omega_\rL^\ra = \omega_\rL{}^{\ra\rb} \chi_{\rL \rb}$
and find that a certain combination of terms must vanish,
\begin{align}
0 = -\ri \omega_\rL^{\ra\rb} \chi_{\rL \rb} - \omega_\rL^{\rb\rc} \chi_{\rL \rc} \nabla_\rb J_\rL^\ra
	+ (\omega_\rL)^\ra{}_{\bar \rb} J_\rL^{\bar \rb}~.
\end{align}
We can decompose $J_\rL^\ra$ as
\begin{align}
J_\rL^\ra = -\frac{\ri}{2} \chi_\rL^\ra + Z_\rL^\ra~.
\end{align}
Since $J_\rL^\ra$ is a Killing vector by assumption and $\chi_\rL^\ra$ is Killing by construction,
$Z_\rL^\ra$ must also be a Killing vector.
Using this expression, one can prove that
\begin{align}
\cL_{Z_\rL} \Omega_\rL^\ra = 0 \quad \implies \quad
\cL_{Z_\rL} (\omega_\rL)^\ra{}_{\bar \rb} = 0~,
\end{align}
and so $Z_\rL^\ra$ is a tri-holomorphic Killing vector.
Since $\Omega_\rL^\ra = \omega_\rL^{\ra\rb} \chi_{\rL \rb}$,
we find the additional condition that $Z_\rL^\ra$ commutes with the homothetic
conformal Killing vector.

All the relevant features of the target space geometries in the critical and non-critical
cases have now been determined.
The interested reader can straightforwardly check that the action is invariant,
$\delta S=0$, by applying techniques similar to those we have used elsewhere in this
appendix.

\section{$\cN=4$ $\to$ (1,1) AdS superspace reduction}
\label{AppD}

In subsection \ref{fromN4toN2} we studied the reduction of all $\cN=4$ AdS
superspaces to the (2,0) one. 
It was observed that the (4,0) case admits only the reduction to (2,0) AdS superspace.
The (3,1) and (2,2) superspaces also admit consistent reductions to (1,1) AdS.
Here we elaborate on the details of the (1,1) reduction procedures.
This analysis is parallel to that of the (2,0) reduction while some differences occur.

We start by noting that, for the (3,1) and (2,2) AdS supergeometries, the conditions
$\cS^{11\bau\bad}=\cS^{12\bau\bau}=\cS^{12\bau\bad}=0$
make the algebra (\ref{1111-1})--(\ref{1111-4})
isomorphic to that of the  (1,1) AdS superspace.
Let us now see how to project the (3,1) and  (2,2) AdS superspaces to (1,1).

The $\cN=2$ projection of a tensor field and of the $\cN=4$ covariant derivatives
is defined as in eqs. (\ref{N2red-1})--(\ref{N2red-3}).
Since the (3,1) and (2,2) derivatives $\big(\cD_a,\,\cD_\a^{1\bau},\,-\cD_\a^{2\bad} \big)$ 
form a closed  algebra, 
which is isomorphic to that of the covariant derivatives of the (1,1) AdS superspace,
one can use the freedom to perform general coordinate, local Lorentz and SU(2) transformations
to chose a gauge in which 
\bea
 \cD_\a^{\1\1} |=\sqrt{ -\ri \frac{\bar \m}{|\m|} } {\frak D}_\a
 ~, \qquad 
\cD_\a^{\2\2}|=-\sqrt{ \ri \frac{\m}{|\m|} } \bar {\frak D}_\a ~.
\eea 
Here ${\frak D}_A = ({\frak D}_a , {\frak D}_\a, \bar{\frak D}^\a)$
denote the covariant derivatives of the (1,1) AdS superspace, eq. \eqref{11AdSderivatives}.
They obey the (anti) commutation 
relations \eqref{11AdSsuperspace}.

Next consider a Killing vector field for one of the $\cN=4$ AdS superspaces, 
\bea
\x = \x^a \cD_a + \x^\a_{i\bai} \cD^{i\bai}_\a~.
\eea
We introduce the $\cN=2$ projections of the transformation parameters involved
\bsubeq
\bea
&
l^a:=\x^a|
~,~~~
l^\a:=\sqrt{ -\ri \dfrac{\bar \m}{|\m|} }\, \x^\a_{1\bar{1}}|~,~~~
\bar{l}^\a=\sqrt{ \ri \dfrac{\m}{|\m|} } \,\x^\a_{2\bar{2}}|~,~~~
\l^{ab}:= \L^{ab}|~;
\label{N2proj-11-1}
\\
&
\ve^\a:=-\sqrt{ \ri \dfrac{\m}{|\m|} }\,\x^\a_{1\bad}|~,~~~~
\bar{\ve}^\a:=\sqrt{ -\ri \dfrac{\bar \m}{|\m|} }\,\x^\a_{2\bau}|~;
\label{N2proj-11-2_0}
\\
&
\ve_\rL=-\dfrac{1}{4S}\L^{22}|
~,~~
T_\rL:=-\ri\L^{12}|~,~~~~
\ve_\rR=-\dfrac{1}{4S}\L^{\bad\bad}|~,~~
T_\rR:=-\ri\L^{\bau\bad}|
~.~~~~~~~~~
\label{N2proj-11-2}
\eea
\esubeq

The parameters $(l^a, \, l^\a,\,\bar{l}_\a,\, \l^{ab}  )$
describe the infinitesimal isometries of  the (1,1) AdS superspace.
This can be easily proven  by $\cN=2$ projection of the equations (\ref{sK-1})--(\ref{sK-2-4}).
The parameters $(\ve^\a,\,\bar{\ve}_\a, \, \ve_\rL,\bar{\ve}_\rL,\,\ve_\rR,\bar{\ve}_\rR,\,T_\rL,\,T_\rR)$
describe the extra supersymmetry and $R$-symmetry transformations 
of either the (3,1) or (2,2) superspace.
The (1,1) projection of the relations 
(\ref{sK-1})--(\ref{sK-2-4}) and \eqref{7.14c0}--\eqref{7.14c}
gives certain constraints on the parameters in (\ref{N2proj-11-2_0})--(\ref{N2proj-11-2}).
Such constraints are different in the (3,1) and (2,2) cases as we are going to describe now.

\paragraph{(3,1)$\,\to\,$(1,1)}

For the reduction from (3,1) to (1,1) AdS superspace, 
we can always choose $w^{i\bai}$ to be
\bea
w^{1\bad}=w^{2\bau}=0~,~~~~~~
\mu:=\ri S(w^{2\bad})^2~,~~~
\mub=-\ri S(w^{1\bau})^2~.
\eea
The constraint on the (3,1) structure group, eq. \eqref{7.14c0}, implies that 
\bea
\ve:=\ve_\rL
~,~~~
\ve_\rR=-\frac{\ri\mu}{|\mu|}\,\bar{\ve}~,~~~
T_\rL=T_\rR:=T~.
\label{3111uuu}
\eea
In the (3,1) case, the (1,1) projection of eqs. \ref{sK-1})--(\ref{sK-2-4}) gives
\bsubeq
\bea
&&
\bar{\frak D}_\a\ve_\b=
-4|\mu|\ve_{\a\b}\,\ve
~,~~~
{\frak D}_\a\ve_\b
=4|\mu|\ve_{\a\b}\,\ve~,
\label{extra3111-1}
\\
&&
{\frak D}_\a\ve=
\frac{\ri}{2} \ve_{\a}
~,~~
\bar{\frak D}_\a\ve=-\frac{\ri}{2}\ve_\a
~,~~~~
{\frak D}_\a T=\bar{\frak D}_\a T=0
~.
\label{extra3111-2}
\eea
\esubeq
These imply the  conditions 
\bea
{\frak D}_\a\Big(\ve_\b-\bar{\ve}_\b\Big)=
\bar{\frak D}_\a\Big(\ve_\b-\bar{\ve}_\b\Big)=0
~.
\label{extra3111-1b}
\eea

\paragraph{(2,2)$\,\to\,$(1,1)}

For the reduction from (2,2) to (1,1) superspace 
we can always choose the constant parameters $l^{ij}$ and $r^{\bai\baj}$ to be
\bea
&l^{12}=r^{\bau\bad}=0~,~~~
l:=l^{11}~,~~
r:=r^{\bau\bau}
~,~~
|l|=|r|=1
~,~~~~
\mu:=-\ri S\bar{l}\bar{r}~,~~
\mub=\ri Slr
~.~~~~~~
\eea
The constraint on the (2,2) structure group, eq. \eqref{7.14c}, implies that 
\bea
\ve_\rL=\bar{l}{\bm\ve}_\rL~,~~~({\bm\ve}_\rL)^*={\bm\ve}_\rL~,~~~
\ve_\rR=\bar{r}{\bm\ve}_\rR~,~~~({\bm\ve}_\rR)^*={\bm\ve}_\rR~,~~~
T_\rL=T_\rR=0
~.
\label{2211uuu}
\eea
In the (2,2) case, the (1,1) projection of eqs. (\ref{sK-1})--(\ref{sK-2-4}) gives
\begin{subequations}
\bea
&
{\frak D}_\a\bar{\ve}_\b=
-4\ve_{\a\b}|\mu|l{\bm\ve}_\rL
~,~~~~~~
{\frak D}_\a\ve_\b
=-4\ve_{\a\b}|\mu|\bar{l}{\bm\ve}_\rR
~,
\label{extra2211-1}
\\
&
{\frak D}_\a{\bm\ve}_\rL=
\dfrac{\ri}{2} \bar{l} \,\bar{\ve}_{\a}~,~~~~~~
{\frak D}_\a{\bm\ve}_\rR=
\dfrac{\ri}{2} l\,\ve_{\a}
~.
\label{extra2211-2}
\eea
\end{subequations}


\begin{footnotesize}

\end{footnotesize}

\end{document}